%% file: sup2.tex
\documentclass[12pt]{book}
\usepackage{makeidx,tsukuba}%
\usepackage[dvips]{graphicx}
\usepackage[usenames,dvipsnames]{color}
\makeauthorindex
\makeindex
\printAuthorIndex

\title{H.E.S.S. contributions  to the \\
\textbf{28th International Cosmic Ray Conference}\\
 Tsukuba, Japan}
\date{July 31 - August 7, 2003}
\begin{document}
\maketitle
\BookTitle{\itshape H.E.S.S. contributions to the 28th International Cosmic Ray Conference}
\parbox{\linewidth}{\rule{0pt}{29cm}}
\input{toc}
\parbox{\linewidth}{\rule{0pt}{29cm}}
{\input{008827_2}}
{\input{009395-1}}
{\input{008971-1}}
{\input{009431}}
{\input{009383-1}}
{\input{009405-1}}
{\input{009303-1}}
{\input{007412-1}}
{\input{008834-1}}
{\input{008971-3}}
{\input{008564-1}}
{\input{008546-2}}
{\input{008949-1}}
{\input{008971-2}}
\end{document}

%% file: toc.tex
{\Huge \bf Contents}\\
\begin{enumerate}
\bf
\item Status of the H.E.S.S. Project \hfill 5
\item Performance of the H.E.S.S. cameras. \hfill 9
\item Observation Of Galactic TeV Gamma Ray Sources \\ With H.E.S.S. \hfill 13 
\item First Results from Southern Hemisphere AGN Observations \\ Obtained with the {H$\cdot $E$\cdot $S$\cdot $S$\cdot $} VHE Gamma-ray Telescopes \hfill 17
\item Study of the Performance of a Single Stand-Alone H.E.S.S. Telescope: Monte Carlo Simulations and Data \hfill 21
\item Application of an analysis method based on a semi-analytical \\ shower model to the first H$\cdot $E$\cdot $S$\cdot $S$\cdot $ telescope. \hfill 25
\item The Central Data Acquisition System of the \\ H.E.S.S.\ Telescope System \hfill 29
\item Mirror alignment and performance of the optical system of the H.E.S.S. imaging atmospheric Cherenkov telescopes \hfill 33
\item Calibration results for the first two H$\cdot $E$\cdot $S$\cdot $S$\cdot$ array telescopes. \hfill 37
\item Arcsecond Level Pointing Of The H.E.S.S. Telescopes \hfill 41
\item A Novel Alternative to UV-Lasers Used in Flat-fielding \\ VHE {\bf $\gamma $}-ray Telescopes \hfill 45
\item Atmospheric Monitoring For The H.E.S.S. Project \hfill 49
\item Implications of LIDAR Observations at the H.E.S.S. Site in\\ Namibia for Energy Calibration  \hfill 53
\item  Optical Observations of the Crab Pulsar using the first H.E.S.S. Cherenkov Telescope \hfill 57
\end{enumerate}

%% file: 008827_2.tex
%
%
%
%

%




\chapter{Status of the  H.E.S.S. Project}

\author{Werner Hofmann, for the H.E.S.S. collaboration\\
{\it Max-Planck-Institut f\"ur Kernphysik, D 69029 Heidelberg, P.O. Box 103989}
}

\section*{Abstract}
H.E.S.S. - the High Energy Stereoscopic System - is a system of four large imaging
Cherenkov telescopes under construction in the Khomas Highland of Namibia, 
at an altitude of 1800~m. With their stereoscopic reconstruction of air showers,
the H.E.S.S. telescopes provide very good angular resolution and background rejection,
resulting in a sensitivity in the 10 mCrab range, and an energy threshold around 100~GeV.
The H.E.S.S. experiment aims to provide precise spectral and spatial mapping in
particular of extended sources of VHE gamma rays, such as Supernova remnants.
The first two
telescopes are operational and first results are reported; the next two
telescopes will be commissioned until early 2004. 

\section{The H.E.S.S. Telescopes}

The H.E.S.S. Cherenkov telescopes are characterized by a mirror area of
slightly over 100\,m$^2$, with a focal length of 15\,m, and use cameras 
with fine pixels of $0.16^\circ$ size and a large field of view of $5^\circ$.

Construction of telescopes is well underway; 
the steel structures of all four telescopes
have been erected and equipped with drive systems; 
two telescopes are fully equipped with mirrors and cameras and take data since
June 2002 and March 2003, respectively. The final two telescopes will
be commissioned early in 2004; all parts, such as mirrors, phototubes etc.
are in hand, and the cameras are under assembly in France. The site
infrastructure is complete and includes a building with the experiment control room, 
offices, and workshops, a residence building, Diesel power generators and a 
Microwave tower linking the site to Windhoek and from there to the internet.

The H.E.S.S. telescopes use an alt-az mount, which 
rotates on a 15~m diameter rail.
The steel structures are designed for high
mechanical rigidity.  Both
 azimuth and elevation are driven by friction drives acting on auxiliary
drive rails, providing a positioning speed of $100^\circ$/min.
Encoders on both axes give 10'' digital resolution; with the additional analogue
encoder outputs, the resolution is improved by another factor 2 to 3.
After initial tests and
a few months of operation of the first telescope, the drives were slightly modified
for smoother operation; the telescope design is now quite mature. 

\begin{figure}[t]
  \begin{center}
    \includegraphics[height=12pc]{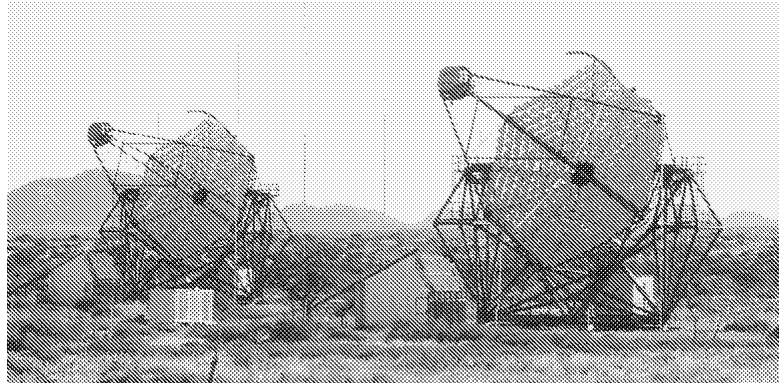}
  \end{center}
  \vspace{-0.5pc}
\caption{The first two H.E.S.S. telescopes, ready to take data.}
\label{f1}
\end{figure}

The mirror of a H.E.S.S. telescope is composed of 380 round facets of 60~cm
diameter; the facets are made of ground glass, aluminized and quartz coated,
with reflectivities in the 80\% to 90\% range.
The facets are arranged in a Davies-Cotton fashion, forming a dish with 107~m$^2$
mirror area, 15~m focal length and $f/d \approx 1.2$.
To allow remote alignment of the mirrors, each mirror is equipped with two
alignment motors with internal resolvers. The alignment procedure uses the image
of a star on the closed lid of the PMT camera, viewed by a CCD camera at the 
center of the dish. The procedure and the resulting point spread function are described 
in detail elsewhere in these proceedings. Due to the superior quality of both the
mirrors and the alignment system, the on-axis point spread function is significantly
better than initially specified. The imaging quality is stable over the elevation 
range from $30^\circ$ to the Zenith. The point spread function varies with 
distance $\theta$ (in degr.) to the optical axis 
as $r_{80} = (0.42^2 + (0.71 \theta)^2)^{1/2}~~~\mbox{[mrad]}$; $r_{80}$ is the
circle containing 80\% of the light of a point source at the height of the
shower maximum. Over most of the field of view, light is well contained within a single
pixel.

Telescope pointing was verified using the images of stars on the camera lid. Without any corrections,
star images were centred on the camera lid with a {\small rms} error of 28''. Using
a 12-parameter model to correct for misalignments of the telescopes axes etc.,
a pointing precision of 8'' {\small rms} is reached. Finally, using a guide telescope
attached to the dish for further corrections, the pointing can be good to 2.5'' {\small rms}. 
H.E.S.S. should therefore be able to locate gamma ray sources to
a few arc-seconds.

The PMT cameras of the H.E.S.S. telescopes provide $0.16^\circ$ pixel size over
a $5^\circ$ field of view, requiring 960 PMT pixels per telescope. The complete
electronics for signal processing, triggering, and digitization is contained in the 
camera body; only a power cable and a few optical fibers connect the camera.
For ease of maintenance, the camera features a very modular construction. 
Groups of 16 PMTs together with the associated electronics form so-called
``drawer''  modules, 60 of which are inserted from the front into the camera body,
and have backplane connectors for power, a readout bus, and trigger lines.
The rear section of the camera contains crates with a PCI bus for readout,
a custom crate for the final stages of the trigger, and the power supplies.
%
%
The camera uses Photonis XP2960 PMTs, operated at a gain of 
$2\times 10^5$. The PMTs are individually equipped with DC-DC converters
to supply a regulated high voltage to the dynodes; for best linearity, the four last dynodes
are actively stabilized.

The key element in the signal recording of the H.E.S.S. cameras is the ARS
(Analogue Ring Sampler) ASIC, which samples the PMT signals at 1 GHz and 
provides analogue storage for 128 samples, essentially serving to delay the
signal until a trigger decision is reached. To provide a large linear dynamic range 
in excess of $10^4$ up to 1600 photoelectrons, 
two parallel high/low gain channels are used for each PMT.
A camera trigger is formed by a coincidence of some number  of
pixels (typically 3-5) within an $8\times 8$ pixel group exceeding an adjustable threshold; typical
operating thresholds are in the range of 3 to 5 photoelectrons. The pixel comparators
generate a pixel trigger signal; the length of the signal reflects the time the 
input signal exceeds the threshold. Since typical noise signals barely exceed the
threshold and result in short pixel trigger signals, the effective resolving time
of the pixel coincidence is in the 1.5 to 2~ns range, providing a high suppression
of random coincidences. At the time of this writing, the two telescopes are
triggered indepedently, and stereo images are combined offline using
GPS time stamps. A central trigger processor controlling 
electronic delays and coincidence logic will soon be installed.
This will allow to impose
arbitrary telescope configurations in the trigger, and to operate the telescopes
either as a single four-telescope system, or as subsystems, up to
four individual telescopes pointed at different objects.

A number of auxiliary instruments serve to monitor telescope performance and
atmospheric quality. These include laser and LED pulsers at the center of a
dish for flatfielding, and infrared radiometers and a lidar system to detect
clouds and characterize aerosol scattering. Details are given elsewhere in these
proceedings.

\section{First data}

After the first telescope was equipped with mirrors in autumn of 2001, the 
camera was installed in May 2002 and first data were taken in June 2002.
As expected for a single telescope, a significant fraction -
roughly half - of the images are caused by muons, either in the form of 
full rings or of short ring segments.

The night-sky background - predicted to be about 100 MHz photoelectron rate per pixel - induces
a noise of 1.2 to 1.5 photoelectrons {\small rms} in the PMT pixels, consistent with
expectations. 

Muon rings are used to verify the overall performance and calibration
of the telescopes. Rings are classified according to their radius -
related to the muon energy - and by the impact parameter, which governs
the intensity distribution along the ring. The observed photoelectron yield agrees
to better than 15\% with expectations, indicating 
that the optical system, the PMTs and electronics calibration are quite
well understood. This tool can be used to monitor the evolution of the
detectors, as explained in greater detail in an accompanying paper.

Another important check for the performance of  the telescope is the trigger rate. 
The rate varies smoothly
with threshold. Even for thresholds as low as four
photoelectrons, trigger rates are governed by air showers rather than by
night-sky noise, which would induce a much faster variation with threshold. 
With a typical threshold of 4-5 photoelectrons,
event size distributions peak around 100 to 150 photoelectrons; 
for the H.E.S.S. telescopes one photoelectron
corresponds approximately to one GeV deposited energy.

Objects observed so far include SN 1006, RXJ 1713-3946, PSR B1706-44,
the Crab Nebula, and NGC 253, PKS 2005-489, PKS 2155-304 as extragalactic
source candidates. Clear signals are detected from the Crab Nebula and
for PKS 2155, confirming the earlier detection by
the Durham telescopes. The spectral shape of the Crab data agrees well
with other measurements; for PKS 2155, a slightly
steeper spectrum is measured. Details are given in other contributions to this
conference.
First stereo data were collected in March 2003, using the first two
telescopes with an offline selection of concident events. 
As expected, muons rings
were found to be absent in coincident events. A parallel trigger
to retain some muon events when stereo coincidence operation begins
is under consideration.

\section{Conclusion}

The first two H.E.S.S. telescopes are operational since June 2002 and
March 2003, respectively, and first results concerning
the technical performance of the telescopes, both for the optical system and
the camera, look encouraging and did not expose major problems. Current schedules
call for completion of the Phase-I four-telescope system in 2004. An expansion of the
system - Phase II - with increased sensitivity is foreseen; the Phase II telescopes and
their arrangement are under study.

\section*{Acknowledgement}

Construction and operation of the H.E.S.S. telescopes
is supported by the German Ministry for Education and Research BMBF.

\endofpaper

%% file: 009395-1.tex
%
%
%
%






\chapter{
Performance of the H.E.S.S. cameras.
}

\author{%
%
%
P.~Vincent$^1$,
J.-P.~Denance$^1$, J.-F.~Huppert$^1$, P.~Manigot$^2$, M.~de~Naurois$^2$, P.~Nayman$^1$, J.-P.~Tavernet$^1$, F.~Toussenel$^1$,
L.-M.~Chounet$^2$, B.~Degrange$^2$, P.~Espigat$^3$, G.~Fontaine$^2$, J.~Guy$^1$, G.~Hermann$^4$, A.~Kohnle$^4$,
C.~Masterson$^4$, M.~Punch$^3$, M.~Rivoal$^1$, L.~Rolland$^1$, T.~Saitoh$^4$,
for the H.E.S.S. collaboration.
\\
{\it 
(1) LPNHE, IN2P3/CNRS Universit\'es Paris VI \& VII, Paris, France\\
(2) LLR, IN2P3/CNRS Ecole Polytechnique, Palaiseau, France\\
(3) PCC, IN2P3/CNRS College de France, Paris, France\\
(4) Max Planck Institut fuer Kernphysik, Heidelberg, Germany} \\
}

\section{Introduction}

The H.E.S.S. experiment is a new generation ground-based atmospheric Che\-ren\-kov detector.
The first phase of this experiment consists of a square array of four telescopes with 120-metre spacing.
Each telescope, equipped with a mirror of 107 m$^2$, has a focal plane at 15 metres where a camera is installed.
Each camera consists of 960 photo-multipliers (PMs), providing a total field of view of 5$^\circ$, with the 
complete acquisition system (analogue to digital conversion, read-out, fast trigger, on-board acquisition)
being contained in the camera. The cameras of the H.E.S.S. telescopes are
currently being installed in the Khomas highlands, Namibia.  The first
telescope has been taking data since June, 2002 and the second since February,
2003. Stereoscopic coincient trigger mode should begin in June, 2003 and full Phase I
operation should be underway early in 2004. 
The performance of the cameras will be presented and their characteristics as measured during data 
taking will be compared with those obtained during the construction phase 
using a test bench. Future upgrades based on experience operating the two first cameras are also discussed.

\section{The cameras of the H.E.S.S. telescopes.}

\begin{figure}[t]
  \begin{center}
    \includegraphics[height=15.pc]{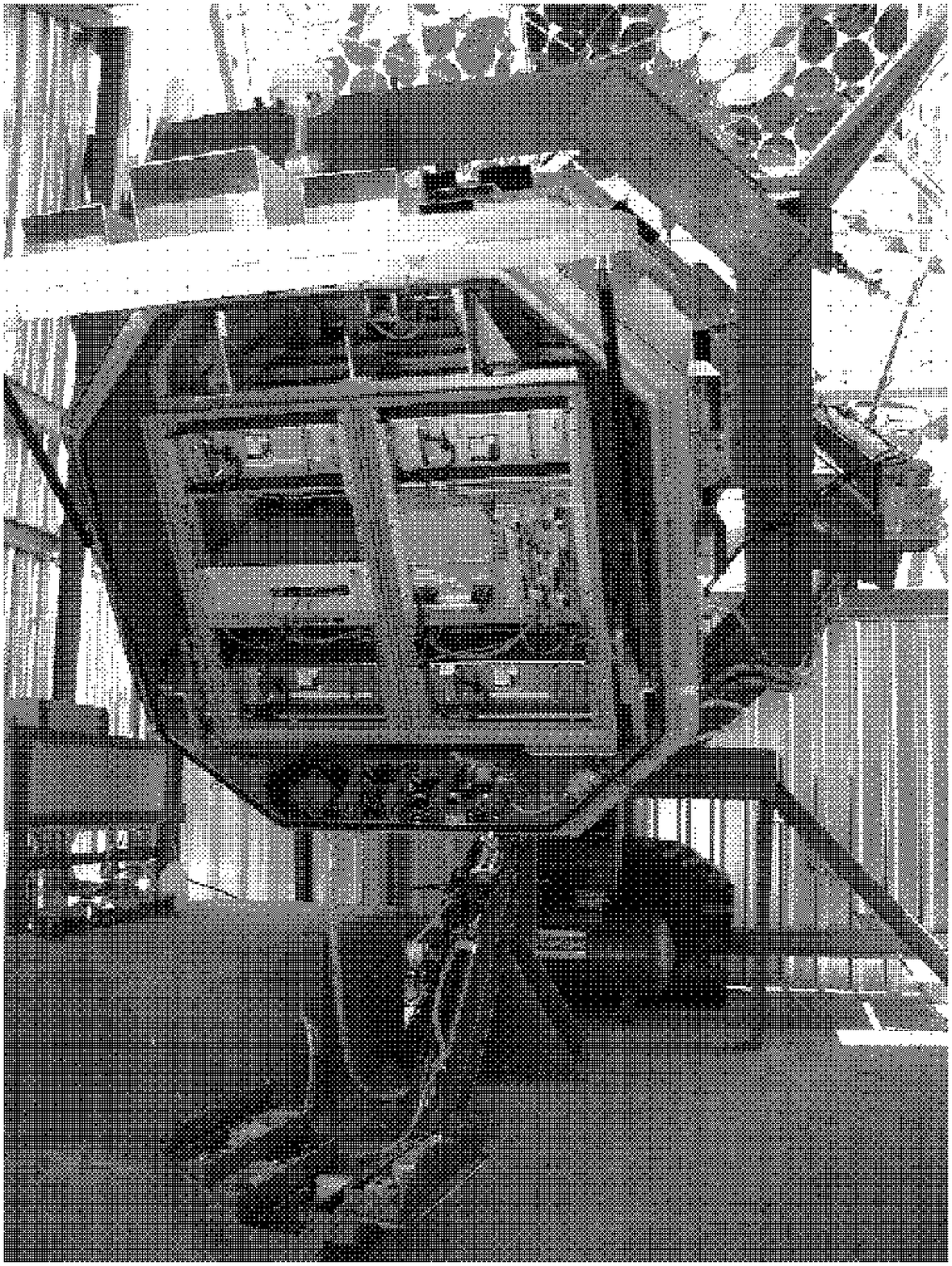}
    \includegraphics[height=15.pc]{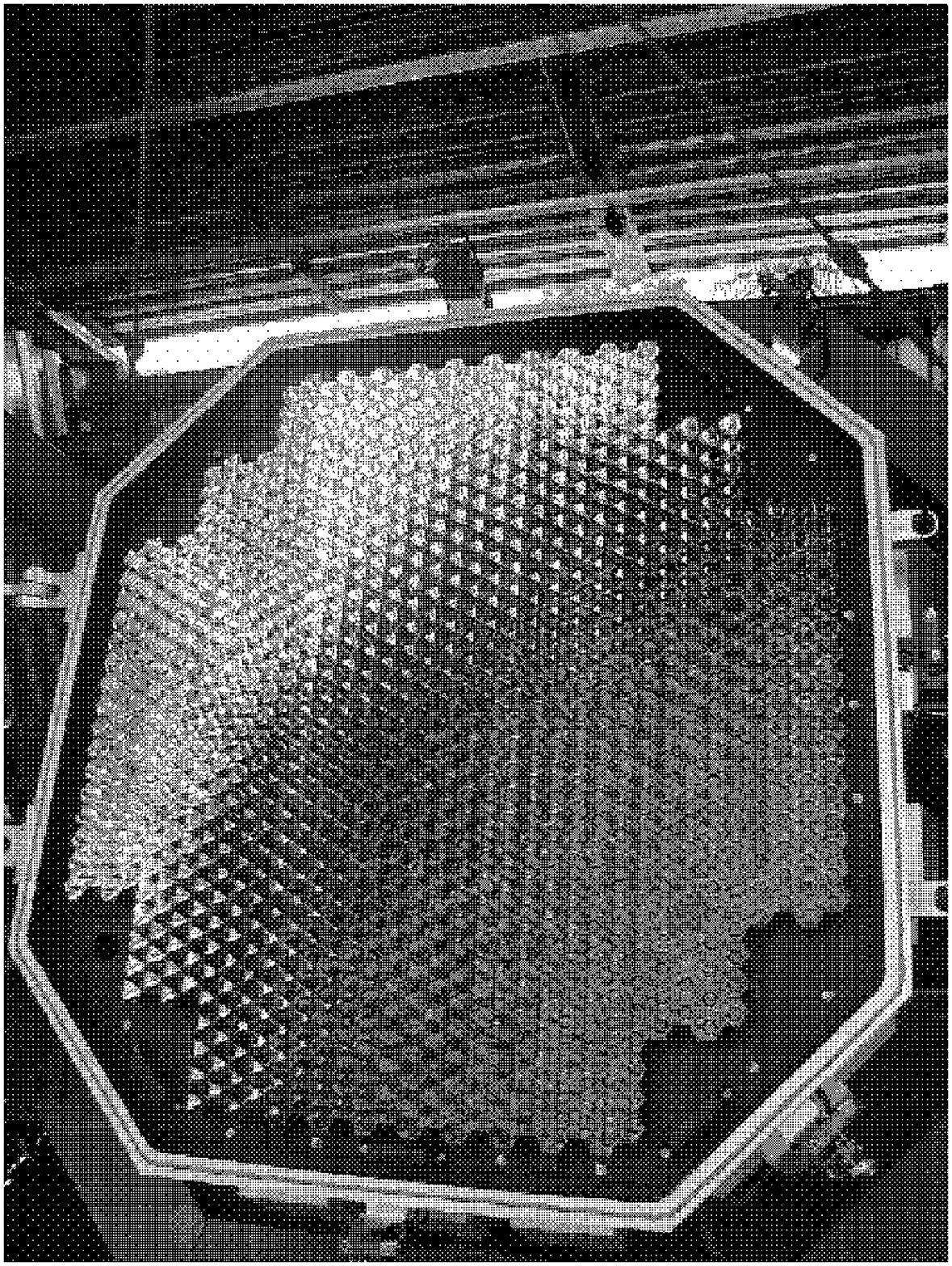}
    \includegraphics[height=15.pc]{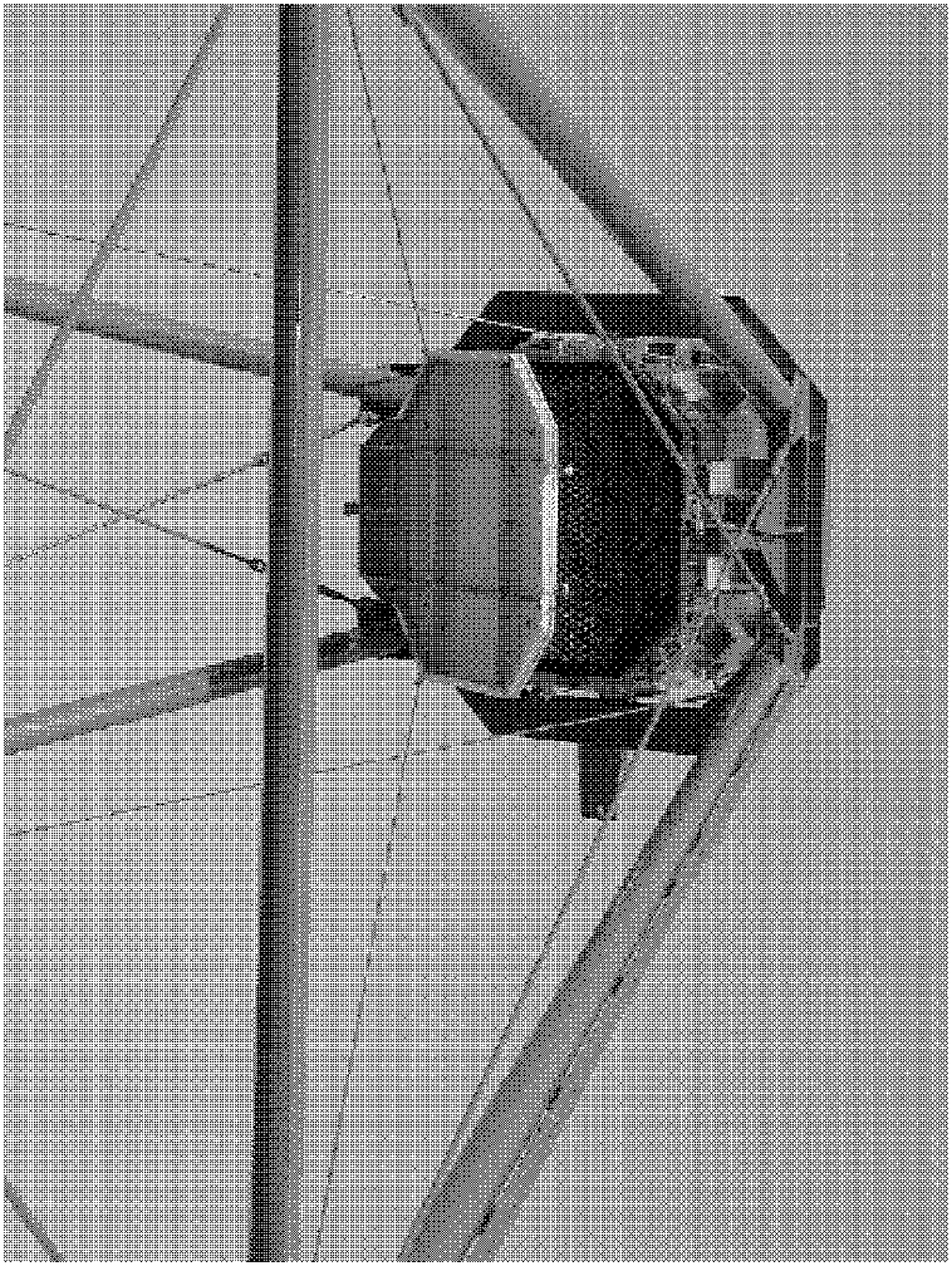}
  \end{center}
  \vspace{-0.5pc}
  \caption{View of the second H.E.S.S. camera. The picture on the left shows the rear with the four power
	supplies, the mixed crate incorporating the custom bus and compact PCI bus, and the network interface. 
	The central picture shows
	the front side of the camera with lid open showing the 960 Winston cones. 
	On the right we see the camera on the telescope.}
\end{figure}

A camera is approximately octagonal, fitting in a cylinder 2 metres in length and 1.6 metres in diameter, and weighing  
about 900~kg (see Fig. 1).
The front part contains 60 interchangable modules (``drawers'') with 16 PMs each, lodged in a ``pigeon-hole'' plate. 
The drawers are held by only two screws and can be easily extracted from the body of the camera to be replaced by a new drawer.
Each drawer communicates with the rest of the electronics through three connectors at the rear, which plug in automatically
when drawers are installed. This conception allows an easy access for tests and repairs of the camera electronics.  
In front of the drawers, a plate in three sections holds individual Winston cones for each PM which concentrate the 
Cherenkov light in the central region of the photo-cathode where the quantum efficiency is at a maximum of about 30\%. 
These cones allow the collection of about 75\% of the photons reflected from the mirror. 
They also considerably reduce the background contribution from albedo
by limiting the PM's field of view to the angular size of the mirror.
In the rear of the camera, an electronics rack is equipped with four power-supply crates, the camera acquisition and control systems,
and the network interface.
This rack can be slid out on rails from the camera body to access the cables and connectors between front and back side of the camera. 
Lastly, only three cables come from the camera to the ground: a copper cable for the current, one optical fibre for the 
network, and an other fibre for communications with central trigger.

\section{The electronics.}

The electronics of the camera consist of a front-end contained in the drawers which
includes the readout and first-level trigger and a second section with the local acquisition system mentioned above. 
Drawers contain 16 PMs, each powered by an active base. These bases provide a high voltage of more than a thousand volts 
calibrated to generate a signal of $2\times10^5$ electrons for each photon converted at the photo-cathode. The PMs use
a borosilicate window and provide a 20--30\% quantum efficiency in the wavelength range 300--700 nm.

The readout channel takes advantage of analogue memories ARS0 (``Analog Ring Sampling memory'') developed for the 
A{\small NTARES} experiment by the CEA/{\small DAPNIA}-SEI. 
These memories sample the signal at 1~GHz and store it in 128 cells while awaiting the trigger decision. 
The pulse from each PM is divided between two channels with different amplification factors. 
A high-gain channel, for low signal amplitudes, gives a dynamic range from 
1 to about 100 photo-electrons. Before this upper limit, at around 16 photo-electrons, the low gain channel can measure a signal
up to 1600 photo-electrons. The overlapping region allows inter-calibration between both channels. The signal from a
triggered event is read from the analogue memory in a window of 16 samples and then digitized with a 12-bit ADC and stored in an 
FPGA chip. The samples can be saved for an analysis of the pulse shape or integrated directly in the FPGA so as to transmit and
save only the total charge in a pixel. The readout-window size is a programmable parameter that can be changed as a function of 
future studies.

The local camera trigger is based on two parameters: the number of photons arriving in a pixel and the identification of a 
concentration of signal in a part of the camera. 
To construct the latter criterion the camera has been divided in 38 sectors of 64 PMs
with logic on cards contained in the rear crate. 
Sectors overlap with their neighbours to prevent local inhomogeneities which would result from a shower image 
arriving in the boundary between two sectors. 
The time needed to build the trigger signal is about 70~ns, which is fast enough permit reading of the signal stored in the ARS0. 
A card dedicated to the trigger management (``GesTrig'') sends a signal through two fanout cards to the 480 analogue memories. 
The memories stop acquiring data, a programmable pointer identifies the region of interest given the trigger-signal formation
time, and the readout of the data starts.
The time until the converted signals are ready in the drawers' FPGAs is measured to be 270$\:\mu$s after the shower's arrival, 
and is remarkably stable.

The interface with central trigger of the multi-telescope system is performed via a local module embedded in the camera. 
This central trigger interface
is connected to the GesTrig trigger manager card and informs the central trigger of the current status
of the camera with a ``busy'' signal. If there is no coincidence with other telescopes, the central trigger returns a ``fast clear'' 
within a couple of $\:\mu$s, which is sent to the GesTrig and thence to the drawers to stop the readout of the analogue memories 
and to reset the drawers.

\section{Data acquisition architecture and performance}

The acquisition system is based on the use of the new Compact-PCI (cPCI) norm that allows 64-bit word transfer at 33 MHz.
A second bus (CustomBUS) within the data acquisition crate is dedicated to the configuration of the sectorization of the trigger.
The drawers are connected to the acquisition by 4 final buses (Box-Bus), and when an event is available for transfer they send a 
request to a card holding FIFO memories (FIFO-card) located on the cPCI bus. 
This card plays the r\^ole of master and controls the transactions on the 4 buses by sending acknowledges to slaves (drawers). 
All buses accept asynchronous transfer, and the full data transfer from the 15 drawers present on each bus is completed after 
$340\:\mu$s. This year, the RIOC-4065 processor from CES has been installed on the first two cameras. This new processor is able 
to perform direct access between a card in the cPCI bus and its own memory and so improve the performance of the acquisition.
The FIFO memories are read-out through the cPCI bus by a DMA chip that transfers the full camera's data in less than 
$140\:\mu$s.
This last time, together with the bus transfer time and the ARS conversion time of the previous section, defines the dead-time
of the acquisition. As the readout time of the FIFO memory is lower than the other times, this task can be parallelized
so that the dead-time for a camera is $610\:\mu$s, corresponding to a maximum acquisition rate of 1.6~kHz which represents
an improvement of a factor of three from the initial camera's performance. 
Under these conditions, at a typical trigger configuration 
of 4 pixels at 5 photo-electrons, the observed counting rate is 250\,Hz. This rate gives a dead-time of about 14\%, 
compared to 30\% with the initial camera.

The data acquisition system (DAS) in the camera is build around the Linux operating system and written in C. This system
controls the behaviour of the overall camera: one card controls the camera lid operation, the 95 fans and 16 temperature sensors;
a GPS card for time stamps receives interruptions from the GesTrig trigger card;
a CAN bus interface controls the four power-supply crates; an I/O card manages the trigger;
a mezzanine card located on the CustomBUS 
receives some serial event data the from central trigger interface (event number \ldots). 
Data are transfered from the local CPU to the central DAS via a 100 Mbits/s network for conversion and storage in ROOT format. 

\section{Conclusion.}

Since the installation of the first camera ``prototype'' several upgrades have been carried out. Experimentally we had strong 
indications that
noise from the switching power-supplies considerably disturbed the data ``traffic'' on the BoxBus. The buses on new camera have 
been modified to remedy this problem and the first prototype has been upgraded accordingly.  
Other upgrades are under study to further increase the acquisition speed, e.g., 
to double the number of Box-buses to gain a further factor two in the data transfer.  
Finally, some tests on DMA transfer may allow the FIFO memory readout time to be reduced to $60\:\mu$s. 
This would not decrease dead time but would leave the CPU free to perform, for example, other monitoring tasks or data compression
(zero-suppression).
In conclusion, the performance of the first two cameras for the Phase I of H.E.S.S. is promising and, and are undergoing continual 
upgrades in order to optimize performance.

\endofpaper

%% file: 008971-1.tex
%
%
%

%
\chapter{Observation Of Galactic TeV Gamma Ray Sources With H.E.S.S.}

\author{Conor Masterson$^1$, for the H.E.S.S. Collaboration$^2$\\
  (1) {\it MPI fuer Kernphysik, P.O. Box 103980, D-69029 Heidelberg, Germany} \\ 
  (2) {\tt http://www.mpi-hd.mpg.de/HESS/collaboration} \\
}

\section*{Abstract}
The first telescope of the H\textsc{.E.S.S.}\ stereoscopic
{Cherenkov\ telescope system started operation in summer 2002. In
  spring 2003 a second telescope was added, allowing stereoscopic
  observations. A number of known or potential TeV gamma-ray emitters
  in the southern sky were observed.  Data on the Crab nebula taken at
  large zenith angles show a clear signal and serve to verify the
  performance and calibration of the instrument. Observations of other
  Galactic sources are also summarized.

\section{Introduction}
The H\textsc{.E.S.S.}\ experiment commenced operations on-site in
Namibia in June, 2002, with the first of four Cherenkov telescopes.
With its high resolution camera (0.16$^\circ$\ pixel size) and large
mirror area (107 m$^2$) the single H\textsc{.E.S.S.}\ telescope is a
sensitive instrument in its own right, comparing favourably with
existing detectors. The large field of view of the detector,
($\approx5^\circ$) makes it a good choice for observations of extended
galactic objects. Observations were made of a number of candidate
$\gamma$-ray\ sources with the single telescope, pending the
installation of the rest of the array. These sources included the Crab
nebula, an established TeV source, as well as a number of other
Galactic sources.

The Crab nebula was discovered at TeV energies in 1989 [6] and is
conventionally used as a standard reference source of TeV
$\gamma$-rays, due to its relative stability and high flux. It was
observed with the first telescope in October and November 2002 for a
total of 4.65 hours (live-time). Due to the latitude of the
H\textsc{.E.S.S.}\ experiment (21$^\circ$\ South), observations were
taken over a zenith angle range of 45$^\circ$\ to 50$^\circ$.

\section{Analysis of Data}

Since the data reported in this paper were taken with the first
H\textsc{.E.S.S.}\ telescope operating in single telescope mode, a
standard analysis of type Supercuts [5] was applied in order to
extract a $\gamma$-ray\ signal. This uses simple selection criteria
based on parameters calculated from the moments of the Cherenkov 
images.
  
Data were taken in {\small ON-OFF} observation mode, with 25-minute
observations of the source accompanied by similar observations of a
control region offset by 30 minutes in Right Ascension from the
source. In order to calibrate the system, a number of artificial light
sources are used, including an array of light emitting diodes on the
inside of the camera lid. These LEDs are used to measure the single
photo-electron gain of the system. Also, a laser mounted on the dish
allows flat-fielding of the camera [3].
  
Images were cleaned using a two-step technique, requiring pixels in
the image to be above a lower threshold of 5 photo-electrons and to
have a neighbour above 10 photo-electrons. Second-moment parameters
were calculated for each cleaned image using the Hillas [1]
definitions and these parameters were used to select candidate
$\gamma$-ray events. The selection criteria were optimized using
Monte-Carlo simulated $\gamma$-ray showers and real background runs at
the same zenith angle range as the observations. The selection cuts
are summarized in table 1, a diagram illustrating the parameter
definitions is shown in figure 1.

\begin{table}[t]
  \begin{minipage}{.55\columnwidth}
    \begin{center}
      {\begin{tabular}{l c l} \hline
        Parameter &Cut  \\
        \hline
        Length                &4.8   &mrad      \\
        Width (lower)         &0.05  &mrad      \\
        Width (upper)         &1.3   &mrad      \\
        Length/Amp.           &0.016 &mrad/p.e. \\
        Distance              &17.0  &mrad      \\
        $\alpha$                 &9.0   &deg.      \\
        \hline
    \end{tabular}}
    \end{center}
    \caption{Optimized $\gamma$-ray selection criteria.}
  \end{minipage}
  \hspace{1pc} 
  \begin{minipage}{.42\columnwidth}
    \begin{center}
      \includegraphics[width=\columnwidth,height=4cm]{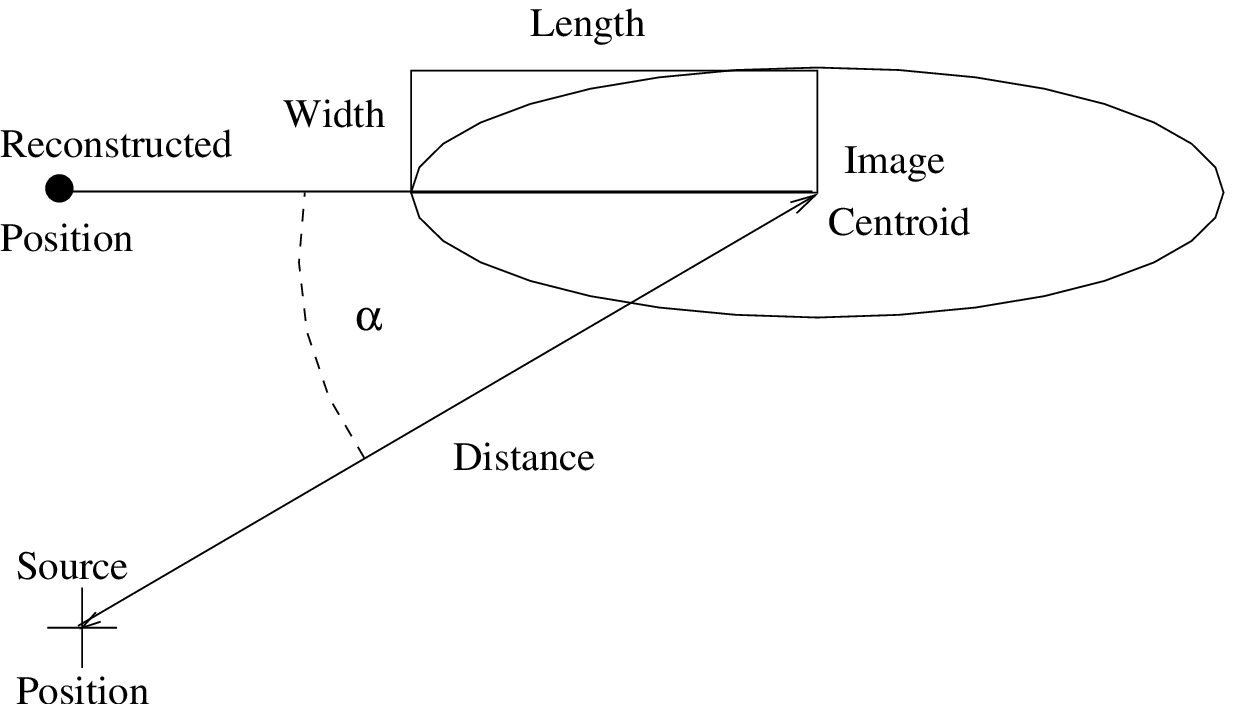}
      {\small {\bf Fig. 1.} Definition of Hillas Parameters}
    \end{center}
  \end{minipage}
\end{table}

\section{Results}

The data from the Crab nebula observations have been analysed using
the above technique, giving a steady rate of 3.6 $\gamma\ 
\textrm{min}^{-1}$\ with a significance of 20.1 $\sigma$ after
applying the above-mentioned selection cuts.  The $\alpha$ parameter
distributions for the {\small ON} and {\small OFF} data are shown in
Figure 2. The two-dimensional skyplot is shown in figure 3.  The
source reconstruction for the skyplot uses a simple single telescope
source reconstruction scheme based on Hillas parameters [4].

\begin{figure}[t]
  \begin{minipage}{.55\columnwidth}
    \begin{center}
      \includegraphics[height=5cm]{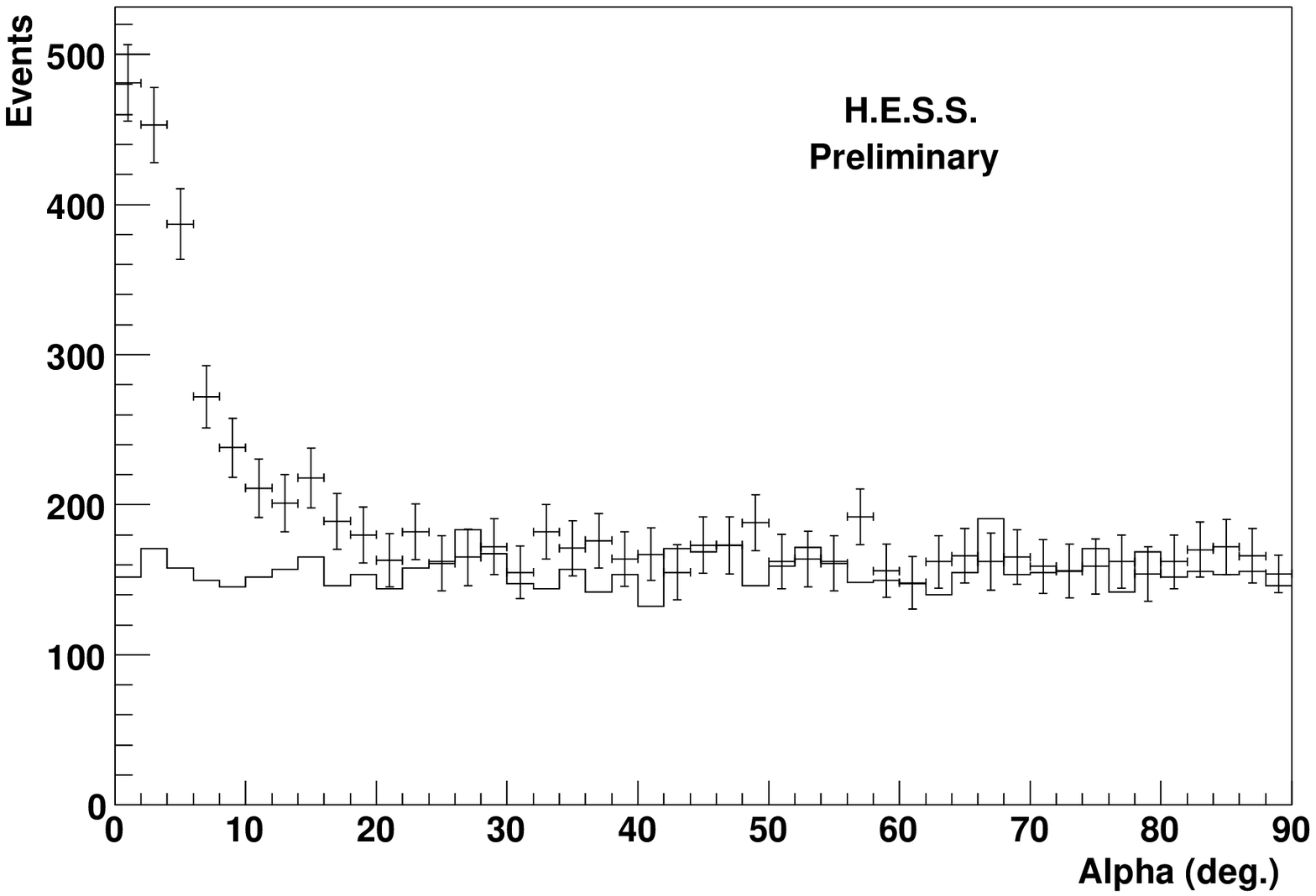}
    \end{center}
    {\small {\bf Fig. 2.}  Alpha plot, 2d plot of $\gamma$-ray\ excess from the Crab nebula,
 the {\small OFF} data is normalized to take account of the exposure time differences.}
  \end{minipage}
  \hspace{1pc} 
  \begin{minipage}{.42\columnwidth}
    \begin{center}
      \includegraphics[height=5cm]{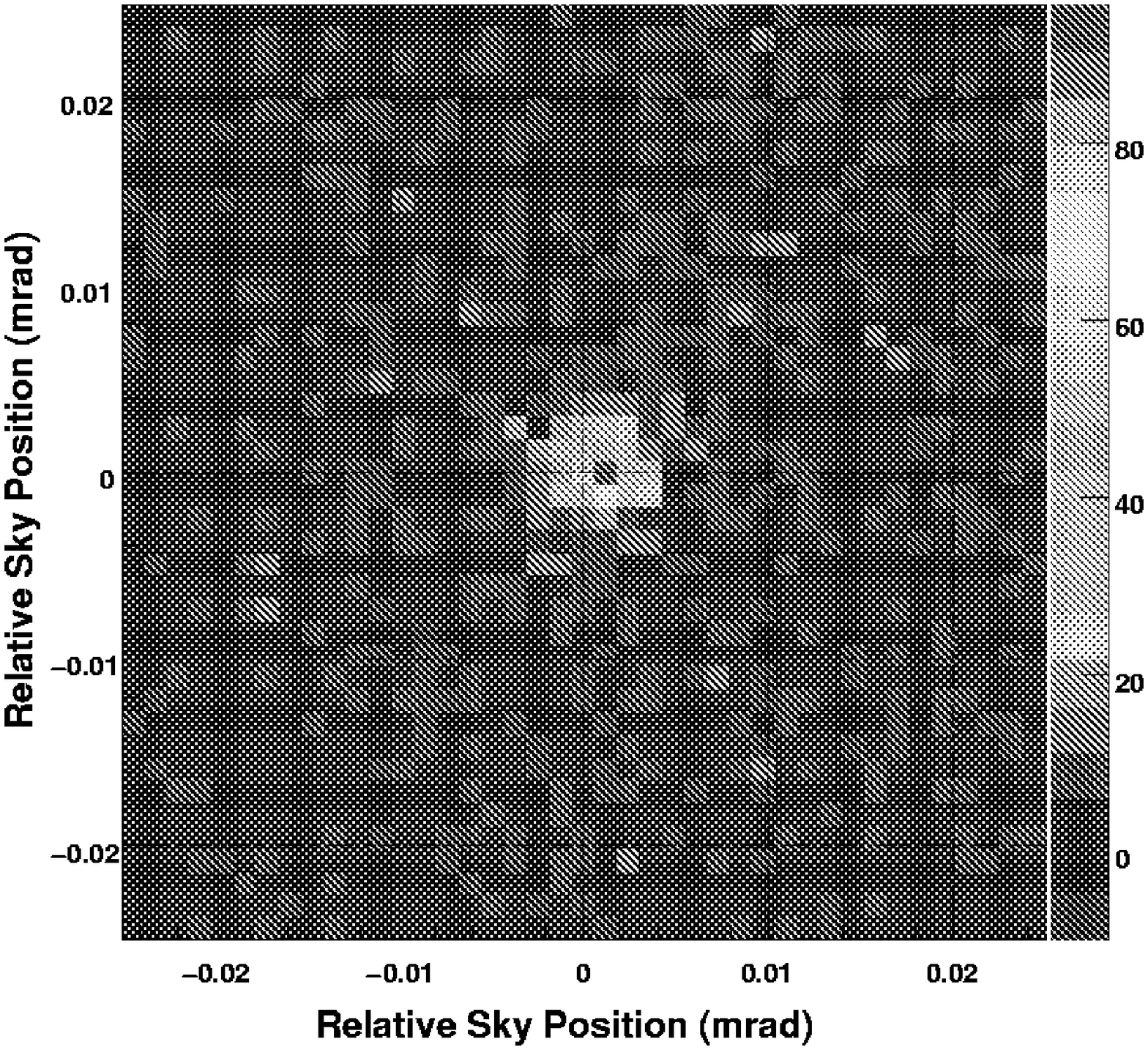}
    \end{center}
    {\small {\bf Fig. 3.}  Reconstructed skyplot of $\gamma$-ray excess in mrad around the source position}
  \end{minipage}
\end{figure}

The effective area for $\gamma$-rays has been estimated for one of the
Monte Carlo simulations described in the accompanying article using
the above selection cuts. The pre- and post-selection effective area
distributions as a function of the true Monte Carlo input energy are
shown in figure 4 for simulated $\gamma$-rays at a zenith angle of
$45^\circ$. The differential $\gamma$-ray rate for a source with a
spectrum similar to that of the Crab is given in figure 5. It can be
seen that the energy threshold after the above selection cuts, as
defined by the peak in the differential rate distribution, is 780 GeV.
The energy threshold before selection cuts is 590 GeV. The fixed cuts
on Hillas parameters described reject most $\gamma$-rays at high
energies, this may be remedied by varying the cuts with image
amplitude, which is currently under study.

\begin{figure}[t]
  \begin{minipage}{.46\columnwidth}
    \begin{center}
      \includegraphics[width=\columnwidth,height=5.3cm]{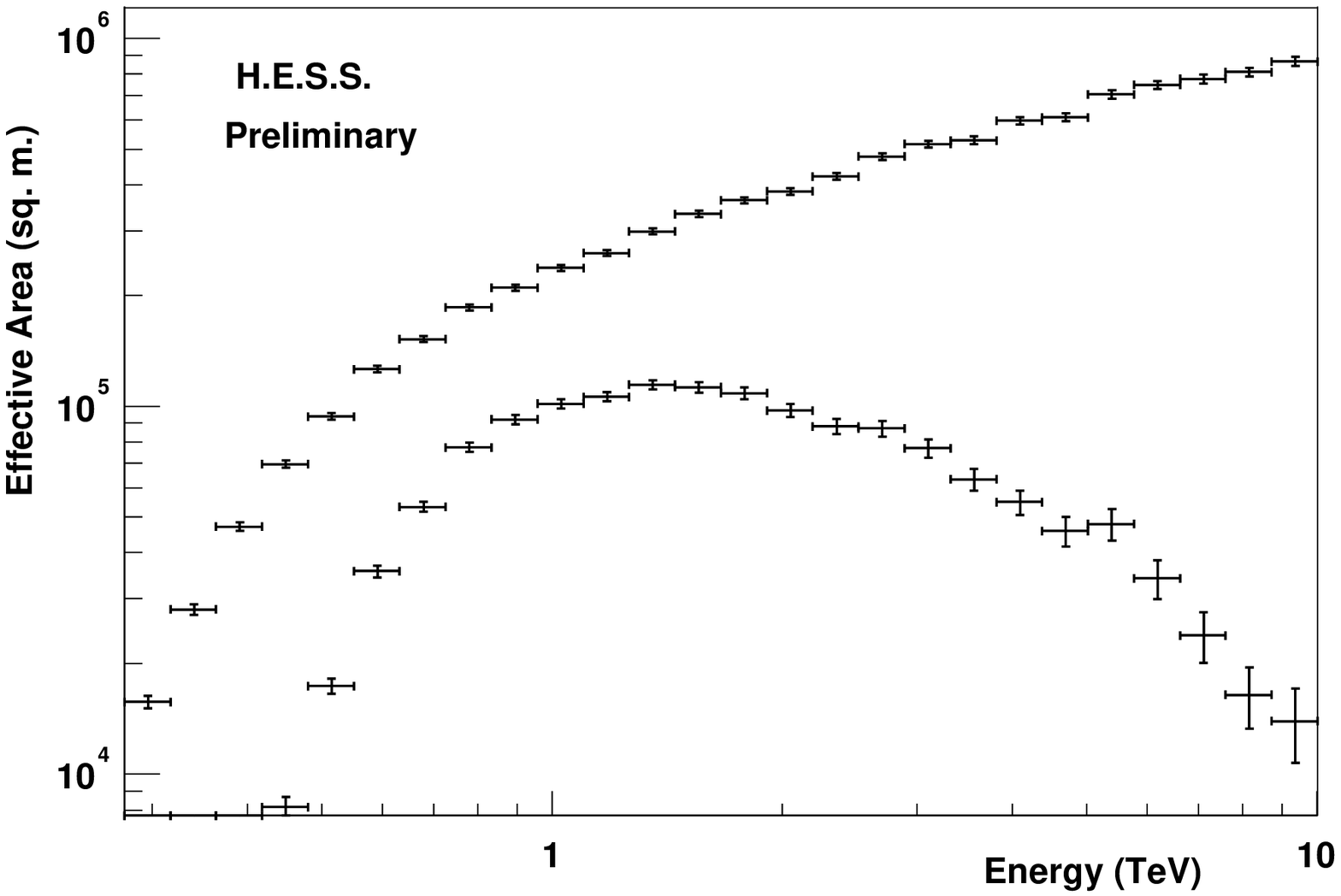}
    \end{center}
  {\small {\bf Fig. 4.}  Effective areas before and after selection cuts as a function
    of true simulated energy}
  \end{minipage}
  \hspace{1pc} 
  \begin{minipage}{.46\columnwidth}
    \begin{center}
      \includegraphics[width=\columnwidth,height=5.3cm]{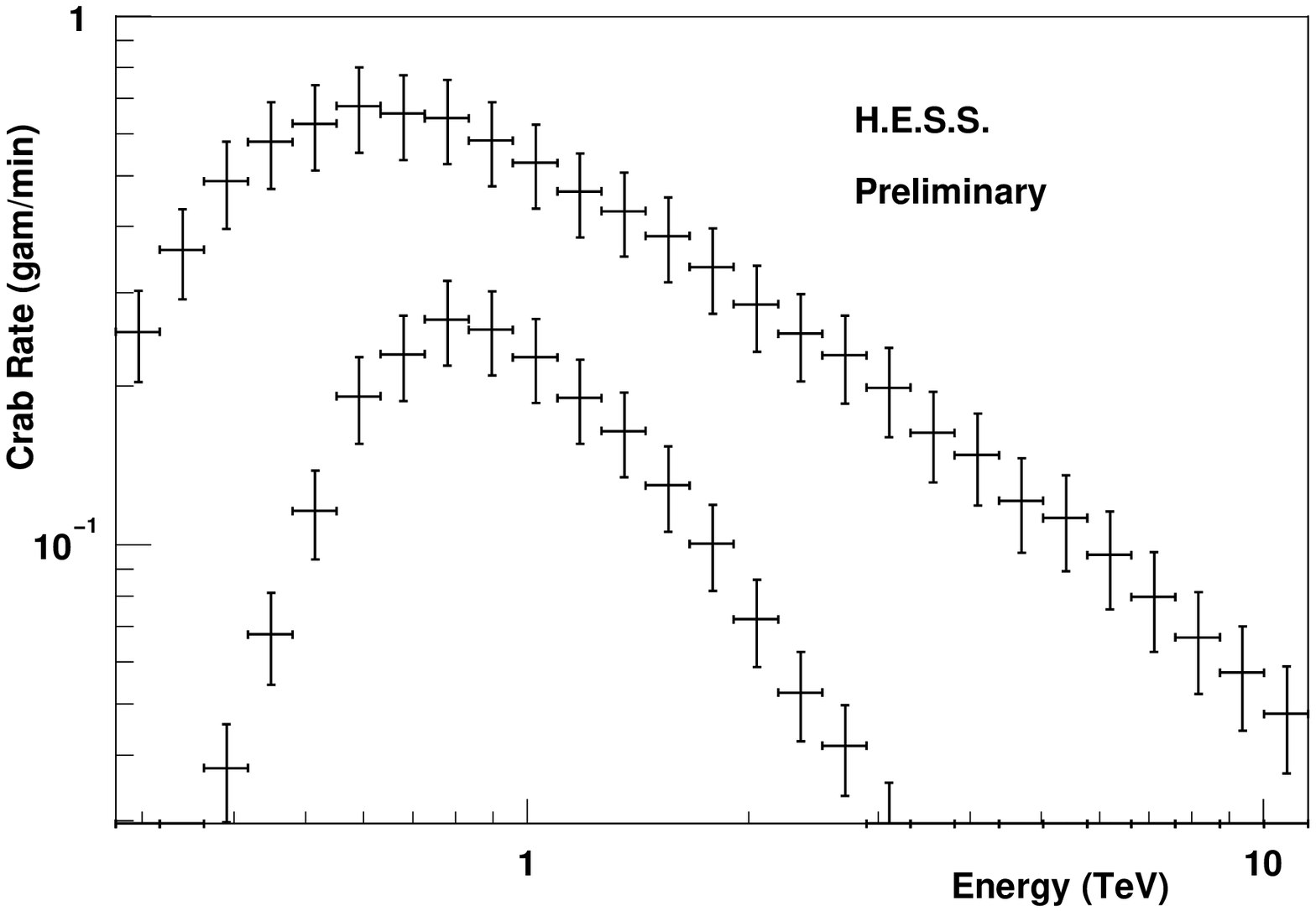}
    \end{center}
  {\small {\bf Fig. 5.}  Differential rate for a source
    with a spectrum similar to the Crab at $45^\circ$\ zenith angle.}
  \end{minipage}
\end{figure}

A preliminary estimation of the integral flux based on one of the
Monte Carlo simulations gives a value of $(2.64 \pm 0.20)\times
10^{-7} $ m$^{-2} $ s$^{-1}$ ($> 1$ TeV), for which
the quoted error include only the statistical errors; no systematic
errors are included. Preliminary analysis of the spectral energy
distribution indicates that the signal follows a power-law form with a
slope not inconsistent with measurements by other instruments.
Uncertainties in the energy threshold and spectral analysis are in
large part due to differing estimates of the collection efficiency for
$\gamma$-rays from different Monte Carlo simulations, which is the
subject of ongoing study.

\section{Observations of other Galactic Sources}

Observations were made of a number of other Galactic sources with the
first H\textsc{.E.S.S.}\ telescope during 2002 and early 2003,
calibration and analysis of these data will be presented in the talk
accompanying this paper. Observations are summarized in table 2, with
live time corrected exposures on the sources and the mean zenith angle
of the observation. The remaining Crab data is currently under
analysis.

\begin{table}[h]
  \begin{center}
    \begin{tabular}{l c c c}
      \hline
      Source &Obs. Time (hrs) &Mean Zenith Angle ($^\circ$) &Type \\
      \hline
      Crab (total)&14.2       &47.7     &Plerion\\
      Vela       &22.4        &28.6     &Plerion\\
      Cen X-3    &29.6        &38.27    &X-Ray Binary\\
      SN1006     &41.0        &23.6     &SNR\\
      Vela Jr    &1.2         &24.9     &SNR\\
      RXJ 1713   &1.2         &16.7     &SNR\\ 
      \hline
    \end{tabular}
    \caption{Summary of galactic Source Observations up to beginning May 2003}
    \label{tab:galobs}
  \end{center}
\end{table}

\section{Conclusions}

A strong signal has been detected from the Crab nebula during the
first few months of operation of the first H\textsc{.E.S.S.}\ 
instrument. Preliminary work suggests that the spectral slope is
consistent with measurements from other instruments, while the
absolute flux normalization is the subject of further study. The
second telescope of the H\textsc{.E.S.S.}\ system has been
commissioned and stereo observations have commenced. Calibration and
analysis of data taken in stereo mode will also be reported on in the
talk accompanying this paper.

\section{References}

\re
1.\ Hillas A.\ 1985, in Proc. 19nd I.C.R.C. (La Jolla), Vol. 3, p. 445. 
\re
2.\ Konopelko A.\ et al., These Proceedings. 
\re
3.\ Leroy N.\ et al., These Proceedings.
\re
4.\ Lessard R. W.\ et al., 2001, Astroparticle Physics, 15, 1. 
\re
5.\ Punch M.\ 1991, in Proc. 22nd I.C.R.C. (Dublin), Vol. 1, p. 464. 
\re
6.\ Weekes T. C.\ et al. 1989, Astrophysical Journal, 342, 379 (1989).

\endofpaper

%% file: 009431.tex
%
%
%
%








\chapter{
First Results
from Southern Hemisphere AGN Observations Obtained with the {H$\cdot$E$\cdot$S$\cdot$S$\cdot$} VHE
Gamma-ray Telescopes}

\author{
A.Djannati-Ata\"\i$^1$, for the {\small H$\cdot$E$\cdot$S$\cdot$S$\cdot$} collaboration$^2$.\\
{\it 
(1) PCC-IN2P3/CNRS College de France Universit\'e Paris VII, Paris, France\\
(2) {\tt http://www.mpi-hd.mpg.de/HESS/collaboration}   
}
}

\section*{Abstract}
The first and second telescopes of the {\small H$\cdot$E$\cdot$S$\cdot$S$\cdot$} stereoscopic system are
operating since June 2002 and February 2003, respectively.
We will present the first results from a number of southern AGN observed
using the first two {\small H$\cdot$E$\cdot$S$\cdot$S$\cdot$} telescopes, which already yield a significant
sensitivity in mono-telescope mode, with a threshold for detection below
that of other Imaging Atmospheric Cherenkov Telescopes. In this paper
we report in particular on the first detection of an AGN by {\small H$\cdot$E$\cdot$S$\cdot$S$\cdot$}: the
BL Lac object {PKS\,2155-304} was seen during July and October 2002 at a total
significance level of 11.9 standard deviations (s.d.).

\section{Introduction}

The BL Lac object {PKS\,2155-304} is one of the brightest nearby blazars
($z = 0.116$) in the optical to X-ray range, and is a highly variable source. 
Numerous multi-wavelength observations (see e.g.\ [3]) have clearly shown the  
synchrotron nature of its emission which extends up to 
hard X-rays (e.g., see results from BeppoSax [2]). 
As its peak synchrotron frequency lies at UV/soft X-rays,
{PKS\,2155-304} is classified as a High frequency-peaked BL Lac object 
or HBL [4].
 
{PKS\,2155-304} was first seen in the GeV {$\gamma$-ray} range by the E{\small GRET} detector 
[10] aboard the satellite C-GRO, 
exhibiting a hard spectrum with a differential index
of $1.71 \pm 0.24$. It was thereby considered as a strong potential
TeV source, despite its redshift for which estimations of the
Extragalactic Background Light (EBL) absorption effects are
significant.     
Chadwick {\it et al.} [1] reported its first detection in the
TeV range with the Mark 6 telescope at Narrabri (Australia) at a
level of 6.8 s.d. and a flux 
above 300\,GeV of $4.2 \pm 2.1 \times10^{-11}\:\mathrm{cm^{-2}\:s^{-1}}$.

We report here on observations made during July and October 2002 on
{PKS\,2155-304} by the 
first {\small H$\cdot$E$\cdot$S$\cdot$S$\cdot$} atmospheric Cherenkov telescope operating at a threshold
energy of $\sim 150\:\mathrm{GeV}$. Its Davies-Cotton reflector has a
focal length  
of 15 metres and an f/d of $1.2$. It is made up of 380 60\,cm diameter
mirror facets, giving an effective reflecting area of
$107\:\mathrm{m^2}$. The camera is equipped with 960 PMs with a pixel size of
$0.16{\ensuremath{^\circ}}$ for a full field of view of $\sim 5{\ensuremath{^\circ}}$ (see [6,11]).
The data-set and the analysis technique are presented in the next
section. The results and detected signal are 
then given, and we conclude with a short discussion.   

\section{Data-set and Analysis}

Observations of {PKS\,2155-304} started shortly after the
first {\small H$\cdot$E$\cdot$S$\cdot$S$\cdot$} telescope became operational in June 2002. The data-set used here
consists of 7 pairs of {\small ON}$-${\small OFF} observation runs 
({\small ON} being at the source position, and {\small OFF} at a control
region displaced in Right Ascension) taken
during four nights from July 15$\mathrm{^{th}}$ to 18$\mathrm{^{th}}$, 2002,
and 15 pairs taken from September 29$\mathrm{^{th}}$ to October
10$\mathrm{^{th}}$, 2002, for an {\small ON} live-time of 
2.18 and 4.7 hours, respectively.         

Raw data, consisting of images of cosmic-ray showers, muons, and
candidate {$\gamma$-rays} are first processed through a calibration chain,
including pedestal subtraction, ADC-to-photoelectron (p.e.) gain scaling,
flat-fielding, bad-channel filtering and image-cleaning (see [7]).
Image shape and orientation parameters, obtained after a simple
moments analysis [5] 
are then used to discriminate against the cosmic-ray background. The
parameters retained for this discrimination are the length (L), 
width (W), distance (D), the ratio of the length to the charge in the image
(LoverS or L/S), and the pointing angle, $\alpha$ --- which is the
angle at the image barycentre between the actual source position and
the reconstructed image axis of the {$\gamma$-ray} candidate.

The cut values given in Table 1 were determined through an
optimisation procedure where a simulated {$\gamma$-ray} spectrum with a
differential index of $-2.8$  was tested against real background
events (available from {\small OFF}-source runs). Simulated {$\gamma$-ray} images were
obtained through full Monte Carlo simulations of showers in the
atmosphere, and of the telescope response (see [9]). 

\begin{table}[b]
 \caption{Cut values: L, W and D are in mrad, L/S in mrad/p.e. and $\alpha$
 in degrees.}
\begin{center}
\begin{tabular}{l|ccccc}
\hline
Parameter	&L 	&W	&D	&L/S	&$\alpha$   \\
\hline
Upper cut & 5.8   & 1.42	&17     & 0.017	& 8{\ensuremath{^\circ}}\  \\
\hline
\end{tabular}
\end{center}
\end{table}

The {$\gamma$-ray} efficiency and the overall background rejection factor obtained
using the above cuts are respectively $\epsilon_{\gamma} = 25 \%$ and
$R_{h} \approx 4400$ with
a corresponding quality factor $QF=\epsilon_{\gamma} \times
    \sqrt R_{h}$ of 16. 
Observations of the Crab, with cuts adapted to its low elevation
transit, yield a rate of $3.6$ \,{$\gamma\ \textrm{min}^{-1}$} and a significance per 
hour of 9.3 ([8]).

\section{Results}

\begin{figure}[t]
  \begin{center}
    \includegraphics[height=13.5pc,width=.47\linewidth]{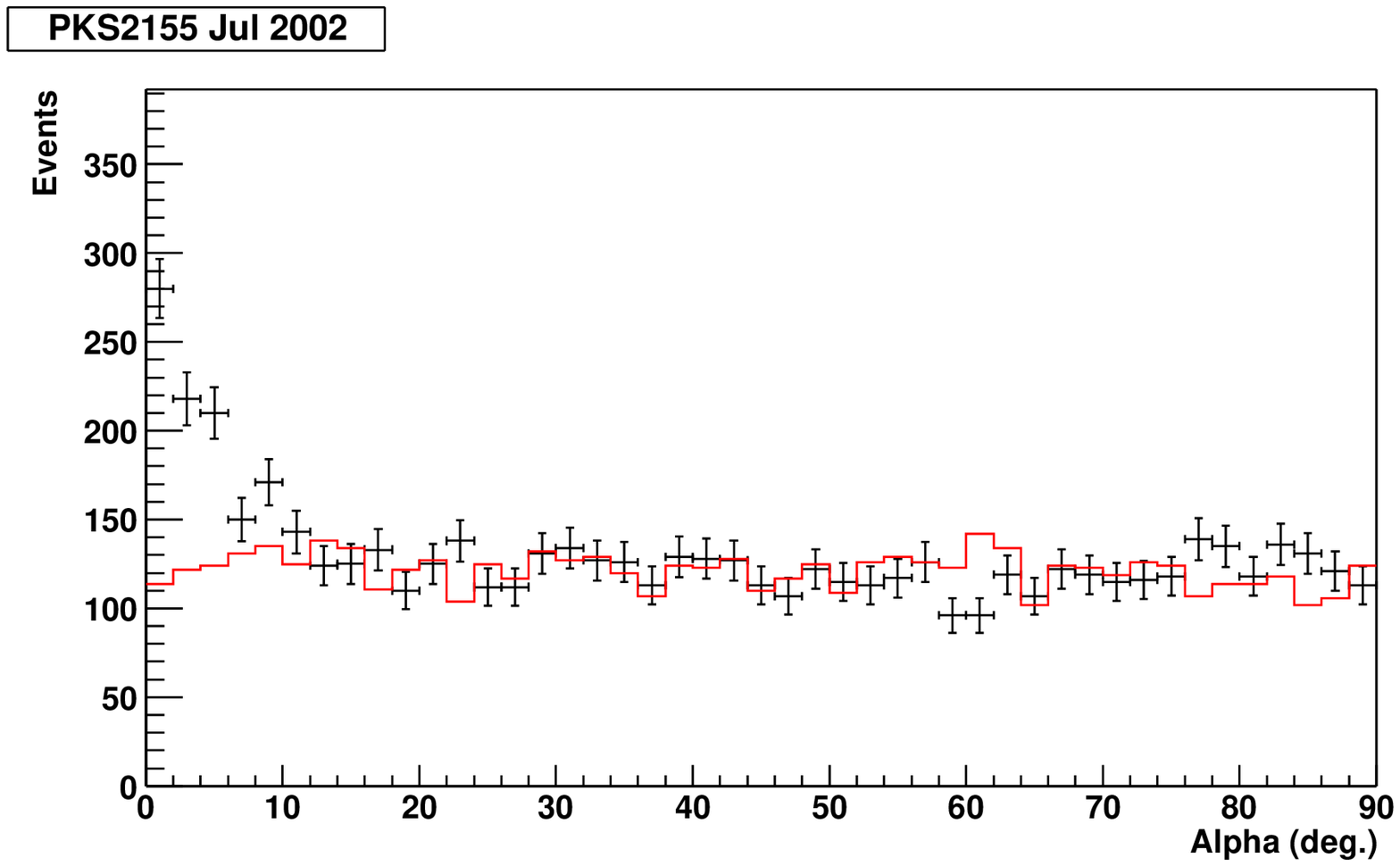}
	\hspace*{.04\linewidth}	
    \includegraphics[height=13.5pc,width=.47\linewidth]{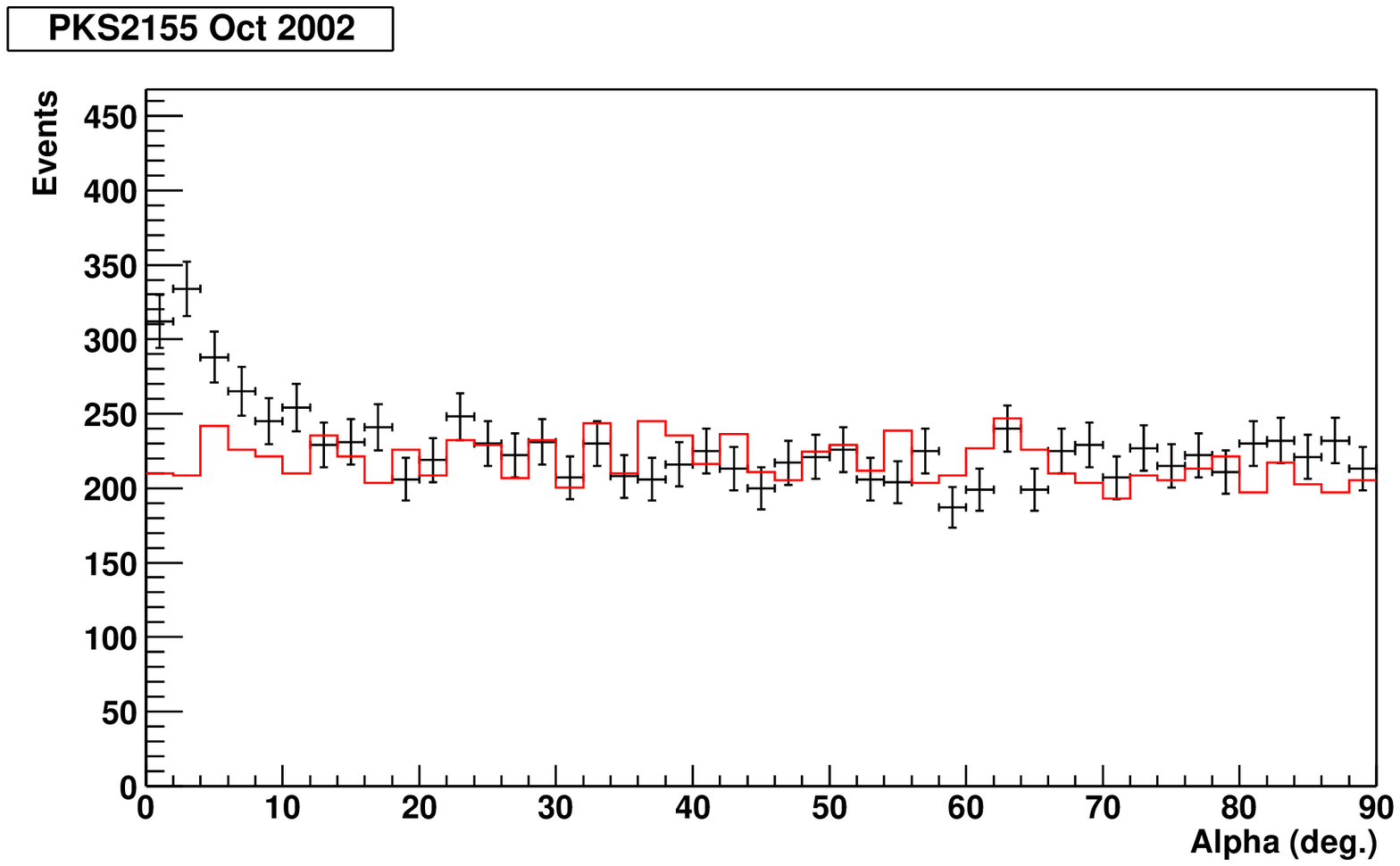}
  \end{center}
  \vspace{-0.5pc}
  \caption{The pointing angle  $\alpha$-plot of {PKS\,2155-304} observations
    for July (left panel) and October (right panel) 2002. The {\small
    OFF}-source distributions have been normalised to the control
    region between $30{\ensuremath{^\circ}}$ and $90{\ensuremath{^\circ}}$.}
\end{figure}

Fig.\ 1 shows the pointing angle $\alpha$-plots of {\small ON} and
{\small OFF}-source cumulated data after selection cuts for the July
and October 2002 data-sets. Excesses of 404 and 337 events,
corresponding to {$\gamma$-ray} rates of 3.1 and  1.2 {$\gamma\ \textrm{min}^{-1}$}, 
are observed at significance levels of 9.9 and 6.6 s.d, respectively for
the two periods. 
Hence the TeV flux of {PKS\,2155-304}, as measured by {\small H$\cdot$E$\cdot$S$\cdot$S$\cdot$}, decreased
significantly over a period of about three months. 


\begin{table}[t]
 \caption{Live-time, number of {\small ON} and {\small OFF} events
 within the cuts, excess, rate and the significance for July and
 October 2002.}
\begin{center}
\begin{tabular}{r|cccccc}
\hline
PKS2155	&Ton &Non 	&Noff	&Excess	& {$\gamma\ \textrm{min}^{-1}$} &Significance \\
\hline
Jul 2002 &2.2h	&1029	&625	& 404  	&3.1 &9.9   \\
\hline
Oct  2002&4.7h	&1444 	&1107	& 337 &  1.2	&6.6   \\
\hline
\end{tabular}
\end{center}
\end{table}

\section{ Discussion and Conclusion}


Observations of {PKS\,2155-304} by the first {\small H$\cdot$E$\cdot$S$\cdot$S$\cdot$} telescope show a clear signal
during July and October 2002, with a total significance of 11.9 s.d.,
and mark definitely 
this source into the still-short list of confirmed extragalactic TeV
sources, together 
with Mkn~421 (z=0.031), Mkn~501 (z=0.034), 1ES1959+650 (z=0.048) and
1ES1426+428 (z=0.129). 

Comparisons of the detected rates during July, 3.1 {$\gamma\ \textrm{min}^{-1}$}, and
October 2002, 1.2 {$\gamma\ \textrm{min}^{-1}$}, show a clear dimming of {PKS\,2155-304} (by a factor
$\sim3$) in the
latter period. Although the comparison of the weekly averaged X-ray
count-rates for 
the two periods, as
monitored by the {\it All Sky Monitor} on board the satellite RXTE,
shows a slightly brighter source in July 2002, it has not been
possible to make quantitative correlations between the X-rays and
{$\gamma$-rays} due to the very faint flux of the former.

This source, at an intermediate redshift (close to that of
1ES1426+428) in the growing catalogue of 
extragalactic TeV {$\gamma$-ray} sources, should provide  
further information on the link between the intrinsic spectrum of AGN 
sources and their absorption by the intervening EBL. 
The {\small H$\cdot$E$\cdot$S$\cdot$S$\cdot$} instrument is well-placed to measure such behaviour, as its low
threshold can allow spectral information to be found in the energy region 
where absorption is almost negligible, while still being sensitive up
to the highest {$\gamma$-ray} energies.
An indication of the spectral behaviour, as compared to {\small H$\cdot$E$\cdot$S$\cdot$S$\cdot$} observations
on the Crab Nebula, will be presented at the conference as well as
results of observations on a number of other AGN which are listed in
Table 3.

\begin{table}[t]
 \caption{ Other extragalactic sources observed by {\small H$\cdot$E$\cdot$S$\cdot$S$\cdot$}}
\begin{center}
\begin{tabular}{l|llc}
\hline
Source 	& Redshift & ExposureTime & Type\\
\hline
PKS0548-322 	& 0.069	& 8	& BL Lac\\
1ES1101-232 	& 0.186 & 6	& BL Lac\\
Mkn~421		& 0.031	& 1.6	& BL Lac\\
M87		& 0.00436& 24	& NLRG\\
PKS2005-489	& 0.071	& 11	& BL Lac\\
\hline
\end{tabular}
\end{center}
\end{table}


\section{References}

\noindent 1. Chadwick, P. M. {\it et al.} 1999, ApJ, 513, 161\\
2. Chiappetti L. {\it et al.} 1999, ApJ, 521, 552\\
3. Edelson R. {\it et al.} 1995, ApJ, 438, 120 \\
4. Giommi P., Padovani P. 1995, ApJ, 444, 567 \\
5. Hillas A. 1985, in Proc. 19nd ICRC, 3, 445 \\
6. Hofmann W. {\it et al.}, {These proceedings}\\   
7. Leroy N. {\it et al.}, {These proceedings}\\
8. Masterson C. {\it et al.}, {These proceedings}\\
9. Konopelko A. {\it et al.}, {These proceedings} \\
10. Vestrand W.T., Stacy J.G., Sreekumar P. 1995, ApJ, 454, L93\\
11. Vincent P. {\it et al.}, {These proceedings}\\
\endofpaper

%% file: 009383-1.tex
%
%
%
%






\chapter{
Study of the Performance of a Single Stand-Alone H.E.S.S. Telescope: 
Monte Carlo Simulations and Data}

\author{%
%
%
A.~Konopelko,$^1$ W.~Benbow,$^1$ K.~Bernl\"{o}hr,$^{1,6}$ V.~Chitnis,$^{2,*}$ 
A.~Djannati-Ata\"{i},$^3$, J.~Guy,$^2$ W.~Hofmann,$^1$ 
I.~Jung,$^1$ N.~Leroy,$^4$ S.~Nolan,$^{5,**}$ J.~Osborne,$^5$ 
M.~Punch,$^3$ S.~Schlenker,$^6$ for the H.E.S.S. collaboration \\
{\it 
(1) Max-Planck-Institut f\"{u}r Kernphysik, Heidelberg, Germany \\
(2) LPNHE, Universit\'{e}s Paris VI - VII, France \\
(3) PCC Coll\`{e}ge de France, Paris, France \\
(4) Laboratoire Leprince-Ringuet (LLR), Ecole polytechnique, Palaiseau, France \\ 
(5) Durham University, U.K. \\
(6) Humboldt Universit\"{a}t Berlin, Germany} \\
$^*$ \hspace{0.3 mm} now at Tata Institute of Fundamental Research, India \\
$^{**}$ now at Purdue University, U.S.A. \\
}

\section{Introduction}

\noindent
The High Energy Stereoscopic System (H.E.S.S.), a system of four 12~m imaging Cherenkov 
telescopes is currently under construction in the Khomas Highland of Namibia [1]. The first 
telescope has been taking data since June 2002. An extended sample of cosmic ray images 
recorded in observations at different elevations, as well as a representative sample of 
$\gamma$-ray showers detected from the Crab Nebula after a few hours of observations at 
about 45$^\circ$ in elevation, along with simulated data, give an opportunity to study the 
telescope performance in detail.

\section{Telescope}

\noindent
The telescope mount holds 380 mirrors of 60~cm each, which results in a 107~$\rm m^2$ 
reflecting area. The mirrors are arranged in the Davies-Cotton design for $f/d\,\,\simeq 1.2$. 
The point spread function is such that the radius containing 80\% of the light is about 
0.4~mrad on-axis and 1.8~mrad for 2.5$^\circ$ off-axis. The point spread function is well-reproduced 
by simulations [2]. The mirror reflectivity varies between 78\% to 85\% in the 
wavelength range from 300 to 600~nm. The telescope reflector focus the light onto a high 
resolution imaging camera. About 11\% of the incident or reflected light is obscured by the 
camera support structure. Winston cones are placed in front of the camera in order to optimize 
the light collection efficiency. Efficiency of Winston cones averaged over wavelength range is 
about 73\%. The imaging camera consists of 960 PMs (Photonis XP2960) of 0.16$^\circ$ each, 
and has a  5$^\circ$ field of view. A typical quantum efficiency of PMs
exceeds 20\% over the wavelength range from 300 to 500~nm and has a maximum efficiency 
of 26\% around 400~nm. The overall detection efficiency, averaged in a 
range from 200~nm to 700~nm, is about 0.06. 

\noindent
The H.E.S.S. site has a mild climate with well-documented optical quality. It is at 1800~m above 
the sea level and is relatively far away (about 100~km) from the nearest city of Windhoek, a 
potential source of light pollution. The illumination of camera PMs by the night sky background 
corresponds on average to a photo-electron rate of 80-200~MHz. 
The estimated contamination of the aerosols above the site is 
rather low. After taking into account the atmospheric absorption the overall detection efficiency 
is about 0.036.

\begin{figure}[t]
  \begin{center}
    \includegraphics[height=15.5pc]{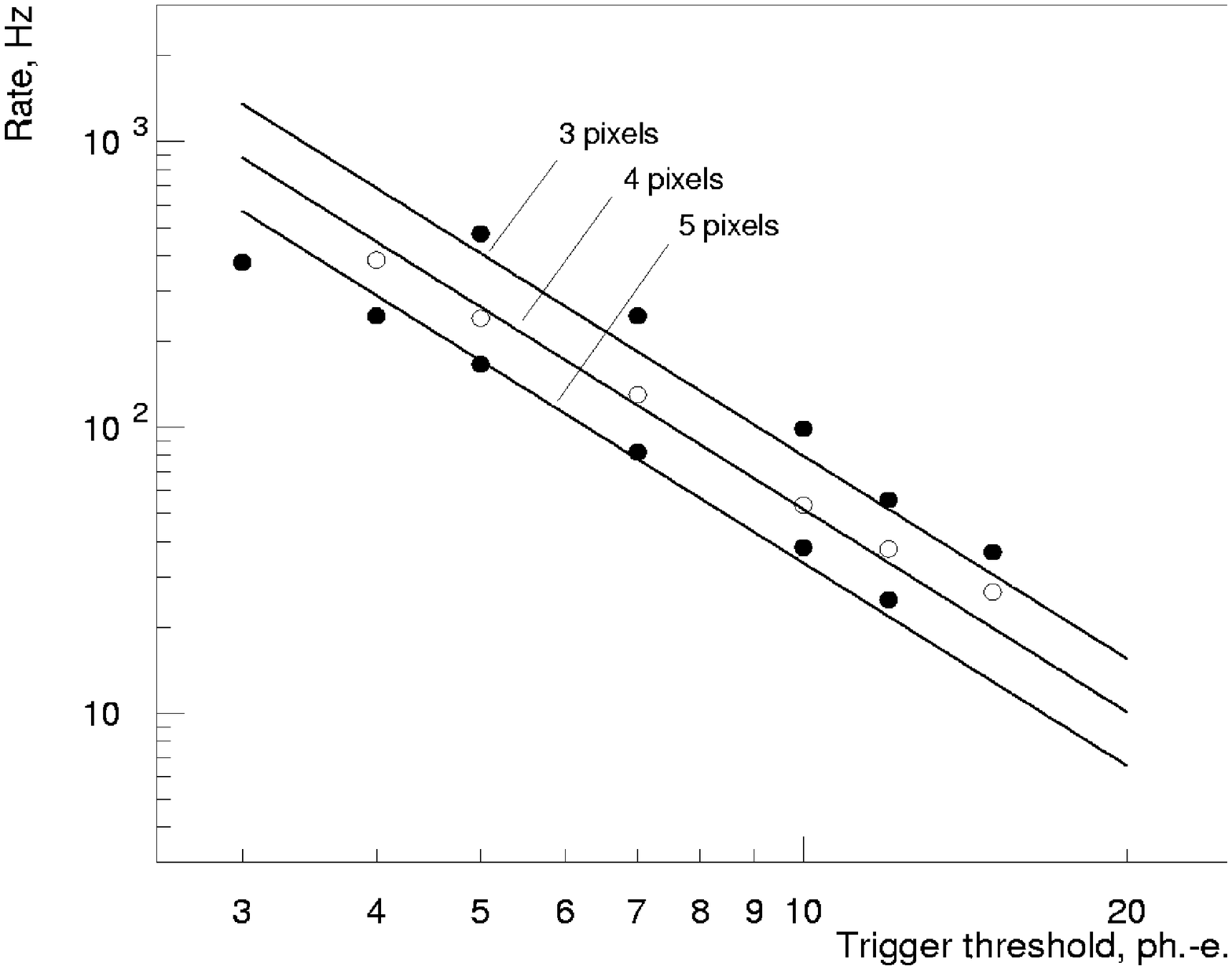}
  \end{center}
  \vspace{-0.5pc}
  \caption{Comparison between the measured (circles) and computed trigger rates (solid lines) for different trigger settings.}
\end{figure}

\begin{figure}[t]
  \begin{center}
    \includegraphics[height=15.5pc]{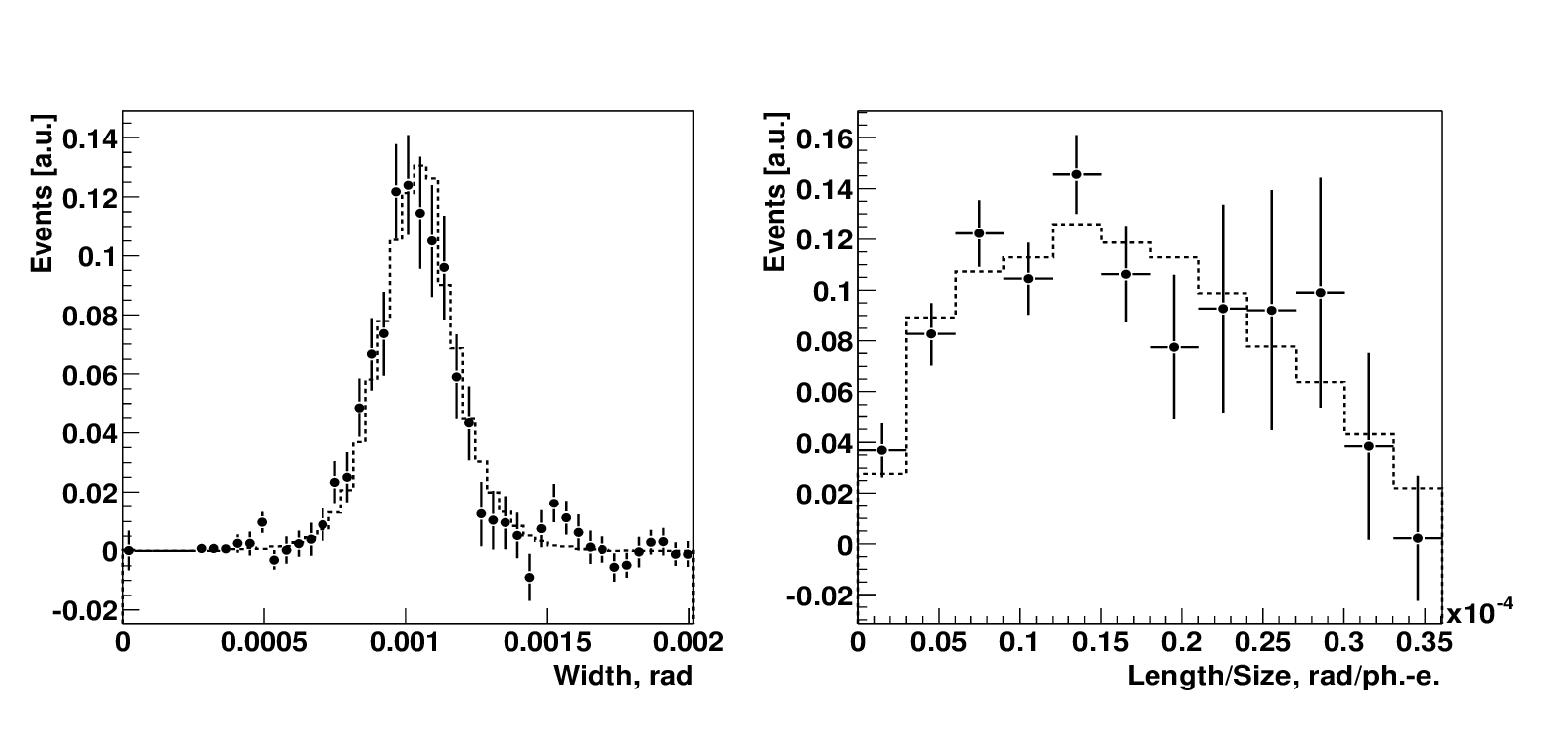}
  \end{center}
  \vspace{-0.5pc}
  \caption{Distributions of image parameters $Width$ and $Length/Size$ for the $\gamma$-rays detected from the 
       Crab Nebula (dots with the error bars) as well as for the simulated $\gamma$-ray images (histogram).}
\end{figure}

\section{Simulations}

\noindent
An extended library of air showers induced by primary $\gamma$-rays, protons, and nuclei was 
generated using a number of Monte Carlo codes available to the H.E.S.S. collaboration, ALTAI, CORSIKA, KASCADE, 
and MOCCA. Possible systematic uncertainties in parameters of the Cherenkov emission 
caused by a specific shower generator were studied in detail. 
Air showers were simulated 
within the energy range from 10~GeV to 30~TeV, and for a number of elevations in a range from 30$^\circ$ 
up to the Zenith. The angle of incidence of cosmic ray showers was randomized over the solid angle around the 
telescope optical axis with a half opening angle of 5$^\circ$ in order to simulate the isotropic 
distribution of arrival directions. Position of shower axis was uniformly randomized around the telescope 
over the area limited typically by a radius of 1000~m.

\noindent
A procedure of simulating the camera response accounts for all efficiencies of the Cherenkov 
light transmission on the way from the telescope reflector to the single camera pixel. A single 
photo-electron response function, measured for a number of PMs, was implemented to model 
the PM output. Simulations trace the propagation time for each individual photon in a shower, 
as well as all delays related to the design of the optical reflector, PM time jitter etc.
An individual photo-electron pulse shape was introduced according to the detailed time profile measured for the 
current electronics setup.  The signal recording procedure conforms to the actual hardware design based 
on 1~GHz ARS ASIC with the signal integration time of 16~ns. The comparator-type trigger scheme 
demands a coincidence of 4 pixels in one of 38 overlapping groups of 64 pixels each but fewer near the 
edges of the camera. The currently used PMs trigger threshold is 140~mV, which roughly corresponds to 
5~photoelectrons. The effective time window for the pixel coincidence was set to 2~ns. 

\section{Event Rate}

\noindent
The telescope trigger rate was measured for a number of pixel coincidences varying from 2 to 7, as 
well as for the different values of adjustable pixel trigger threshold within a range from 3 to 15 photoelectrons.
A ten-minute technical run was taken for each trigger setup at an elevation about 70$^\circ$. The simulated 
trigger rates reproduce the measured rates for all trigger setups with an accuracy of typically 25\% (see Figure~1). 
For the default trigger setup (4-fold pixel coincidence with a pixel signal above 5~ph.-e.) a dead-time 
unfolded telescope event counting rate is about 255~Hz at an elevation of 80$^\circ$. Given the read-out time 
of 1.5~ms during these early measurements it corresponds to a dead time of 40\%. The Monte Carlo predicted 
rate is 253$\pm$18(stat)$\pm$53(syst)~Hz. Air showers from cosmic ray nuclei provide about 27\% of the total 
event rate. The muon 
rate represents a substantial fraction of the telescope rate. Despite the fact that muon images mimic very 
effectively the images from the $\gamma$-ray air showers, they offer a powerful tool for the telescope 
calibration using muon rings [3]. The measured and computed event rate at an elevation of 45$^\circ$ are 
$\simeq 200$~Hz and 208~Hz, respectively.
        
\section{Image Analysis}

\noindent
Recorded images have been cleaned using a standard two-level tail cut procedure [4]. The tail cut values 
were chosen as 5 and 10~ph.-e. For each image a set of second-moment parameters was calculated. 
Distributions of the image parameters for cosmic ray showers and for $\gamma$-ray showers 
extracted from the Crab Nebula data are in a 
good agreement with the simulations of these populations (see Figure~2). 
A set of optimum analysis cuts for 
elevation of 45$^\circ$ is : $Alpha<\,\,9^\circ$; $Distance<$ 17~mrad; 0.05~mrad $<\,Width\,<$ 1.3~mrad; 
$Length<$ 4.8~mrad; $Length/Size<1.6\times 10^{-2}$~mrad/ph.-e.; $Size<10^5$~ph.-e. This set of cuts results in an acceptance for 
$\gamma$-ray showers at the level of 30\% and cosmic ray rejection of 0.02\%. It corresponds to
the quality factor of about 20.       

\section{Performance}

\noindent
Applying the analysis cuts listed above to the Crab Nebula data taken mostly at an elevation of 43$^\circ$  
one can get a $\gamma$-ray rate of about $3.57\pm 0.18$~min$^{-1}$. This rate is consistent 
within 30\% with the $\gamma$-ray rate of $2.9\pm \rm 0.18(stat) \pm 0.9 (syst)$~min$^{-1}$  derived from the 
simulations, assuming the Crab Nebula energy spectrum as measured by HEGRA [5]. The background data
rate after applying the analysis cuts is 2.5$\pm$0.1~min$^{-1}$. It is also consistent with the expectation based 
on the simulations, which is 2.3$\pm$0.1(stat)$\pm$0.6(syst)~min$^{-1}$. The single H.E.S.S. telescope 
can see the signal from the Crab Nebula at the level of 9$\sigma$ after one hour of observations. A
Crab-like source can be detectable at an elevation of 70$^\circ$ after one hour of observations at the 11$\sigma$ level. 
The energy threshold for detected $\gamma$-rays is about 180~GeV at an elevation of 70$^\circ$ and rises to 
550~GeV at an elevation of 45$^\circ$.         

\vspace{1 mm}

\noindent
{\bf References}

\vspace{1 mm}

\re
1.\ Hofmann W.\ 2003, these proceedings
\re
2.\ Cornils R., et al. 2003, 
The optical system of the H.E.S.S. IACTs, Part II, accepted for publication in {\it Astroparticle Physics} 
\re
3.\ Leroy N., et al. 2003, these proceedings
\re
4.\ Punch M.\ 1994, Proc. Workshop ``Towards a Major Atmospheric Detector III'', Tokyo, Ed. T. Kifune, p. 163 
\re
5.\ Aharonian F., et al. 2000, ApJ, 539:317-324. 

\endofpaper

%% file: 009405-1.tex
\chapter{
Application of an analysis method based on a semi-analytical shower model to
the first {\small H$\cdot$E$\cdot$S$\cdot$S$\cdot$} telescope.}

\author{%
%
%
M. de Naurois$^1$, J. Guy$^1$,  A. Djannati-Ata\"\i$^2$ , J.-P. Tavernet$^1$ 
for the {\small H$\cdot$E$\cdot$S$\cdot$S$\cdot$} collaboration$^3$. \\
{\it (1) LPNHE-IN2P3/CNRS Universit\'es Paris VI \& VII, Paris, France} \\
{\it (2) PCC-IN2P3/CNRS Coll\`ege de France Universit\'e Paris VII, France} \\
{{\it (3) }\tt http://www.mpi-hd.mpg.de/HESS/collaboration}
}

\section*{Abstract}
\vspace{-0.3em}
The first {\small H$\cdot$E$\cdot$S$\cdot$S$\cdot$} telescope has been in operation on-site in Namibia since June,
2002.  With its fine-grain camera ($0.16^\circ$ pixelization) and large mirror
light-collection area ($107\mathrm{m}^2$), it is able to see more detailed structures in
the Cherenkov shower images than are characterized by the standard moment-based
(Hillas) image analysis.  Here we report on the application of the analysis
method developed for the CAT detector (Cherenkov Array at Themis) which has
been adapted for the {\small H$\cdot$E$\cdot$S$\cdot$S$\cdot$} site and telescopes.  The performance of the method
as compared to the standard image analysis, in particular regarding background
rejection and energy resolution, is presented.  Preliminary comparisons
between the predicted performance of the method based on Monte Carlo simulation
and the results of the application of the method to data from the Crab Nebula
are shown.

\section{Introduction}
\vspace{-0.3em}
In order to take advantage of the fine pixelization of the CAT camera, a new analysis method for 
Imaging Atmospheric Cherenkov Telescopes was developed [3]. The comparison
of the shower images with a semi-analytical model was used to successfully discriminate between
$\gamma$-ray and hadron-induced showers and to provide an energy measurement with a precision 
of the order of $20\%$, without the need for stereoscopy.
The {\small H$\cdot$E$\cdot$S$\cdot$S$\cdot$} experiment, in operation in Namibia since June 2002, combines the advantages 
of the different previous-generation telescopes: large mirror, fine-pixel camera and stereoscopy.
In this paper, we present the improvements made to the CAT analysis in the framework of {\small H$\cdot$E$\cdot$S$\cdot$S$\cdot$} (operating
in single telescope mode).

\vspace{-0.3em}
\section{Model generation}
\vspace{-0.3em}
Hillas [2], studied the mean development of electromagnetic showers. 
We used his parametrization to construct a model of shower development, which we feed 
into a detector simulation to take into account instrumental effects. After this procedure, we obtain for
each zenith angle $\theta$, primary energy $E$ and impact parameter $\rho$ the predicted intensity
in each pixel of the camera.
Model images have been generated for $30$ values between $50\,\mathrm{GeV}$ and $10\,\mathrm{TeV}$,
zenith angles up to $60^\circ$, and impact parameters up to $300\,\mathrm{m}$ from the telescope.
A multi-linear interpolation method is used to compute the pixel intensity
for intermediate parameters. The model generation has been extensively tested against 
simulation and agrees within $10\%$ up to $10\, \mathrm{TeV}$.

\vspace{-0.3em}
\section{Event reconstruction}
\vspace{-0.3em}
The event reconstruction is based on a maximum likelihood method which uses all available pixels in the camera.
The probability density function of observing a signal $S$, given an expected amplitude $\mu$, 
a fluctuation of the pedestal $\sigma_p$ (due to night sky background and electronics) 
and a fluctuation of the single photoelectron signal (p.e.) $\sigma_s\approx 0.4$ (PMT resolution) is given by
\begin{equation}
P(S|\mu,\sigma_p,\sigma_s) = \sum_{n=0}^\infty \frac{e^{-\mu} \mu^n}{n!\sqrt{2\pi(\sigma_p^2 + n \sigma_s^2)}} \exp \left(
- \frac{(S - n)^2}{2(\sigma_p^2 + n \sigma_s^2)} \right) 
\end{equation}

\vspace{-3em}
\begin{figure}[h!]
\begin{minipage}[b]{0.5\linewidth} 
\noindent The likelihood
\begin{equation}
{\cal L} = 2 \sum_{\mathrm{pixel}} \log \left[P_i(S_i|\mu,\sigma_p,\sigma_s)\right]
\end{equation}

\noindent is then maximized to obtain the primary energy, the target direction $T$ and the impact point $I$. This five parameter
fit can be reduced to four parameters $E$, $\rho$, $\phi$ (azimuthal angle in the camera) and $d$ (angular 
distance of the shower barycenter to the primary direction, see fig. 1), using the alignement of the image centre 
of gravity with $\mathrm{TI}$.
\end{minipage}
\hspace{0.1cm} 
\begin{minipage}[b]{0.45\linewidth}
\begin{center}
\vspace{2em}
\includegraphics[width=7pc]{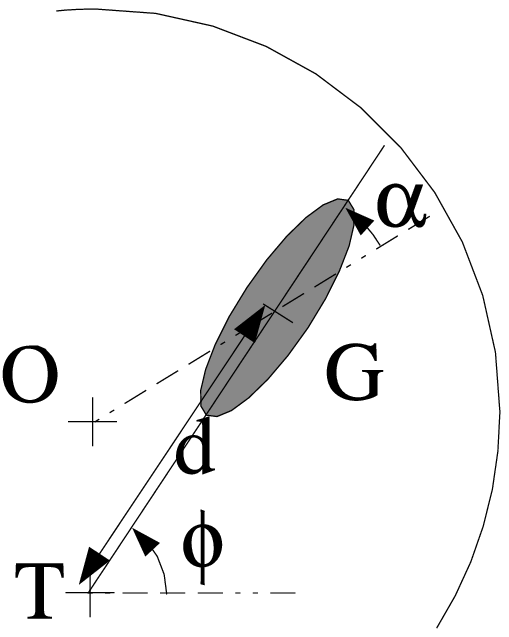}
\vspace{-1em}
\end{center}
\caption{Definition of geometrical parameters used for the shower reconstruction. O is the center 
of the camera, G the image barycenter and T the reconstructed target direction.}
\end{minipage}
\end{figure}
\vspace{-1em}


\vspace{-0.3em}
\section{Signal extraction}
\vspace{-0.3em}
\noindent The following cuts are used in signal extraction
\begin{itemize}
\vspace{-0.7em}
\item A cut on the ratio of the shower length $L$ to its amplitude $S$, designed to reject small muon images : $L/S \leq 1.6\times 10^{-2} \mathrm{mrad}\ \mathrm{p.e.}^{-1}$
\vspace{-0.7em}
\item A geometrical cut of the distance mismatch $\left|\delta D\right| \leq 5\  \mathrm{mrad}$, where $\delta D= \left|\mathrm{TG}\right|-\left|\mathrm{OG}\right|$. This cut selects $\gamma$-rays originating from the center of the field of view and is orthogonal to the commonly used $\alpha$ orientation angle.
\vspace{-0.7em}
\item A goodness of fit ${\cal G} < 0.07$ defined from the likelihood distribution as function of the number of operating pixels $N_{\mathrm{dof}}$ as ${\cal G} =  (\left<{\cal L}\right> - {\cal L})/N_{\mathrm{dof}}$, where the average likelihood and its RMS are obtained by integration of an analytical approximation of eq. 1:
\begin{equation}
\left<{\cal L}\right> = -  \sum_{\mathrm{pixel}} \left(1 + \log 2\pi + \log\left(\mu(1+\sigma_s^2) + \sigma_p^2\right)\right), \quad \sigma_{\cal L}^2 = 2
\end{equation}
\end{itemize}

The distribution of $\cal G$ for simulated $\gamma$-rays and real hadrons is shown in fig. 2 together with the likelihood's average and RMS. 
The distribution for $\gamma$-rays is compatible with an expected mean of $0.0$ and has a slightly larger RMS than the expected 
value $\sqrt{2/N_{\mathrm{dof}}} \approx 5\times 10^{-2}$. A cut ${\cal G} \leq 0.07$ keeps $77\%$ of the $\gamma$-rays and rejects $82\%$ of the hadrons.

\begin{figure}[t]
\begin{center}
\includegraphics[width=17pc]{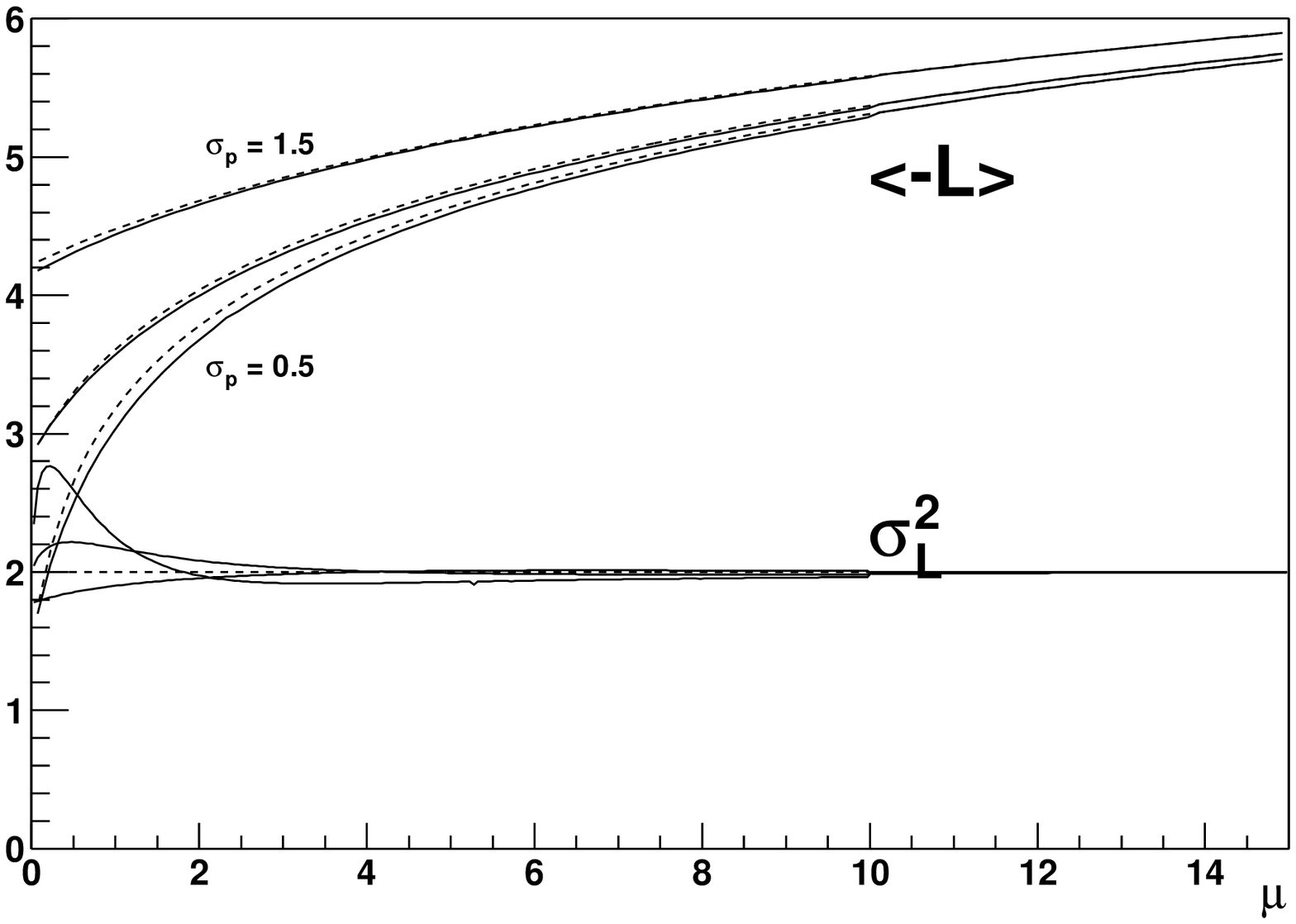} \hfill \includegraphics[width=17pc]{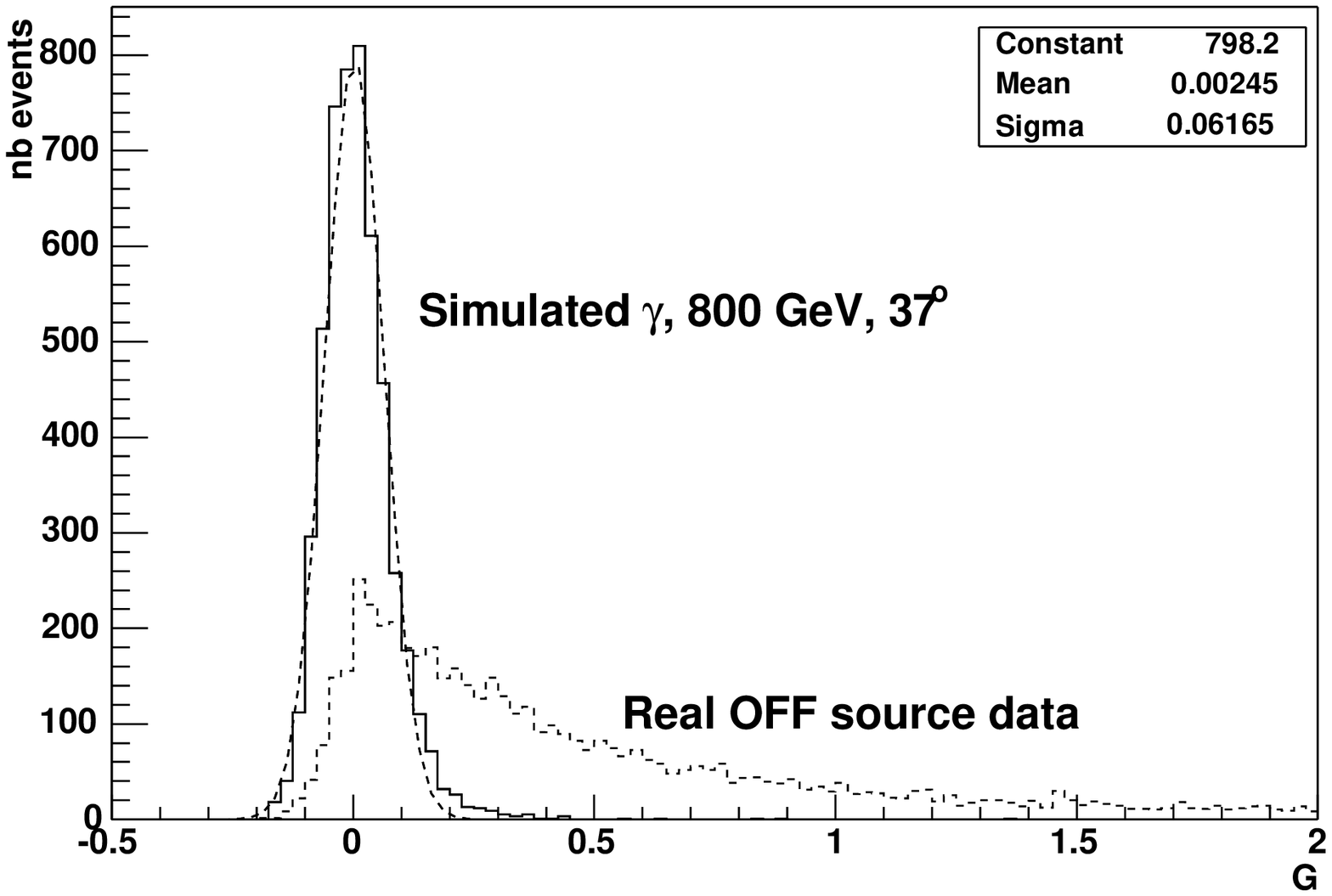} 
\end{center}
\vspace{-0.5pc}
\caption{Left: Likelihood curves for different night sky backgrounds. Solid curves are numerical integrations, dashed curves are simple
analytical approximations (eq. 3). Right: Goodness of fit ($\cal G$) distribution after $L/S$ and $D$ cuts, for
simulated $\gamma$ and real OFF source events.}
\vspace{-1.5em}
\end{figure}

\section{Results}
\vspace{-0.3em}
We have analysed $20$ pairs on the Crab Nebula, corresponding to $4.65$ hours (live-time corrected) of data. 
Fig. $3$ shows the comparison between the standard {\small H$\cdot$E$\cdot$S$\cdot$S$\cdot$} analysis [4] and this work. 
The model analysis alone produces an $\alpha$ plot extending up to $180^\circ$.
The significance in the $\alpha < 9^\circ$ is better in the hillas analysis ($21\,\sigma$ against $16.9\,\sigma$), mainly because
the hadron rejection is not yet fully optimized for the model analysis, but the $\alpha$ resolution is much better 
in the Model analysis ($2.7^\circ$ against $4.2^\circ$), thus providing a better
signal to background ratio in the first bins.

More interesting is the background rejection capability of a {\it combined analysis}. The lower plot of fig. 3 shows the $\alpha$ 
distribution of the events passing both the Hillas and Model analysis cuts. The signal over background ratio is increased by a factor 
of more than $3$, reaching the value of $4.8$ for $\alpha < 9^\circ$ wheras less than $15\%$ of the $\gamma$-rays are lost.
This also results in a net increase in significance up to $25.2\,\sigma$.
The complementarity of the hadron rejection capabilities of both analyses is a very powerful instrument for finding faint sources,
and was successfully used to detect the blazar PKS 2155-304 in October 2002 at the level of $7.4\,\sigma$ [1].

This analysis also provides energy and shower impact measurements with respective resolutions of about $20\%$ and $20\ \mathrm{m}$ 
at $800\,\mathrm{GeV}$. 

\begin{figure}[t]
  \begin{center}
    \includegraphics[width=35pc]{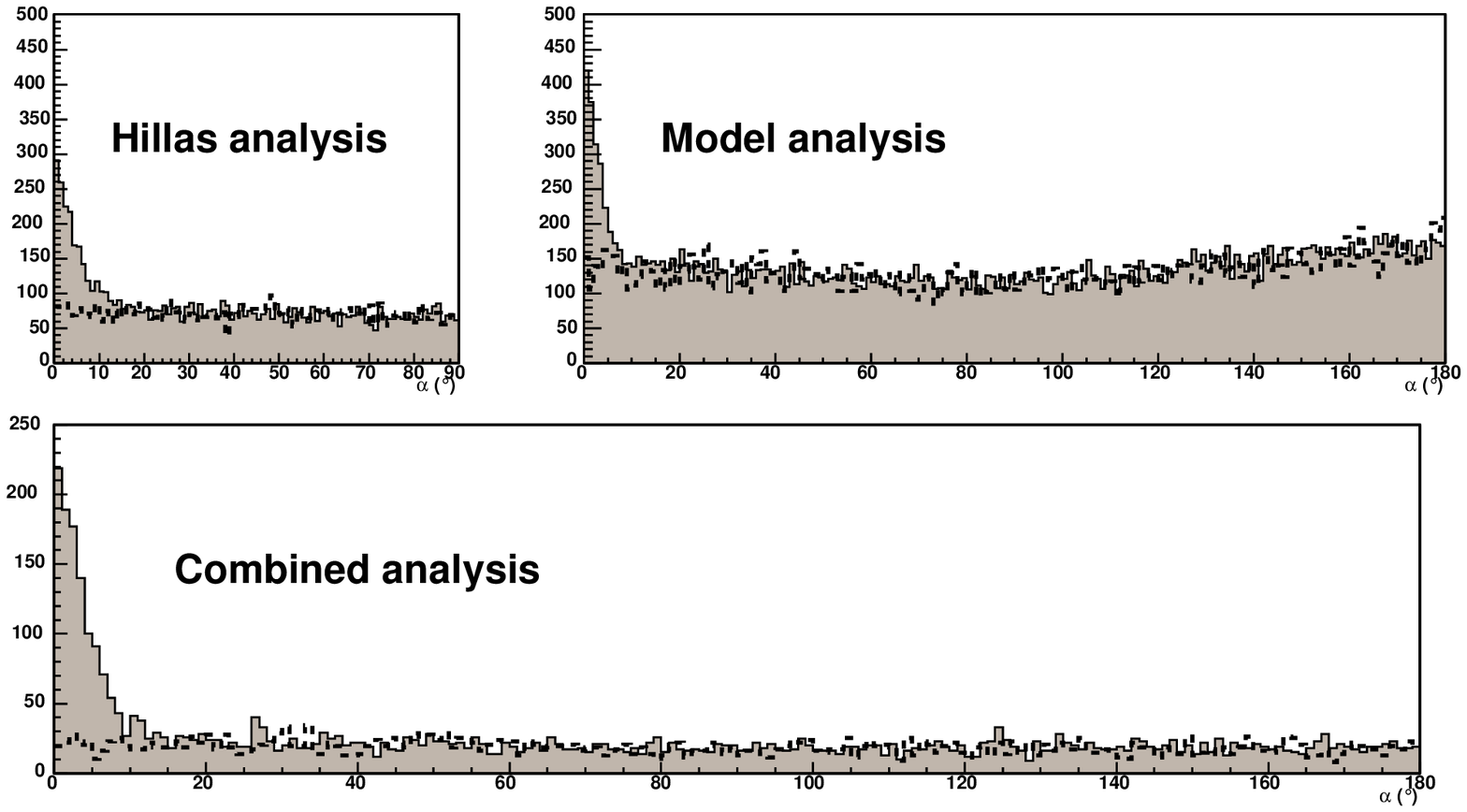}
  \end{center}
  \vspace{-0.5pc}
  \caption{Results of the model analysis. Solid lines: ON source. Dashed lines: OFF source. {\it top left:} $\alpha$ plot with standard {\small H$\cdot$E$\cdot$S$\cdot$S$\cdot$}
analysis based on hillas parameters. {\it top right:} Corresponding $\alpha$ plot for the model analysis. {\it bottom:} Effect of the 
combination of the cuts on the background. }
\vspace{-1.5em}
\end{figure}


\vspace{-0.3em}
\section{Conclusion}
\vspace{-0.3em}
We have developed a powerful analysis for {\small H$\cdot$E$\cdot$S$\cdot$S$\cdot$} based on the comparison of shower images with a semi-analytical model.
This analysis provides a better $\alpha$ angle measurement than the standard analysis (base on Hillas parameters), 
as well as good energy and  shower impact resolution. Moreover, the combination of both analysis provides an additional
background rejection factor of about $3$, which leads to an important increase in significance. This combination
method has been successfully used to detect the blazar PKS 2155-304 at the level of respectively $13\,\sigma$ and $7.4\,\sigma$
in July and October 2002. Further results  will be presented at the conference.

\vspace{-0.3em}
\section{References}
\vspace{-0.3em}
\re
1.\ Djannati-Ata\"\i A, These proceedings
\re
2.\ Hillas A.M..\ 1982, J. Phys. G 8, 1461
\re
3.\ Le Bohec S {\it et al.}, NIM A416 (1998) 425
\re 
4.\ Masterson C, These proceedings
\endofpaper

%% file: 009303-1.tex
\chapter{
The Central Data Acquisition System of the H.E.S.S.\ Telescope System
}

\author{%
%
%
C.~Borgmeier$^1$, N.~Komin$^1$, M.~de~Naurois$^2$, S.~Schlenker$^1$,
U.~Schwanke$^1$ and C.~Stegmann$^1$, for the H.E.S.S.\ collaboration \\
{\it (1) Humboldt University Berlin, Department of Physics, Newtonstr. 15, D-12489 Berlin, Germany\\
(2) Laboratoire de Physique Nucleaire et des Hautes Energies, 4 place Jussieu, T33 RdC, 75252 Paris Cedex 05, France } \\
}

\section*{Abstract}

This paper gives an overview of the central data acquisition (DAQ)
system of the H.E.S.S.\ experiment. The emphasis is put on the chosen
software technologies and the implementation as a distributed system
of communicating objects. The DAQ software is general enough for
application to similar experiments.

\section{Introduction}

The High Energy Stereoscopic System (H.E.S.S.) is an array of imaging
\v{C}erenkov telescopes dedicated to the study of non-thermal phenomena
in the Universe. The experiment is located in the Khomas Highlands of
Namibia. At the end of Phase~I, the array will consist of four
telescopes, two of which are already operational and being used for
data-taking at the time of writing.

Each telescope in the array is a heterogeneous system with several
subsystems that must be controlled and read out.  The telescope
subsystems comprise a camera with 960 individual photo-multiplier
tubes, light pulser systems for calibration purposes, a source
tracking system, an IR radiometer for atmosphere monitoring, and a CCD
system for pointing corrections. Common to the whole array is a set of
devices for the monitoring of atmospheric conditions, including a
weather station, a ceilometer and an all-sky radiometer.

\section{DAQ System Requirements}

The DAQ system provides the connectivity and readout of all the
systems mentioned above. It takes over run control, the recording of
event and slow control data, error handling, and monitoring of all
subsystems. The remoteness and small bandwidth connection of the
H.E.S.S.\ site imply that the DAQ system must be stable and easily
operated.

The main data stream is produced by the cameras which generate events
with a size of 1.5\,kB. At the design trigger rate of 1\,kHz, this
yields a maximum data rate of 6\,MB/s for four telescopes, resulting
in roughly 100\,GB of data per observation night. The data rates from
the other subsystems are significantly smaller.

On the hardware side the requirements are met by a Linux PC farm with
a fast Ethernet network. The details are described in [1].

\section{DAQ Software}

The H.E.S.S.\ DAQ system is designed as a network of distributed
\hbox{C\hskip-0.1ex\raise0.2ex\hbox{\small++}} and \emph{Python} 
objects, living in approximately 100 multi-threaded processes for the
H.E.S.S.\ Phase~I configuration.

For inter-process communication the \emph{omniORB} [2] implementation
of the CORBA protocol standard is used which provides language
bindings for \hbox{C\hskip-0.1ex\raise0.2ex\hbox{\small++}} and
Python. CORBA allows to call methods of objects in remote processes
and to pass data objects. The transport and storage of objects needs a
serialization mechanism which is provided by the \emph{ROOT Data
Analysis Framework}~[3]. Both the ROOT-based H.E.S.S.\ data format and
the ROOT graphics and histogram classes are used online and off\-line
allowing a seamless integration of data analysis code and hence fast
feedback.

The base classes for all DAQ applications are provided by a central
library. Derived classes, implementing the base class interfaces,
control for example a hardware component or handle different types of
data streams. Each DAQ process contains a {\tt StateController} object
which implements inter-process communication and run control state
transitions.

The DAQ system distinguishes four different process categories as
shown in Fig.~1. The {\bf Controllers} directly interact with the
hardware and read out the data. Each hardware component is controlled
by one Controller process. The Controllers push the data to
intermediate {\bf Receivers}, which perform further processing and
store the data. The Receivers also provide an interface that allows
other processes to sample processed data. {\bf Readers} actively
request data from the Receivers at a rate different from the actual
data-taking rate. The data or derived quantities are then available
for display and monitoring purposes. {\bf Manager} processes are not
involved in the data transport but control the data-taking.

\begin{figure}[ht]
  \begin{center}
    \includegraphics[width=34.5pc]{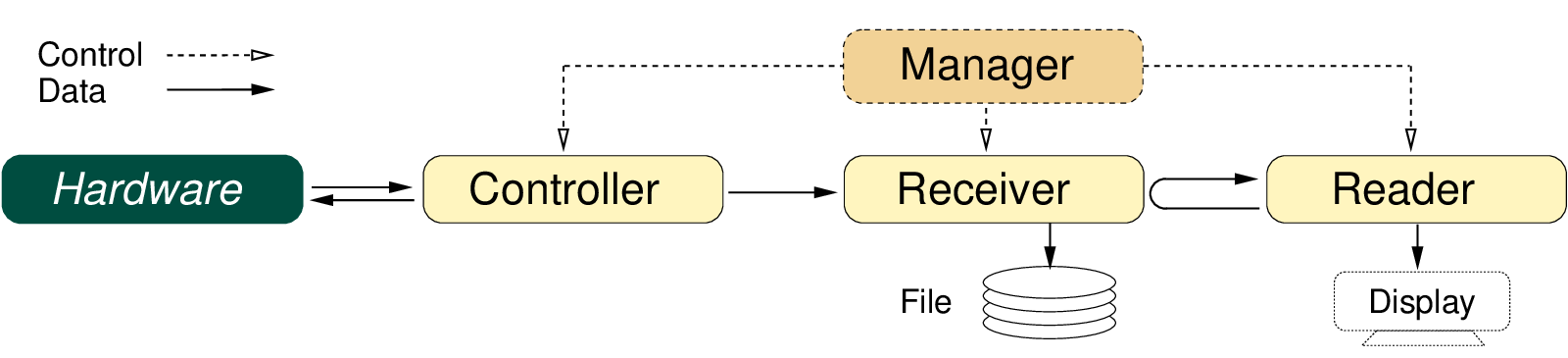}%
  \vspace{-0.4pc}
  \caption{Types of processes in the DAQ system}
  \label{fig:daq_processes}
  \end{center}
\end{figure}

\section{DAQ Configuration}

Different data-taking configurations of the array correspond to
different \emph{run types}. A run type is defined by a set of required
Controllers, Receivers, Readers, and Managers.  Examples for run types
are the observation run type with all available subsystems, various
calibration run types for the cameras, and dedicated run types for
testing and data-taking with specific subsystems, e.g.\ the tracking
system. An actual run is given by its type and a set of parameters.
 
All run type definitions, run parameters, the configuration of the DAQ
system, and observation schedules are stored in a MySQL database
acting as central information source and logging facility for the DAQ
system.

Setting up the required processes for a specific run type is
simplified by combining related processes in groups that are called
{\em contexts}. An example for a context is {\tt CT1} which comprises
all Controllers accessing the hardware of telescope number~1. Every
context contains one Manager that controls the other processes in the
context. At startup, the Manager reads the processes in its context
from database tables and launches them. The Manager serves as an
intervention point to all processes in the context, passes on state
transitions, and takes over error handling in predefined ways.

For data-taking a central DAQ Manager reads the observation schedule
from the database and determines the actual sequence of runs according
to the availability of contexts. Runs that require different contexts
can be processed in parallel. The central DAQ Manager launches the
required context Managers and initiates the run preparation.  After
the preparation of a run the starting and stopping of the data-taking
is taken over by a dedicated Manager which controls the participating
contexts.

The shift crew interacts with the DAQ system via a central control GUI
providing an access point to the system and direct monitoring of the
states of the different processes (cf.~Fig.~2 (left)). Data
monitoring information is shown by a variety of different
displays. The displays are generated by instances of a generic Reader
which are configurable via database tables allowing a simple and
flexible setup of the quantities to be displayed. Fig.~2 (right) shows
some examples of monitoring displays.

\begin{figure}[t]
  \begin{center}
    \includegraphics[width=34.5pc]{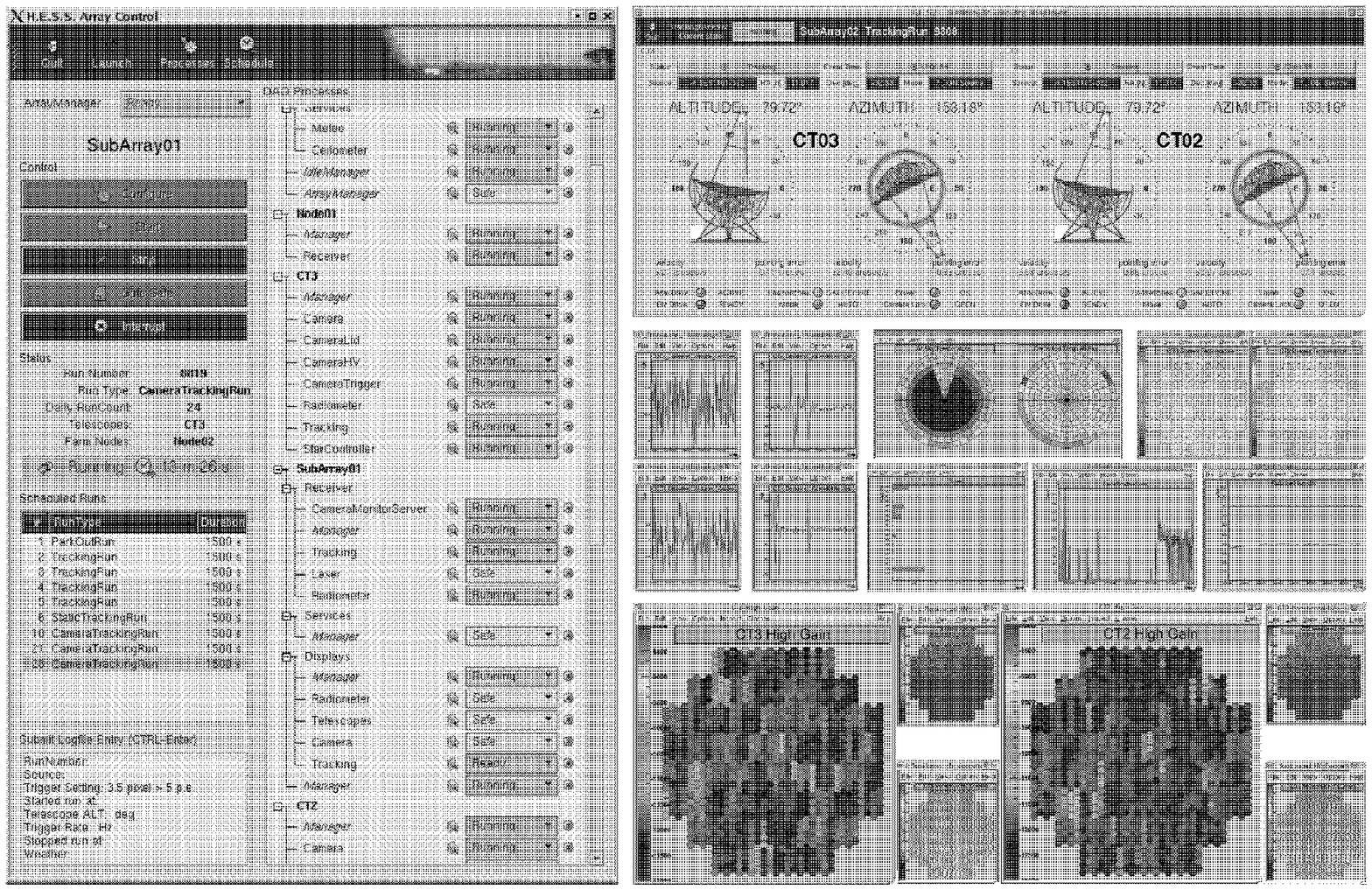}
    \vspace{-0.5pc} 
    \caption{ Central control GUI (left) and examples
    of monitoring displays (right)} \label{daq_GUI}
   \end{center}
\end{figure}

\section{Summary and Outlook}

The described system is in operation since the start of the
observation program with the first telescope. The setup based on
database tables proved its flexibility when integrating new subsystems
into the data-taking. Processing parallel runs was exercised in the
commissioning phase when observing with the first telescope while
aligning the mirrors of the second telescope. The system is now taking
data with two telescopes and is expected to scale well to the full
Phase~I array. \\

\noindent
\emph{Acknowledgments.} This work was supported by the Bundesministerium f\"ur Bildung und Forschung under the contract number 05 CH2KHA/1.

\section{References}

\re
1.\ Borgmeier C.\ et al.\ 2001, Proceedings of ICRC 2001: 2896
\re
2.\ S.L.\ Lo and S.\ Pope, The Implementation of a High Performance 
   ORB over Multiple Network Transports, Distributed Systems
   Engineering Journal, 1998.\\
   See also {\tt http://omniorb.sourceforge.net}
\re
3.\ R.\ Brun and F.\ Rademakers, ROOT -- An Object Oriented Data
   Analysis Framework, Proceedings AIHENP'96 Workshop, Lausanne,
   Sep.\ 1996, Nucl.\ Inst.\ \& Meth.\ in Phys.\ Res.\ A 389 (1997)
   81--86.\\
   See also {\tt http://root.cern.ch}

\endofpaper

%% file: 007412-1.tex
\chapter{Mirror alignment and performance of the optical system
of the H.E.S.S. imaging atmospheric Cherenkov telescopes}

\author{%
Ren\'e~Cornils,$^1$ 
Stefan~Gillessen,$^2$
Ira~Jung,$^2$
Werner~Hofmann,$^2$\\ and
G\"otz~Heinzelmann,$^1$
for the H.E.S.S. collaboration\\
{\it (1) Universit\"at Hamburg, Institut f\"ur Experimentalphysik,
Luruper Chaussee 149, D-22761 Hamburg, Germany}\\
{\it (2) Max-Planck-Institut f\"ur Kernphysik,
P.O. Box 103980, D-69029 Heidelberg, Germany}}

\section*{Abstract}
The alignment of the mirror facets of the H.E.S.S.\ imaging atmospheric
Cherenkov telescopes is performed by a fully automated alignment system using
stars imaged onto the lid of the PMT camera.
The mirror facets are mounted onto supports which are equipped with two
motor-driven actuators while optical feedback is provided by a CCD camera
viewing the lid.
The alignment procedure, implying the automatic analysis of CCD images and
control of the mirror alignment actuators, has been proven to work reliably.
On-axis, 80\% of the reflected light is contained in a circle of less than
1mrad diameter, well within specifications.

\section{Introduction}

H.E.S.S.\ is a stereoscopic system of large imaging atmospheric Cherenkov
telescopes currently under construction in the Khomas Highland of Namibia [5].
The first two telescopes are already in operation while the complete phase~1
setup, consisting of four identical telescopes, is expected to start operation
early 2004.
The reflector of each telescope consists of 380 mirror facets with 60\,cm
diameter and a total area of 107\,$\mathrm{m}^2$.
For optimum imaging qualities, the alignment of the mirror facets is crucial.
A fully automated alignment system has been developed, including motorized
mirror supports, compact dedicated control electronics, various algorithms
and software tools [1-3].
The specification for the performance of the complete reflector requires the
resulting point spread function to be well below the size of a pixel of the
Cherenkov camera.

\section{Mirror alignment technique}

The adjustable mirror unit consists of a support triangle carrying one fixed
mirror support point and two motor-driven actuators.
A motor unit includes the drive motor, two Hall sensors shifted by $90^\circ$
sensing the motor revolutions and providing four TTL signals per turn, and a
55:1 worm gear.
The motor is directly coupled to a 12\,mm threaded bolt, driving the actuator
shaft by 0.75\,mm per revolution.
One count of the Hall sensor corresponds to a step size of 3.4~$\mu$m, or
0.013\,mrad tilt of the mirror.
The total range of an actuator is about 28\,mm which corresponds to 6.15$^\circ$
tilt of the mirror facet.

The alignment uses the image of an appropriate star on the closed lid of the
PMT camera.
The required optical feedback is provided by a CCD camera at the center of the
dish, which is viewing the lid as illustrated in Fig.\,1 (left).
Individual mirror facets are adjusted such that all star images are combined
into a single spot at the center of the PMT camera.
The basic algorithm is as follows: a CCD image of the camera lid is taken.
The two actuators of a mirror facet are then moved one by one, changing the
location of the corresponding spot on the lid.
These displacements are recorded by the CCD camera and provide all information
required to subsequently position the spot at the center of the main focus.
This procedure is repeated for all mirror facets in sequence.

It is -- to our knowledge -- the first time that such a technique is used to
align the mirrors of Cherenkov telescopes.
The major advantages of this approach are evident: a natural point-like source
at infinite distance is directly imaged in the focal plane, and the alignment
can be performed at the optimum elevation.
\begin{figure}[t]
\begin{center}
\includegraphics[width=5.5cm]{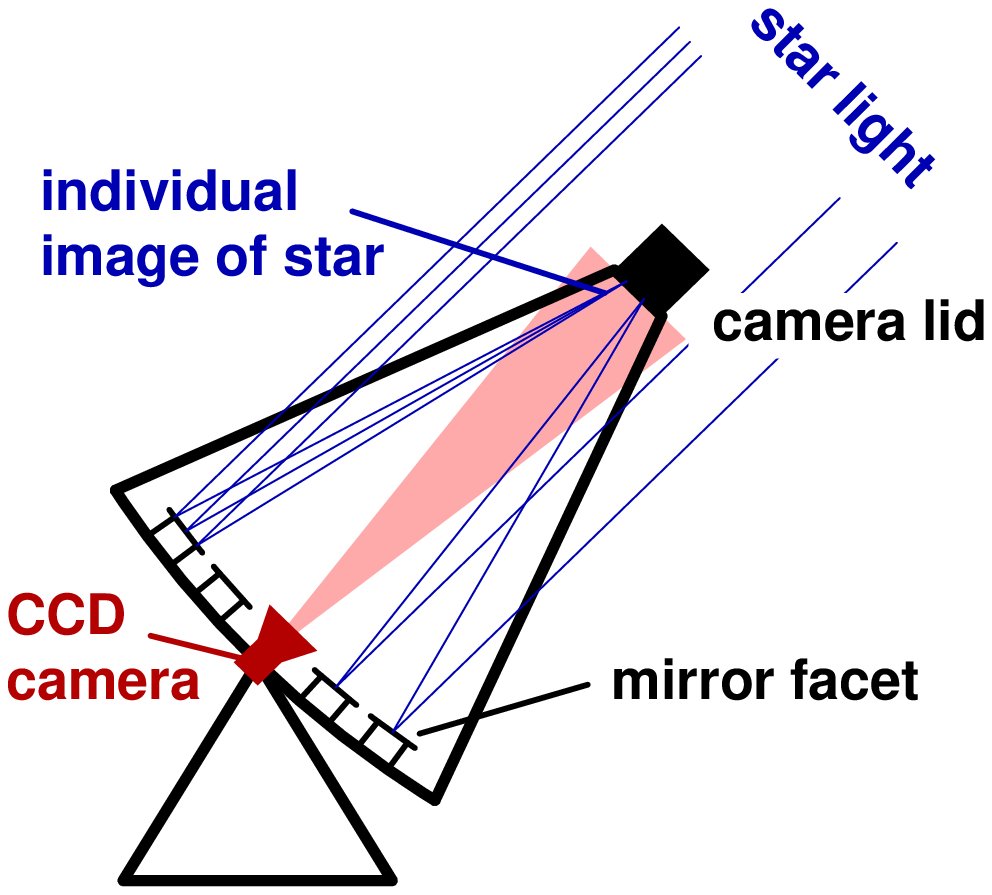}
\hspace*{1cm}
\includegraphics[width=6.2cm]{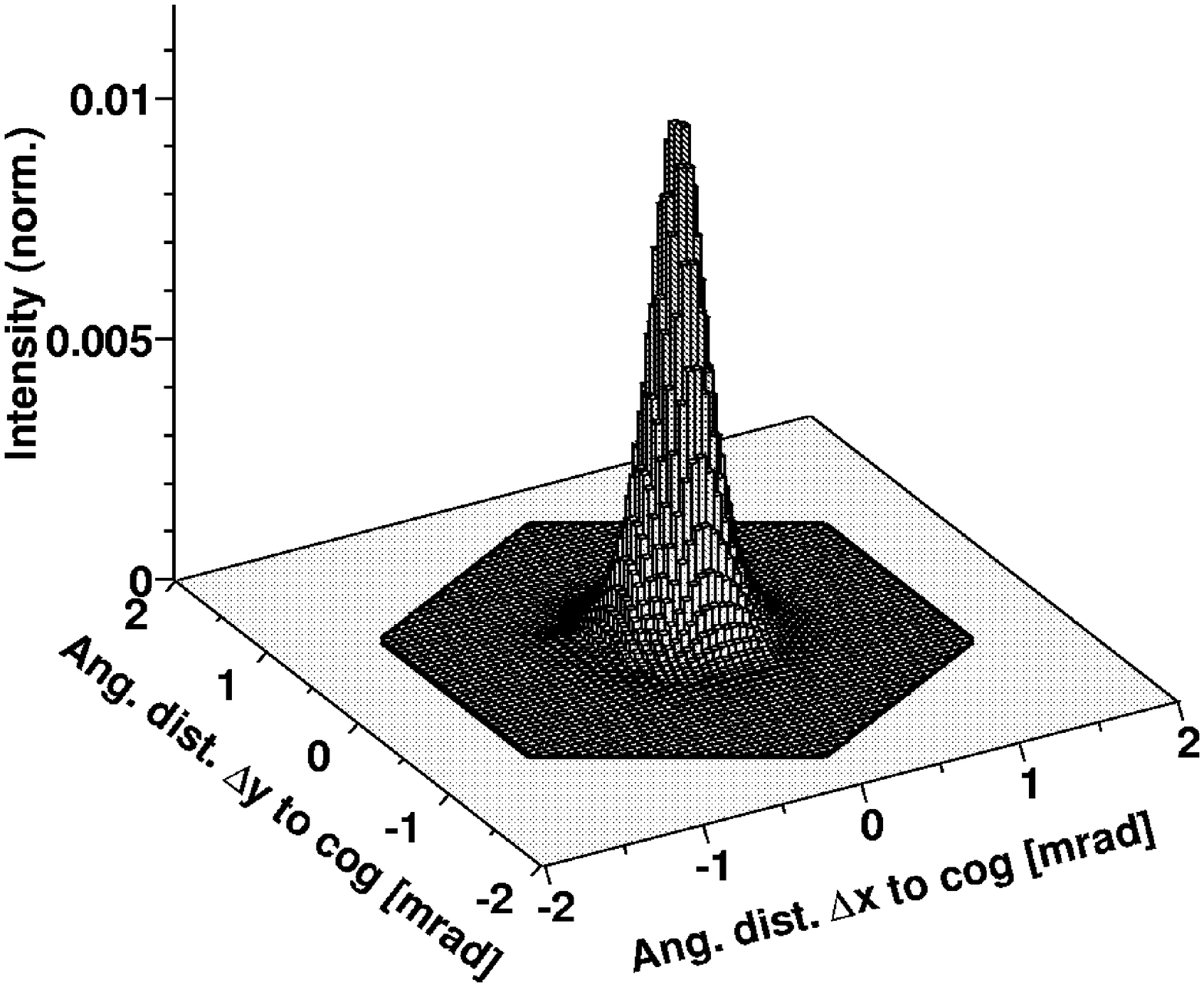}
\end{center}
\vspace{-0.7pc}
\caption{\emph{Left:}
Mirror alignment technique.
The telescope is pointed towards an appropriate star whereupon all mirror
facets generate individual images of this star in the focal plane.
Actuator movements change the location of the corresponding image which is
observed by a CCD camera.
\emph{Right:}
On-axis intensity distribution of a star on the camera lid after alignment.
The hexagonal border indicates the size of a pixel of the PMT camera defined by
the shape of the Winston cones.}
\label{fig::alignment}
\end{figure}

\section{Point spread function}

Fig.\,1 (right) shows a CCD image of the image of a star on the camera lid after
the alignment of all mirror facets in relation to the size of a PMT pixel
(0.16$^\circ$ diameter).
The intensity distribution represents the on-axis point spread function for
telescope elevations within the range used for the alignment
(55$^\circ$--75$^\circ$).
The distribution is symmetrical without pronounced substructure and the width
of the spot is well below the PMT pixel size.

To parameterize the width of the intensity distributions, different quantities
are used: the rms width $\sigma_{proj}$ of the projected (1-dimensional)
distributions and the radius $r_{80\%}$ of a circle around the center of gravity
of the image, containing 80\% of the total intensity.
On the optical axis, the point spread function is characterized by the values
$\sigma_{proj} = 0.23$\,mrad and $r_{80\%} = 0.41$\,mrad
(requirements: $\le$0.5 and $\le$0.9\,mrad, respectively).
This is an excellent result.

\subsection{Variation of the point spread function across the field of view}

\begin{figure}[t]
\begin{center}
\includegraphics[width=5.8cm]{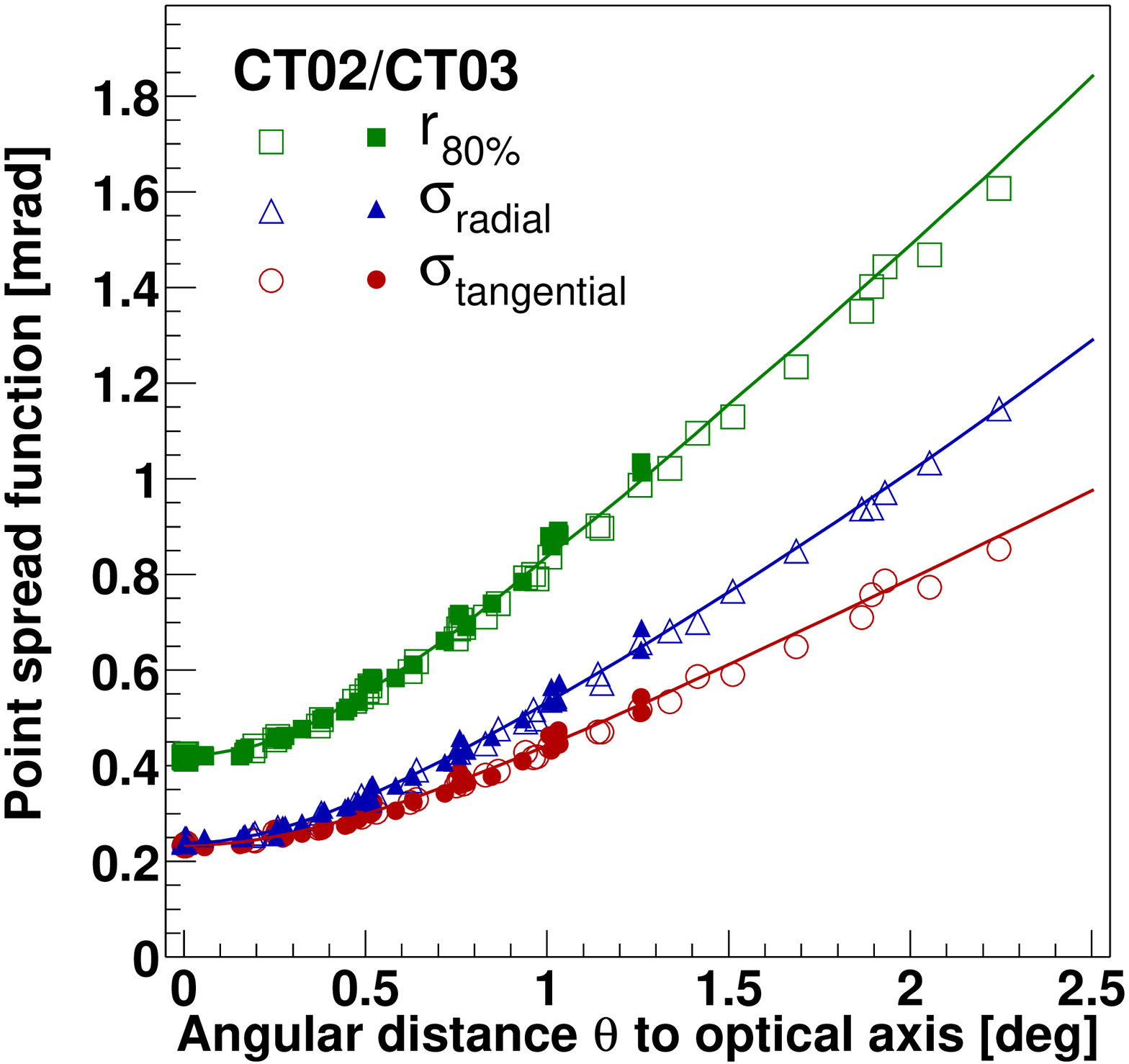}
\hspace*{1cm}
\includegraphics[width=5.8cm]{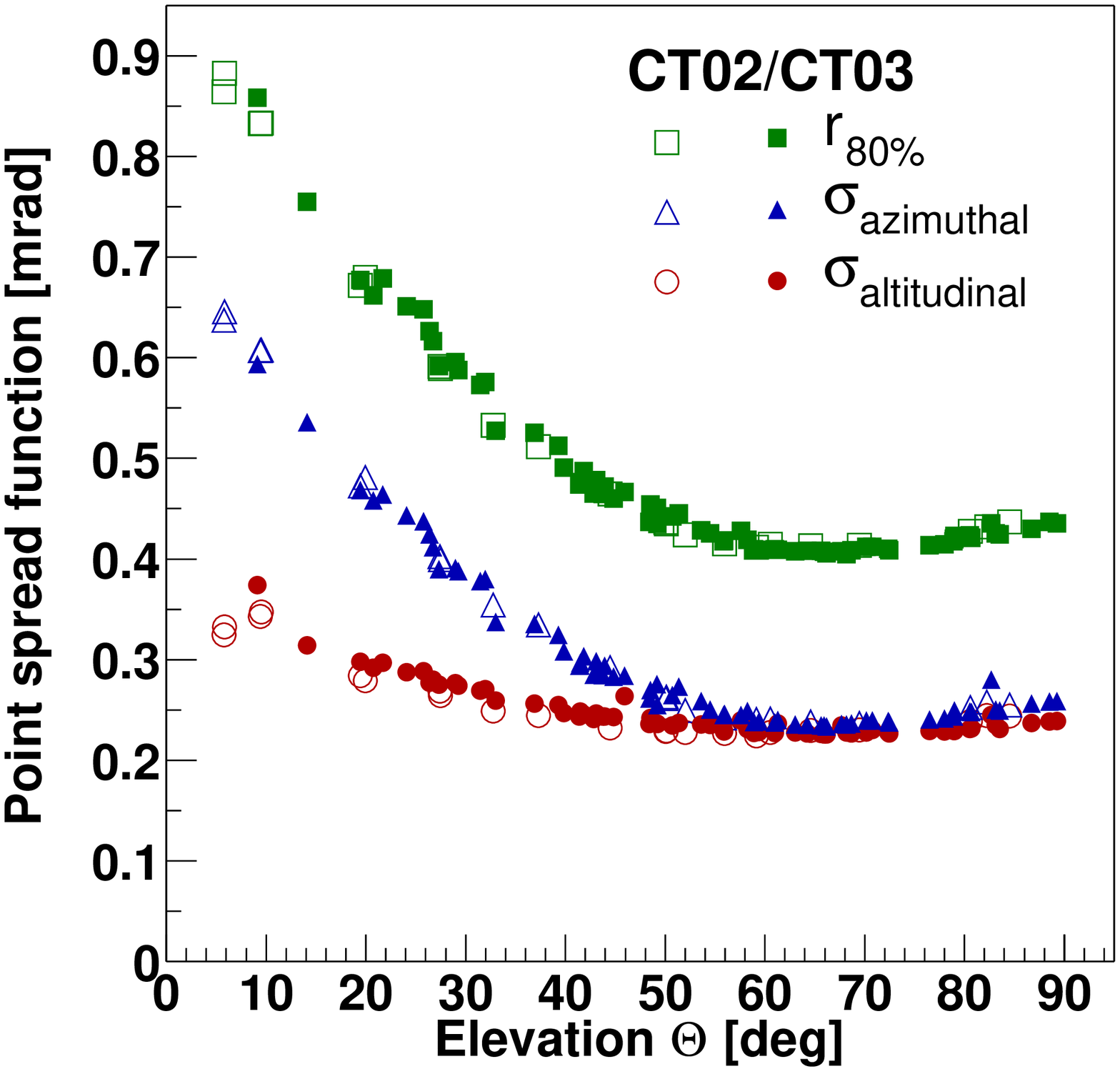}
\end{center}
\vspace{-0.7pc}
\caption{Point spread function of the first two operational H.E.S.S.\ telescopes
(CT03 and CT02).
\emph{Left:}
Width of the point spread function as a function of the angular distance
$\theta$ to the optical axis at elevations around $65^\circ$.
\emph{Right:}
Width of the point spread function as a function of telescope elevation
$\Theta$.}
\label{fig::psf}
\end{figure}
Optical aberrations are significant in Cherenkov telescopes due to their
single-mirror design without corrective elements and their modest $f/d$ ratios.
At some distance from the optical axis, the width of the point spread function
is therefore expected to grow linearly with the angle $\theta$ to the optical
axis.
For elevation angles around $65^\circ$, where the mirror facets were aligned,
Fig.\,2 (left) summarizes the spot parameters as a function of the angle
$\theta$.
Besides $r_{80\%}$, the rms widths of the distributions projected on the
radial ($\sigma_{radial}$) and tangential ($\sigma_{tangential}$) directions
are given.
The measurements demonstrate that the spot width primarily depends on $\theta$;
no other systematic trend has been found and the width $r_{80\%}$ is well
described by
\begin{equation}
r_{80\%} = (0.42^2+0.71^2\theta^2)^{^1/_2} \quad [\textrm{mrad}].
\end{equation}

To verify that the measured intensity distribution is quantitatively understood,
Monte Carlo simulations of the actual optical system were performed, including
the exact locations of all mirrors, shadowing by camera masts, the measured
average spot size of the mirror facets, and the simulated precision of the
alignment algorithm.
The results are included in Fig.\,2 (left) as solid lines, and are in good
agreement with the measurements:
\begin{equation}
r_{80\%} = (0.42^2+0.72^2\theta^2)^{^1/_2} \quad [\textrm{mrad}].
\end{equation}

\subsection{Variation of the point spread function with telescope pointing}

At fixed elevation, no significant dependence of the point spread function on
telescope azimuth was observed.
In contrast, a variation with elevation is expected due to gravity-induced
deformations of the telescope structure.
Fig.\,2 (right) illustrates how the spot widths $r_{80\%}$,
$\sigma_{azimuthal}$, and $\sigma_{altitudinal}$ change with elevation $\Theta$.
The width $r_{80\%}$ is to a good approximation described by
\begin{equation}
r_{80\%} = (0.41^2+0.96^2(\sin\Theta - \sin 66^\circ)^2)^{^1/_2}
\quad [\textrm{mrad}].
\end{equation}
For elevations most relevant for observations, i.e.\ above $45^\circ$, the spot
size $r_{80\%}$ varies by less than 10\%.
At $30^\circ$ it is about 40\% larger than the minimum size but still well
below the size of the PMT pixels.
A detailed analysis of the deformation of the support structure [2,4] revealed
that the stiffness is slightly better than initially expected from finite
element simulations.

\section{Conclusion}

The mirror alignment of the first two H.E.S.S.\ telescopes was a proof of
concept and a test of all technologies involved:
mechanics, electronics, software, algorithms, and the alignment technique
itself.
All components work as expected and the resulting point spread function
significantly exceeds the specifications.
Both reflectors behave almost identical which demontrates the high accuracy of
the support structure and the reproducibility of the alignment process.

\section*{References}

\re
1.\ Bernl\"ohr~K., Carrol~O., Cornils~R., Elfahem~S.\ et al.\ 2003,
The optical system of the H.E.S.S.\ imaging atmospheric Cherenkov telescopes,
Part~I, accepted
\re
2.\ Cornils~R., Gillessen~S., Jung~I., Hofmann~W.\ et~al.\ 2003,
The optical system of the H.E.S.S.\ imaging atmospheric Cherenkov telescopes,
Part~II, accepted
\re
3.\ Cornils~R., Jung~I.\ 2001, Proc.\ of the 27th Int.\ Cosmic Ray Conf.,
eds.\ Simon~M., Lorenz~E., and Pohl~M.\ (Copernicus Gesellschaft),
vol.~7, p2879
\re
4.\ Cornils~R., Gillessen~S., Jung~I., Hofmann~W., Heinzelmann~G.\ 2002,
to appear in The Universe Viewed in Gamma-Rays,
(Universal Academy Press)
\re
5.\ Hofmann~W.\ 2003, these proceedings

\endofpaper

%% file: 008834-1.tex
%
%
%
%






\chapter{Calibration results for the first
two {\small H$\cdot$E$\cdot$S$\cdot$S$\cdot$} array telescopes.}

\author{
N.Leroy,$^1$ O.Bolz,$^2$ J.Guy,$^3$ I.Jung.,$^2$ I.Redondo,$^1$ L.Rolland,$^3$ J.-P.Tavernet,$^3$ K.-M.Aye,$^4$ P.Berghaus,$^5$
K.Bernl\"ohr,$^2$ P.M.Chadwick,$^4$ V.Chitnis,$^3$ M.de\,Naurois,$^3$ A.Djannati-Ata\"\i,$^5$ P.Espigat,$^5$ G.Hermann,$^2$
J.Hinton,$^2$ B.Khelifi,$^2$ A.Kohnle,$^2$ R.Le\,Gallou,$^4$ C.Masterson,$^2$ S.Pita,$^5$ T.Saitoh,$^2$ C.Th\'eoret,$^5$ P.Vincent$^3$ for the {\small H$\cdot$E$\cdot$S$\cdot$S$\cdot$} collaboration$^6$.\\
{\it 
(1) LLR-IN2P3/CNRS Ecole Polytechnique, Palaiseau, France \\
(2) Max Planck Institut f\"ur Kernphysik, Heidelberg, Germany\\
(3) LPNHE-IN2P3/CNRS Universit\'es Paris VI \& VII, Paris, France \\
(4) Department of Physics, University of Durham, Durham, United Kingdom \\
(5) PCC-IN2P3/CNRS College de France Universit\'e Paris VII, Paris, France \\
(6) {\tt http://www.mpi-hd.mpg.de/HESS/collaboration}
}
}
\vspace{-0.3cm}
\section*{Abstract}
\vspace{-0.3cm}
The first two telescopes of the {\small H$\cdot$E$\cdot$S$\cdot$S$\cdot$} stereoscopic system have been installed 
in Namibia and have been operating since June 2002 and February 2003 respectively.
Each camera [2] is equipped with 960 PMs with two amplification channels per
pixel, yielding both a large dynamic range up to 1600 photo-electrons 
and low electronic noise to get high resolution on single photo-electron signals. 
Several parallel methods have been developed to determine and monitor the various 
calibration parameters using LED systems, laser and Cherenkov events.
Results including pedestals, gains, flat-fielding and night sky 
background estimations will be presented, emphasizing the use 
of muon images for absolute calibration of the camera and mirror global 
efficiency, including lower atmosphere effects.
These methods allow a precise monitoring of the telescopes and have shown 
consistent results and 
a very good stability of the system since the start of operation.

\section{Introduction}
\vspace{-0.3cm}
The {\small H$\cdot$E$\cdot$S$\cdot$S$\cdot$} detector performance can be monitored with calibration
data obtained each night of Cherenkov observation.
Methods of calibration and monitoring using LED and laser systems, and also 
Cherenkov events from muon rings and arcs are presented here.

\section{``Classical'' calibration}
\vspace{-0.3cm}
At the initial calibration step, the ADC to photo-electron($\gamma$e) coefficient (ADC$\gamma$e) is determined 
using an LED system providing a $\sim 1\gamma$e pulsed signal; the $\gamma$e distribution
follows a Poisson distribution with an average value of $1\gamma$e.
The single $\gamma$e spectrum is described by a sum of Gaussian
functions normalized by the Poisson probability to have from 1 to $n$ $\gamma$e; the
pedestal being represented by a Gaussian function weighted by the probability of 
having zero $\gamma$e (fig. 1(a)).
The gain of every pixel is monitered showing good stability apart from
those PMs whose base has been damaged by a bright illumination (see 
figure 1(b)). 

The relative pixel efficiencies are measured using
a laser located at the centre of the mirror which provides 
a uniform illumination in the focal plane. The ADC$\gamma$e are then 
flat-fielded with these relative efficiencies. 
To acquire information on electronic noise (typically $0.18\gamma$e)
some data are also taken with the lid closed and the high voltages on.
Such runs provide baseline parameters to take into
account temperature dependencies in the electronics response, for example
the pedestal position is shifted by 10 ADC counts/$^{\circ}\mathrm{C}$.

\begin{figure}[t]
\vspace{-0.5cm}
 \begin{minipage}[b]{0.7\linewidth} 
 \centering
 \includegraphics[height=4.6cm]{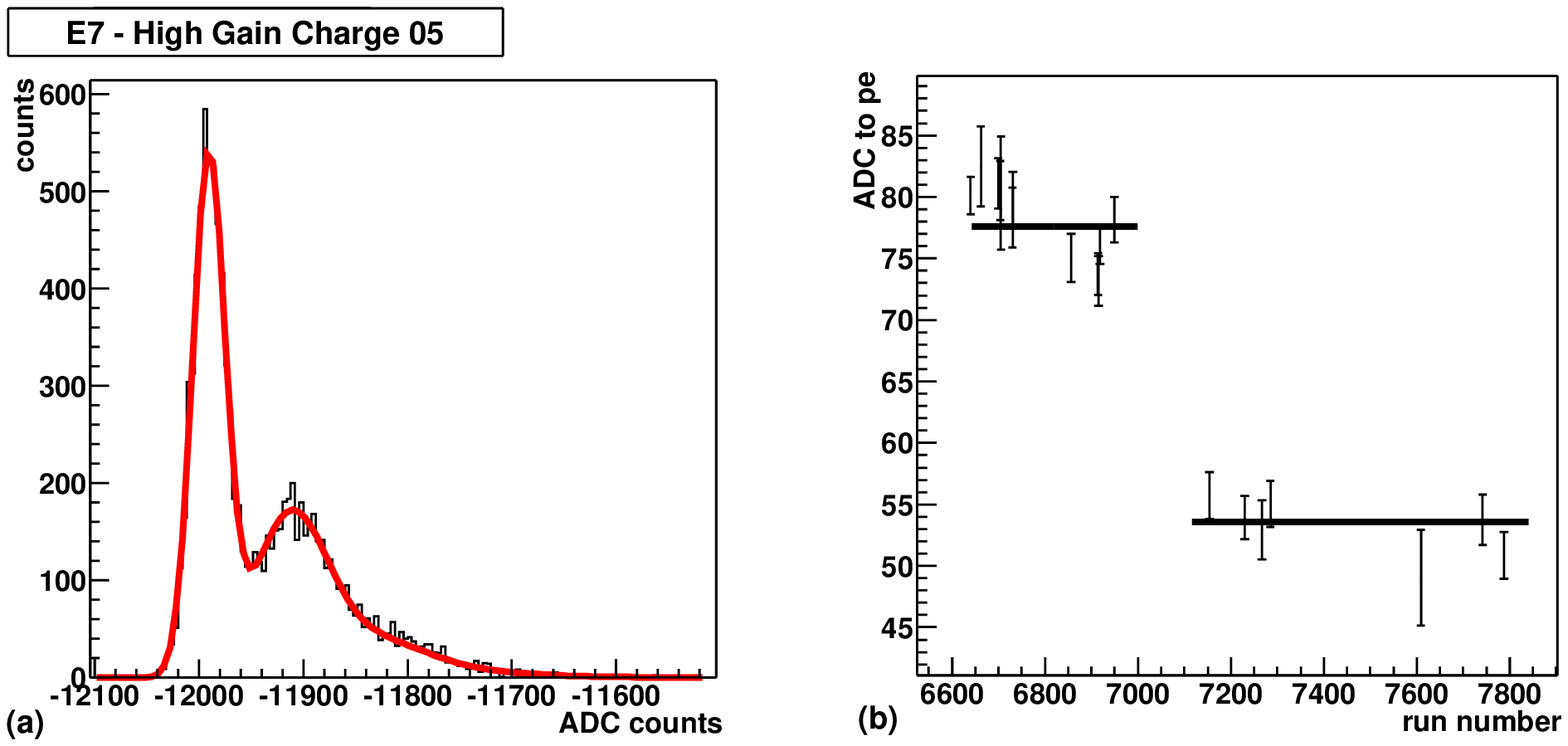}
 \end{minipage}
\hspace{0.1cm} 
 \begin{minipage}[b]{0.27\linewidth}
 \caption{(a) Example of gain determination in high gain channel. (b) Gain of one damaged pixel 
 with run number; the gain decrease is due to a too bright illumination.}
 \end{minipage}
\end{figure}

Finally, some calibration parameters are determined for each Cherenkov run.
First the pedestal position
is determined every minute of acquisition to take account
the above-mentionned temperature dependance. 
Then the Night Sky Background (NSB)
value for each pixel is determined by using the HVI (High Voltage Intensity)
shift or the
pedestal charge distribution. HVI represents the sum of the anode
and divider currents; a baseline value is determined in the runs with closed lid.

In addition, the pixels
to be excluded from the analysis are identified, for example those
with a star in the field of view, with high voltage switched off or unstable. 
Also ARS readout chips [2] (each serving 4 channels)
with incorrect read-out settings are searched for. 
After the detector commissioning, the mean number of pixels excluded from the analysis 
is $\sim$ 40 (4\%).

\section{Calibration with muon rings}
\vspace{-0.3cm}
Another useful tool for the calibration of Cherenkov telescopes is provided by muons 
produced in hadronic showers which cross the mirror,
whose Cherenkov light is emitted at low altitude
(up to 600m above each {\small H$\cdot$E$\cdot$S$\cdot$S$\cdot$} telescope). 
The intensity of the muon images can be used to measure the absolute global 
light collection efficiency of the  telescope.

\begin{figure}[t]
\vspace{-0.5cm}
 \begin{minipage}[b]{0.25\linewidth} 
 \centering
 (a)\hspace{-0.28cm}
 \includegraphics[height=5cm]{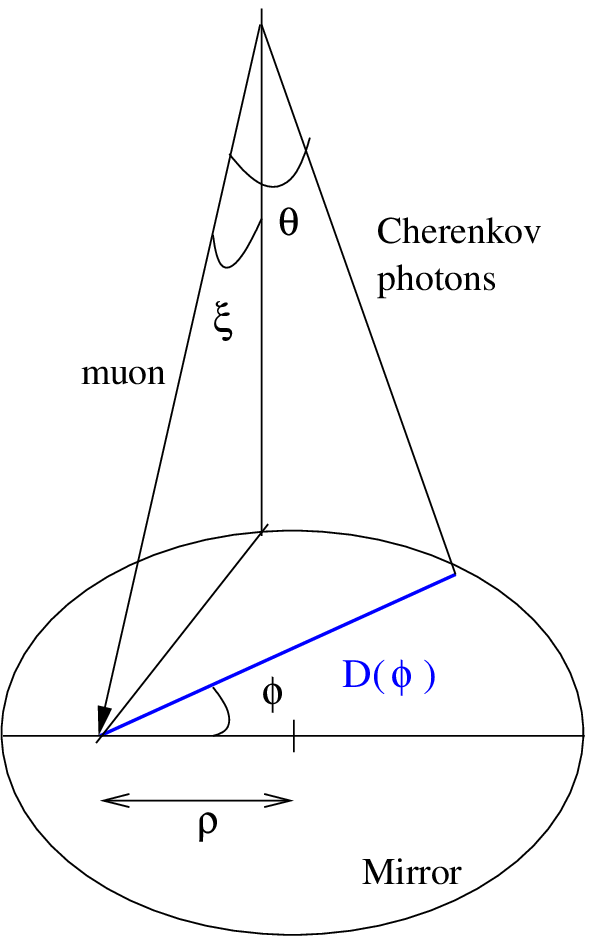}
 \end{minipage}
 \begin{minipage}[b]{0.4\linewidth} 
 \centering
 (b)\hspace{-0.18cm}
 \includegraphics[height=5cm]{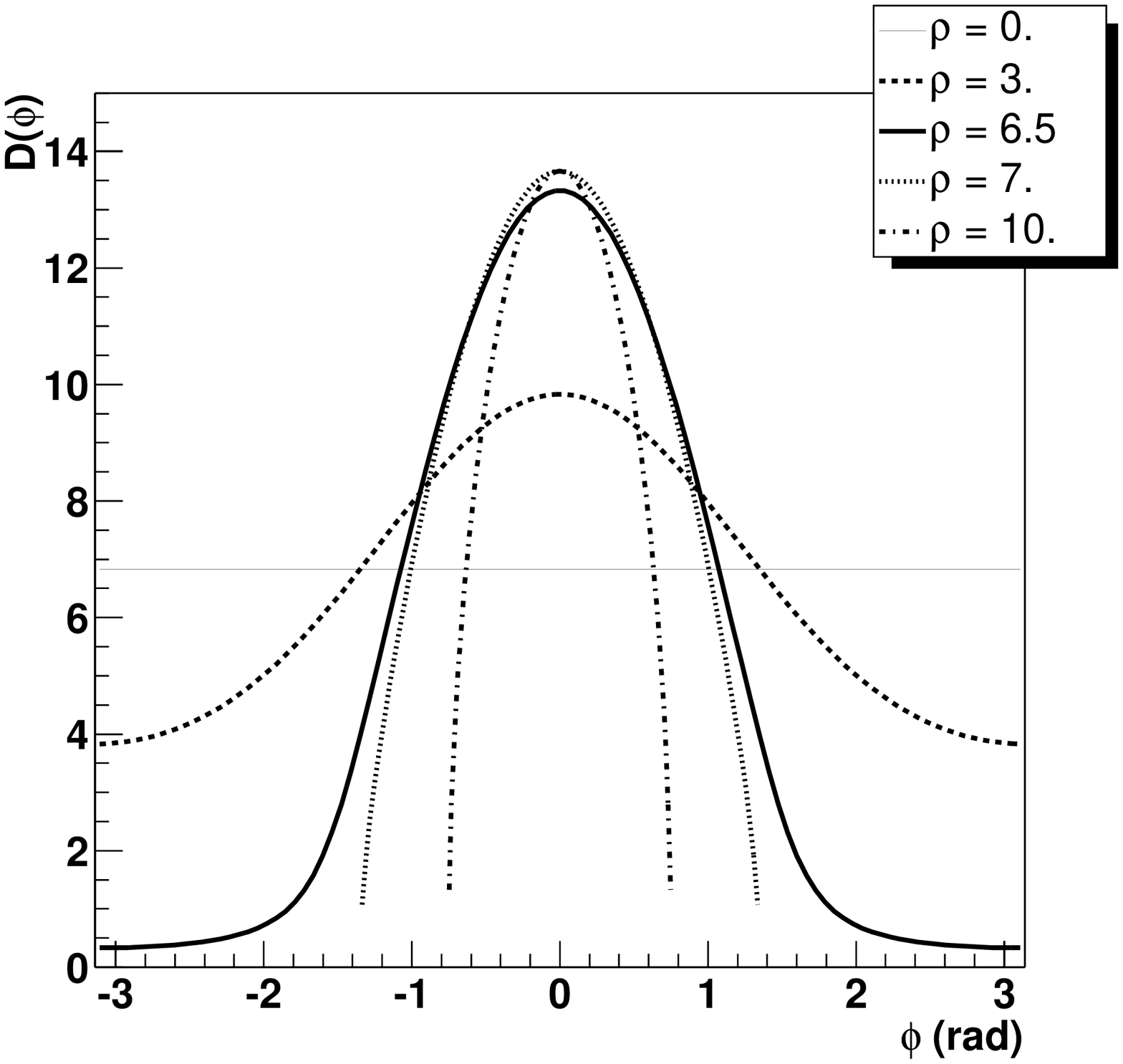}
 \end{minipage}
 \begin{minipage}[b]{0.3\linewidth}
  \caption{(a) Geometry of the Cherenkov emission from a muon near the mirror: $\xi$ is the inclination of the muon to the vertical,
$\rho$ the impact parameter, and $\theta$ the Cherenkov angle.
(b) $D(\phi)$ for different impact parameters.}
 \end{minipage}
\end{figure}

The two principal advantages 
of using muons are : 1) an easily modelled Cherenkov signal is used, and 2) the 
calibration includes all detector elements in the propagation.
For a muon impacting the telescope, the number of $\gamma e$ detected in the 
camera can be expressed [1] as 
\begin{equation}
\frac{\mathrm{d}N}{\mathrm{d}\phi}=\epsilon \frac{\alpha I}{2}\sin(2\theta)D(\phi)
\end{equation}
where $I$ is the integrated photon wavelength, 
 $\epsilon$ the average collection efficiency, $\theta$ the Cherenkov angle of the muon,
$\alpha$ the fine structure constant, $\phi$ the azimuthal angle of the pixel
in the camera and $D(\phi)$ is a geometrical factor representing the length of the 
chord defined by the intersection between the mirror surface (assumed circular and ignoring gaps between mirror tiles) and 
the plane defined by the muon track and the Cherenkov photon (see figure 2 (a) and (b)).

\begin{figure}[b]
\vspace{-0.5cm}
 \begin{minipage}[b]{0.37\linewidth} 
 \centering
 (a)\hspace{-0.18cm}
 \includegraphics[height=4.6cm]{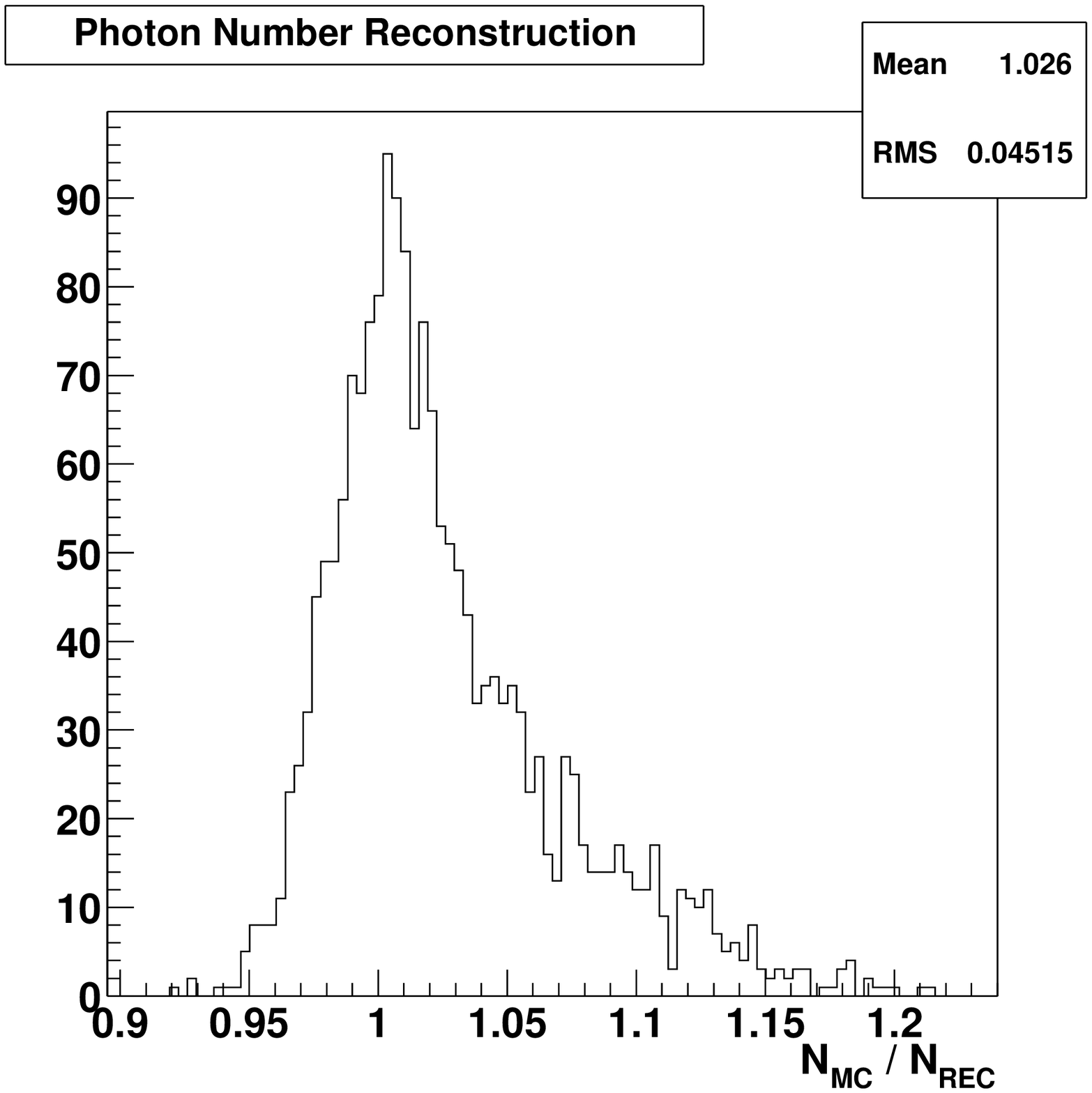}
 \end{minipage}
 \hspace{-.5cm}
 \begin{minipage}[b]{0.4\linewidth} 
 \centering
 (b)\hspace{-0.16cm}
 \includegraphics[height=4.6cm]{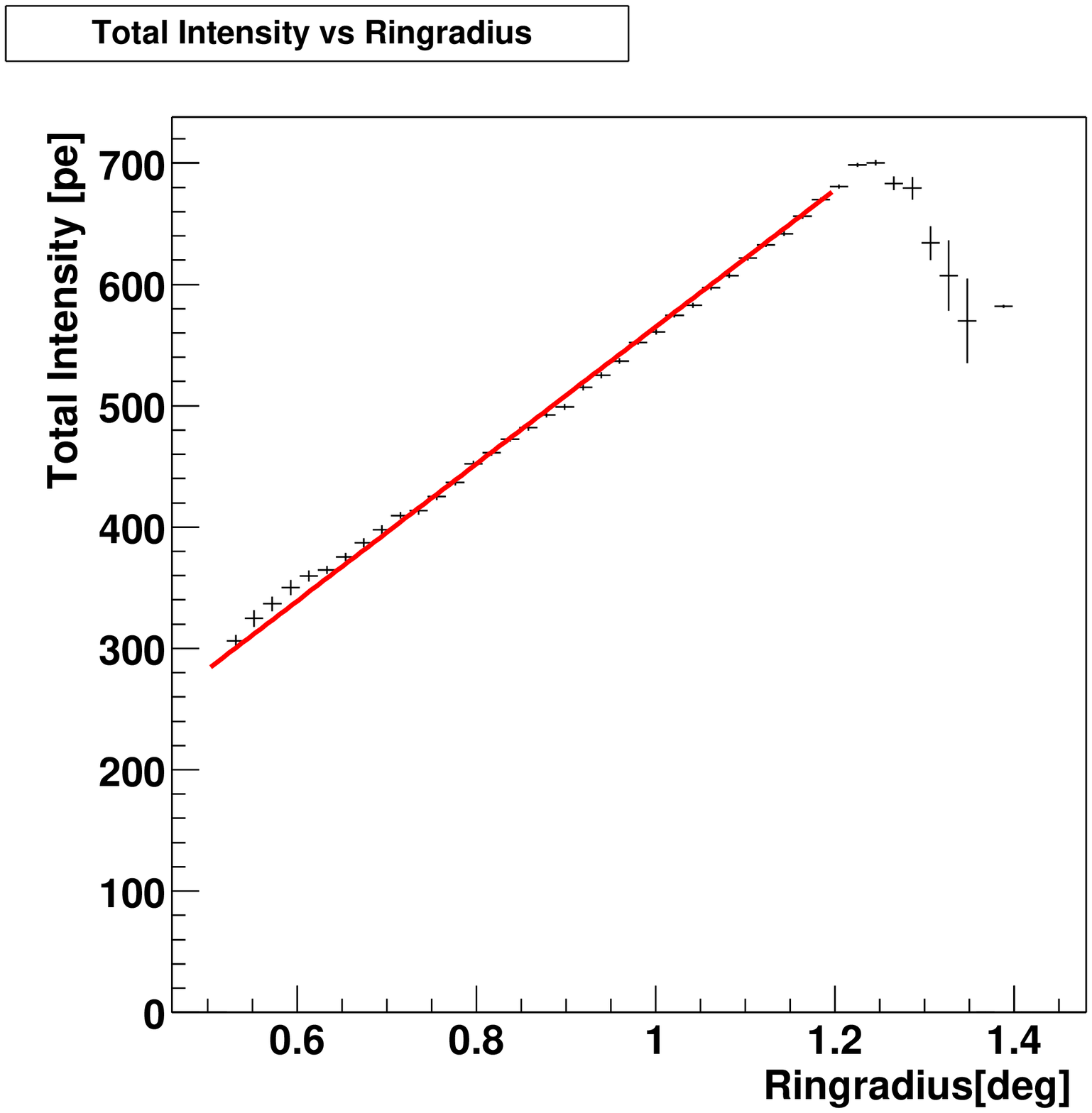}
 \end{minipage}
 \hspace{-1.cm}
 \begin{minipage}[b]{0.3\linewidth}
  \caption{(a) Distribution of the ratio between MC photons and 
  reconstructed photons with the muon analysis.
  (b) Total intensity vs. Cherenkov angle on Crab data. The solid
line is the dependance in radius from eqn. 1.}
 \end{minipage}
\end{figure}

The Cherenkov emission modelling includes the geometry of the ring 
(centre position, radius, width), impact parameter, and light collection efficiency.
These parameters are determined by a $\chi^2$ minimization.

This method has been tested by simulating muons falling near the telescope. 
Figure 3(a) shows that the number of photons generated from the MC simulation
is well reconstructed by the muon analysis.
Thus, muon data can be very useful to test the simulation of the {\small H$\cdot$E$\cdot$S$\cdot$S$\cdot$} instrument.
The model is also able to provide a good reconstruction of real data.
In figure 3(b) the solid line shows the expected dependance 
with $\theta$ (muon ring radius) from eqn. 1. The data
points beyond radius 1.2 are due to a small number of misreconstructed rings (4\% of the muon events).

Figure 4(a) shows the evolution of the collection efficiency for Cherenkov photons 
between 200 and 700 nm for observation runs from complete rings ($\sim$1~Hz).
The variations are less than 10\% and 
it is possible to see the effects of hardware changes. 
No significant correlations with zenith and azimuth angles are observed.

The presence of a large number of muon arcs ($\sim \mathrm{10\,Hz}$) 
in every data run allows us to determine the light collection efficiency of each 
individual pixel 
relatively to the rest of the camera on a run-by-run basis. 
 The relative efficiencies are determined from the mean of the 
residuals between data and the model fit in each pixel.  The residuals follow a Gaussian 
distribution with small tails due to bad or incorrectly calibrated pixels.
The RMS of these efficiencies is $\sim7$\%, consistent with laser 
measurements. 
This method, which also includes incomplete rings in order to increase the
available statistics,  provides a very sensitive monitoring tool.
The figure 4(b) shows the monitoring of two PMs, the damaged channel was
overexposed to light.

\begin{figure}[t]
\vspace{-0.3cm}
 \begin{minipage}[b]{0.45\linewidth}
 \centering
 (a)
 \includegraphics[height=4.6cm]{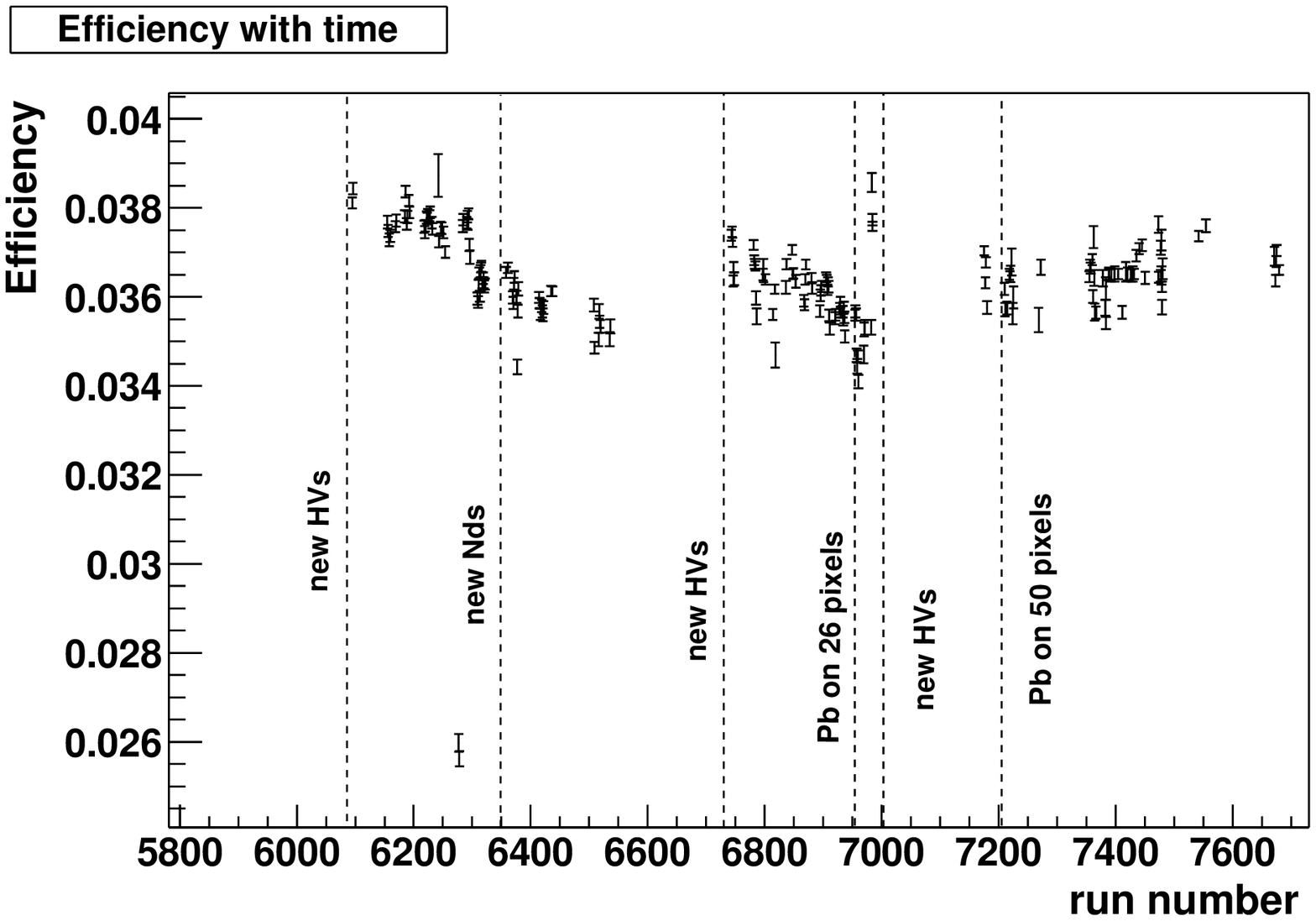}
 \end{minipage}
 \hspace{0.7cm}
 \begin{minipage}[b]{0.45\linewidth}
 \centering
 (b)
 \includegraphics[height=4.6cm]{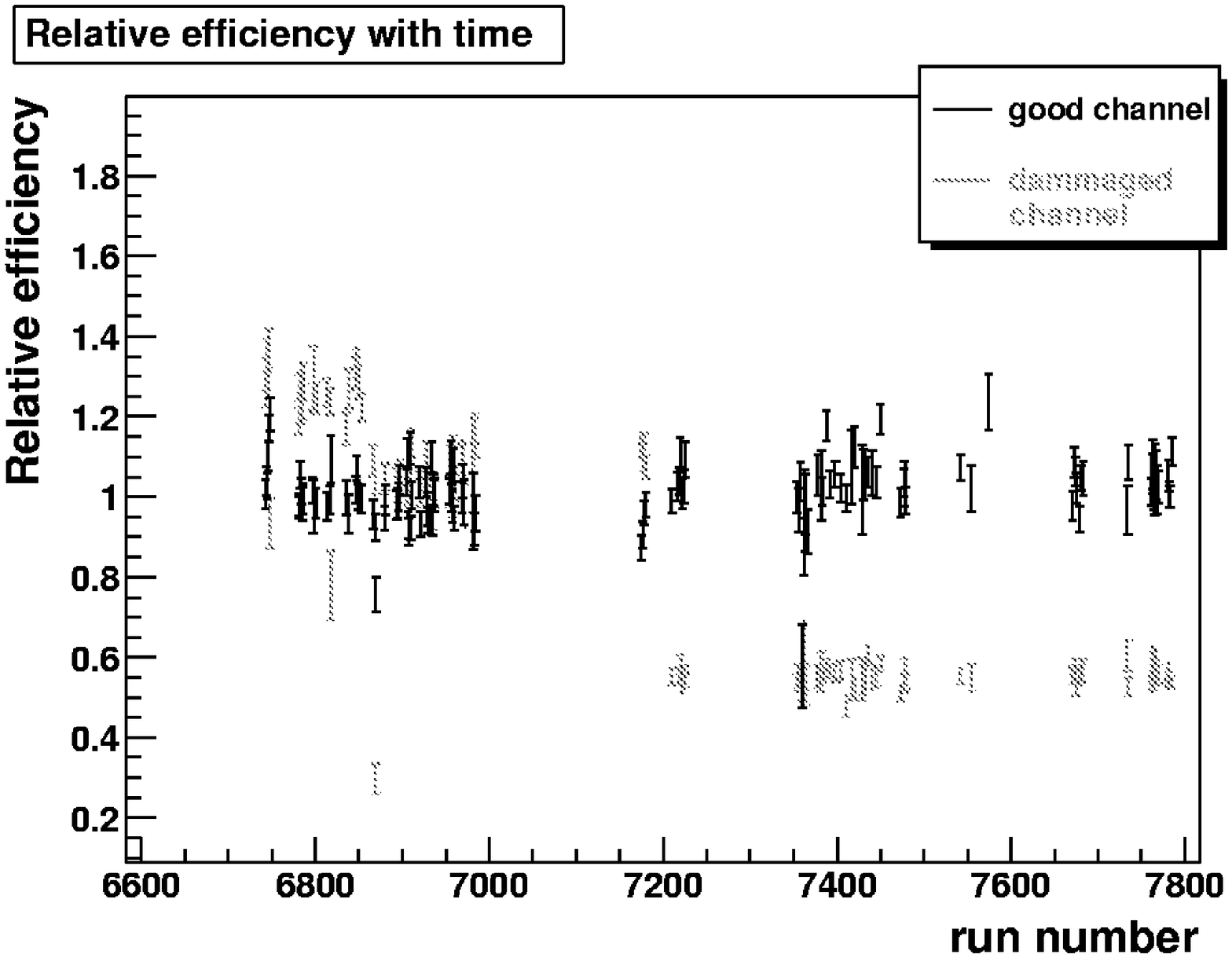}
 \end{minipage}
\caption{
(a) Collection efficiency of Cherenkov light from muon analysis with 
time. Some hardware changes are indicated. The run numbers are from October
to December 2002.
(b) Monitoring of the relative efficiencies of two channels.
}
\vspace{-.6cm}
\end{figure}

\section{Conclusion}
\vspace{-0.3cm}
The calibration methods used by the {\small H$\cdot$E$\cdot$S$\cdot$S$\cdot$} experiment utilize LED and laser systems
dedicated to this purpose and also Cherenkov images 
from local muons selected from the collected data. These independent methods allow us 
to monitor detector response on the few percent level, 
and additionally provide information for the 
selection of runs of good quality. 

\section{References}
\vspace{-0.3cm}
\re
1.\ Vacanti G. {\it et al} \ 1994, Astroparticle Physics 2, 1
\re
2.\ Vincent P. {\it et al} \ 2003, These proceedings

\endofpaper

%% file: 008971-3.tex

%
%
%

\chapter{Arcsecond Level Pointing Of The H.E.S.S. Telescopes}

\author{
Stefan Gillessen$^1$ for the H.E.S.S. collaboration$^2$\\
{\it (1) MPI f\"ur Kernphysik,
P.O. Box 103980, D-69029 Heidelberg,
Germany\\ (2)}
{\tt http://www.mpi-hd.mpg.de/HESS/collaboration}
}

\section*{Abstract}
Gamma-ray experiments using the imaging atmospheric Cherenkov
technique have a relatively modest angular resolution of typically
0.05 to 0.1 degrees per event. The centroid of a point-source emitter,
however, can be determined with much higher precision, down to a few
arcseconds for strong sources. The localization of the Crab TeV source
with HEGRA, for example, was dominated by systematic uncertainties in
telescope pointing at the 25 arcsecond level. For H.E.S.S. with its
increased sensitivity it is therefore desirable to lower the
systematic pointing error by a factor of 10 compared to HEGRA. As the
exposure times are on a nanosecond scale it is not necessary to
actively control the telescope pointing to the desired accuracy, as
one can correct the pointing offline. We demonstrate that we can
achieve the desired 3 arcseconds pointing precision in the analysis
chain by a two step procedure: a detailed mechanical pointing model is
used to predict pointing deviations, and a fine correction is derived
using stars observed in a guide telescope equipped with a CCD chip.

\section{Introduction}
Each H.E.S.S. telescope has a stiff support structure
that bears the alt/az-mounted primary mirror
and the camera in the primary focus. In combination with the drive system the
construction allows the detector to point with an accuracy of 
$\sim$60~arcseconds to any position in the sky.
However, the centroid of TeV point sources can be determined
to an accuracy of few arcseconds. 
Thus the systematic pointing error should be lowered to the same level.

Aside from an improved instrumental sensitivity
this also has astrophysical applications.
M87 [1] is one example, where arcsecond level pointing allows
for the determination of whether the TeV-emission is coming from the centre of the
galaxy or from the jet, as the separation is $\sim$10 arcseconds. 
A high pointing resolution can also contribute
to the distinction between pulsar and nebula
emission for plerionic sources hosting a pulsar. In addition a possible future
TeV detection of the galactic centre requires the localization of the emission
to a few arcseconds.

\section{Methods and Limits}
Due to mechanical imperfections of a telescope, its 
pointing is not fully determined by the axes' positions. There are pointing errors
due to: 
\begin{itemize}
\item Reproducible mechanical errors, {\it e.g.} 
the bending of the structure under gravity
or imperfectly aligned axes
\item Irreproducible effects, such
as wind loads or obstacles on the drive rails
\end{itemize} 
Reproducible errors can be determined once, then predicted and corrected in the future. 
Irreproducible effects can only be corrected by the observation of some known 
reference - preferably star light. In H.E.S.S. the approach is to 
predict reproducible errors ($1^{st}$ step) and then employ
a fine correction based on the observation of stars in parallel to
TeV observations ($2^{nd}$ step). Unfortunately the camera 
(basically a phototube array)
cannot be used to determine the correction 
as the point spread function of the dish
is comparable in size with the pixels of the array
\footnote{
If the point spread function was much bigger than a pixel
the light distribution of a star could be fit over several pixels,
if it was much smaller transits of stars from one
pixel to another could be used.}
[2].

Therefore a CCD camera is mounted on the 
dish. It observes the images of stars on the closed lid of the
camera [2]. The such recorded positions in the focal plane are related to
the phototubes using LEDs mounted in the corners of the array.
By observing many bright stars, 
the mispointing as a function of altitude and azimuth is sampled. 
Afterwards a detailed mechanical model,
whose parameters describe the reproducible mechanical errors, is fit
to the data. It allows in turn the first step correction: Given
a telescope pointing the model returns the expected mispointing.
A nice side-effect is, that the mechanical meaning of the fit parameters
can be used to monitor the mechanical stability of the telescopes.

The fine correction is determined using a guide telescope mounted 
parallel to the optical axis.
It registers stars with a second CCD camera (``SkyCCD").
After identification using a pattern match algorithm the stars' 
positions are predicted using a similar mechanical model, but with different
offsets and focal length. The observed positions will differ from the 
prediction due to the mispointing (including irreproducible
effects and reproducible errors not represented by the model). Thus the difference
measures the mispointing and it is therefore the fine correction to be applied.
 
\begin{table}[th]
 \caption{Limits of the Pointing Asccuracy}
\begin{center}
\begin{tabular}{|l|c|c|}
\hline
Principle Limits & Measured Values & Sum\\
\hline
Accuracy of the reference LED positions & 1.4'' & \\
Reproducibility of  the mechanical deformations & 1.4'' & \\
Stability of the tracking system & 1.3 '' & \\
Accuracy of star position determination & 0.8 ''& 2.5'' \\
\hline
\hline
Limits for Mechanical Model &  &  \\
\hline
Positioning accuracy of the drive system  & 3.5'' & \\
Deformations of the drive rails' shapes & 3.5'' &  \\
Principle limits (above) & 2.5'' & \\
Inaccuracy of the shaft encoders & 2 x 1.5'' & 5.9'' \\
\hline
\end{tabular}
\end{center}
\end{table}

Table 1 lists the effects that limit the
pointing accuracy when using these methods and 
shows their respective contributions as measured. 
The mechanical model should achieve a pointing resolution
of $\sim$6 arcseconds (2D rms). Using the fine correction
should further improve it to 2.5 arcseconds. 

\section{Results}

The pointing of the first two
telescopes has been studied. As the results are similar only data for
the first telescope is presented. An example of measured 
mispointing vectors as a function of 
altitude and azimuth is given in Fig.~1.
Without any correction the raw mechanical pointing accuracy is 28~arseconds
(2D rms) as Fig.~2 shows. After application of the model, the residual is 
8~arcseconds, approximately in agreement with the 
limit for the mechanical model (Table 1).

\begin{figure}[ht]
  \begin{center}
    \includegraphics[height=13pc]{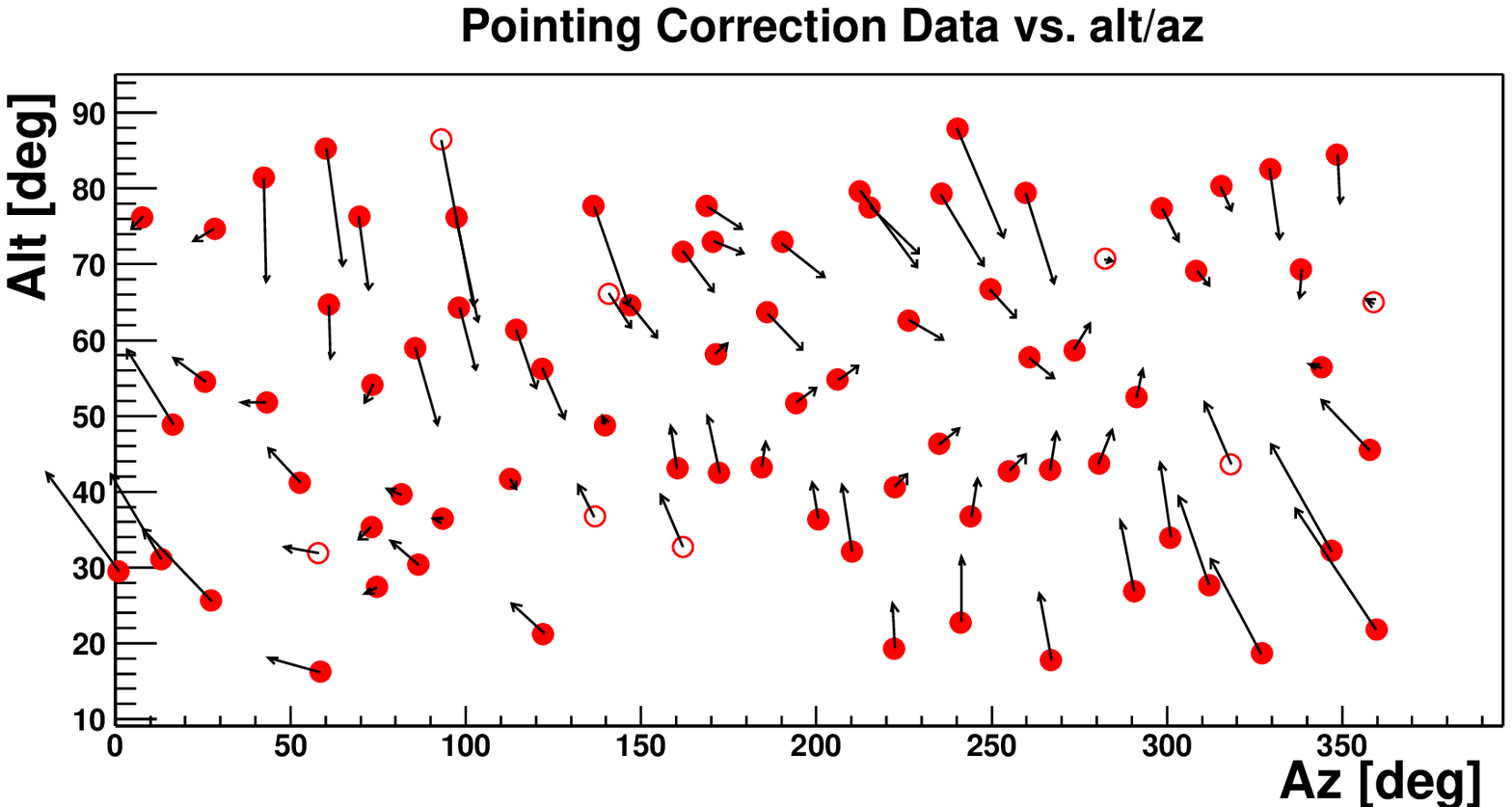}
  \end{center}
  \vspace{-0.5pc}
  \caption{Measured mispointing for the first telescope as a function of altitude and azimuth. Arrows
           are artificially enlarged}
\end{figure}

\begin{figure}
  \begin{center}
    \includegraphics[height=11pc]{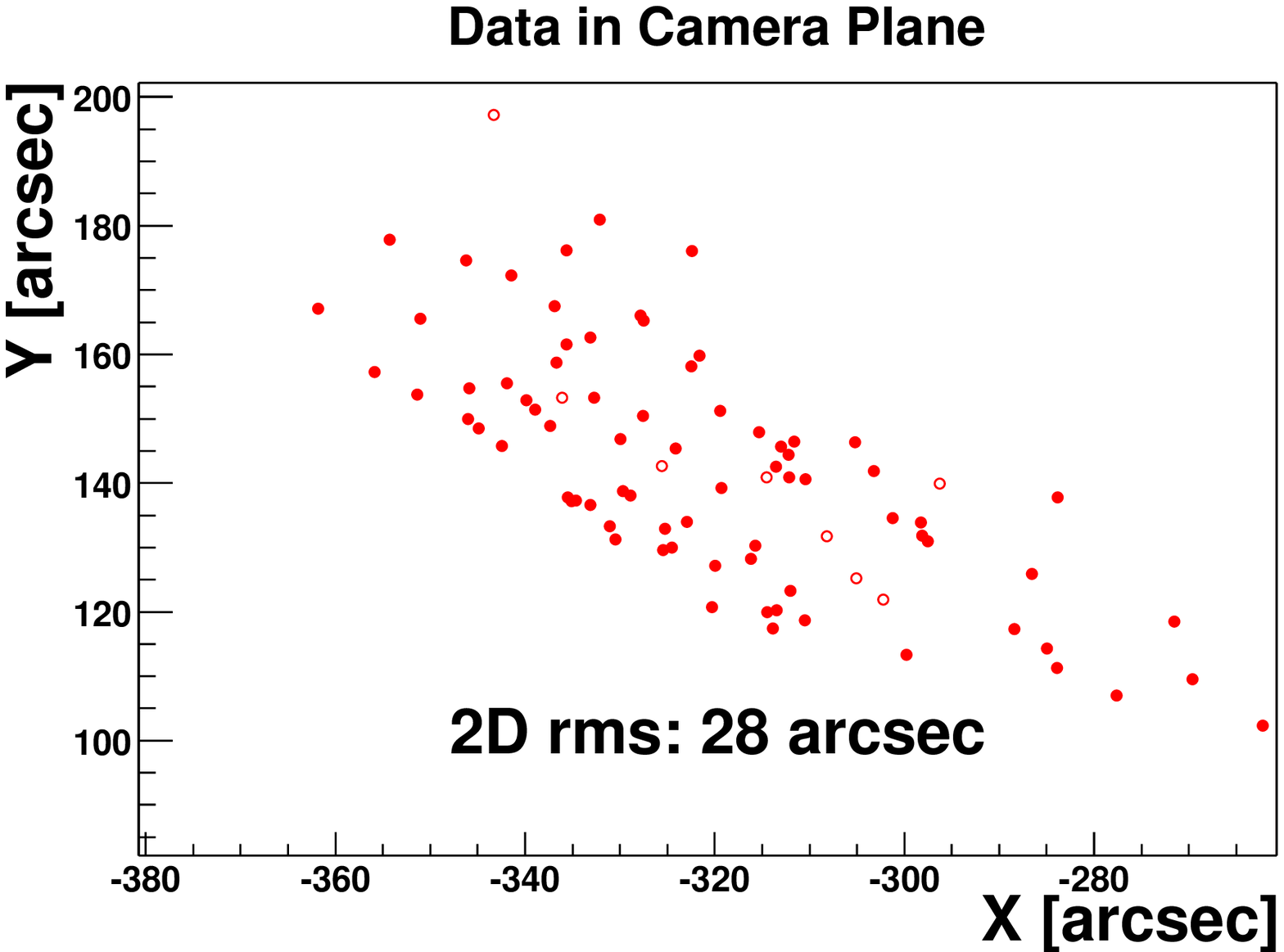}
    \includegraphics[height=11pc]{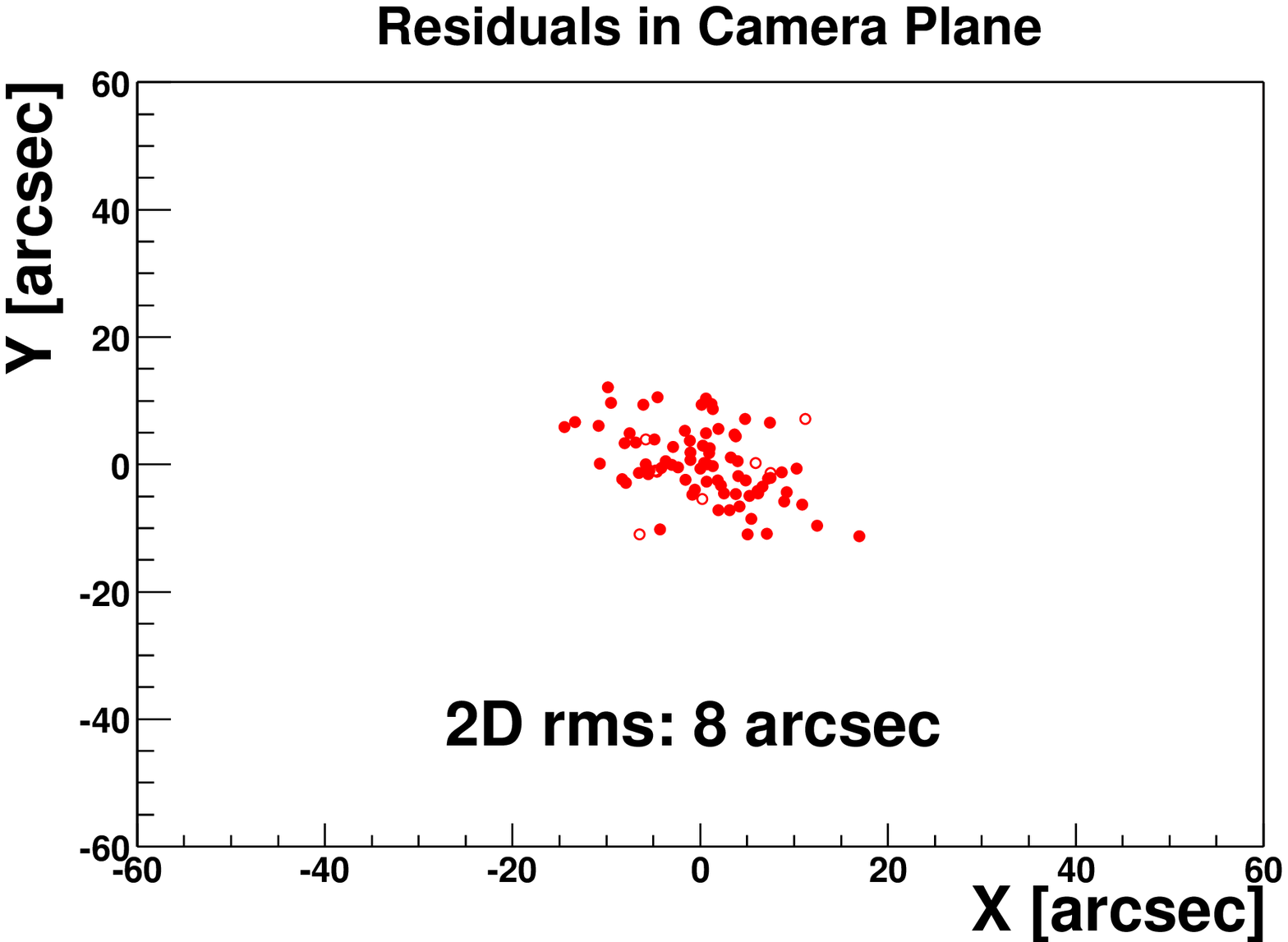}
  \end{center}
  \vspace{-0.5pc}
  \caption{Left: Raw positions 
           in the focal plane before the 
           application of the model\newline Right: After application of the fit
	   (same scale)}
\end{figure}

In order to test the fine correction, the SkyCCD has recorded the same stars
that are used for the determination of the mechanical model. 
The same kind of model is used to describe
these positions in the SkyCCD, resulting in similar residuals. 
If the residuals in the focal plane are actually due to 
mispointing, they must agree point by point with the residuals from the SkyCCD fit.
A correlation plot for two residuals is given in Fig.~3, clearly
showing the expected correlation. Thus the SkyCCD residual can be used to
measure the focal plane residual, allowing for the desired fine correction. After its
application a 2D rms of $\sim$2.5~arcseconds remains, 
which agrees with the principle limit from Table 1.
\begin{figure}[h]
  \begin{center}
    \includegraphics[height=11pc]{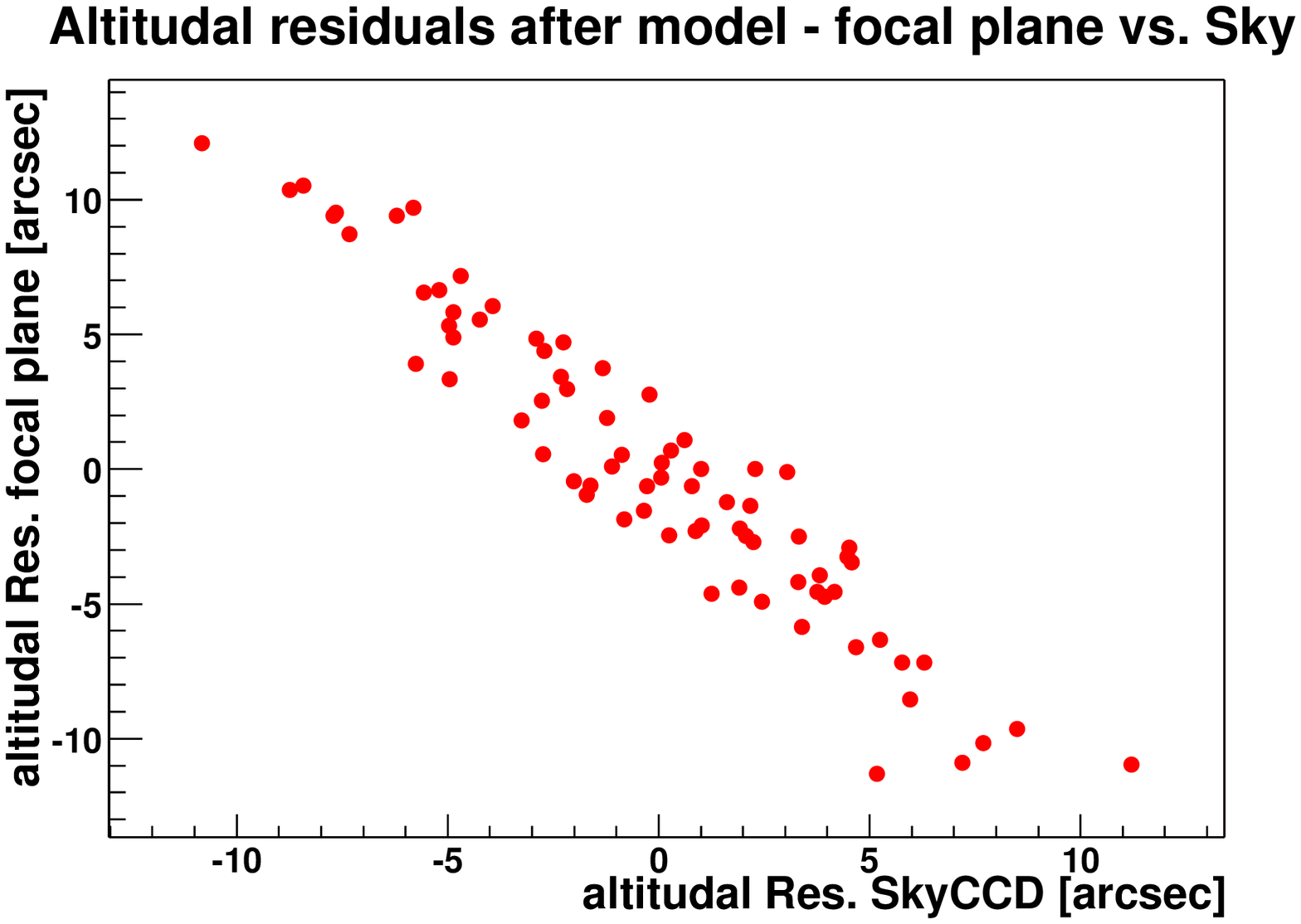}
    \includegraphics[height=11pc]{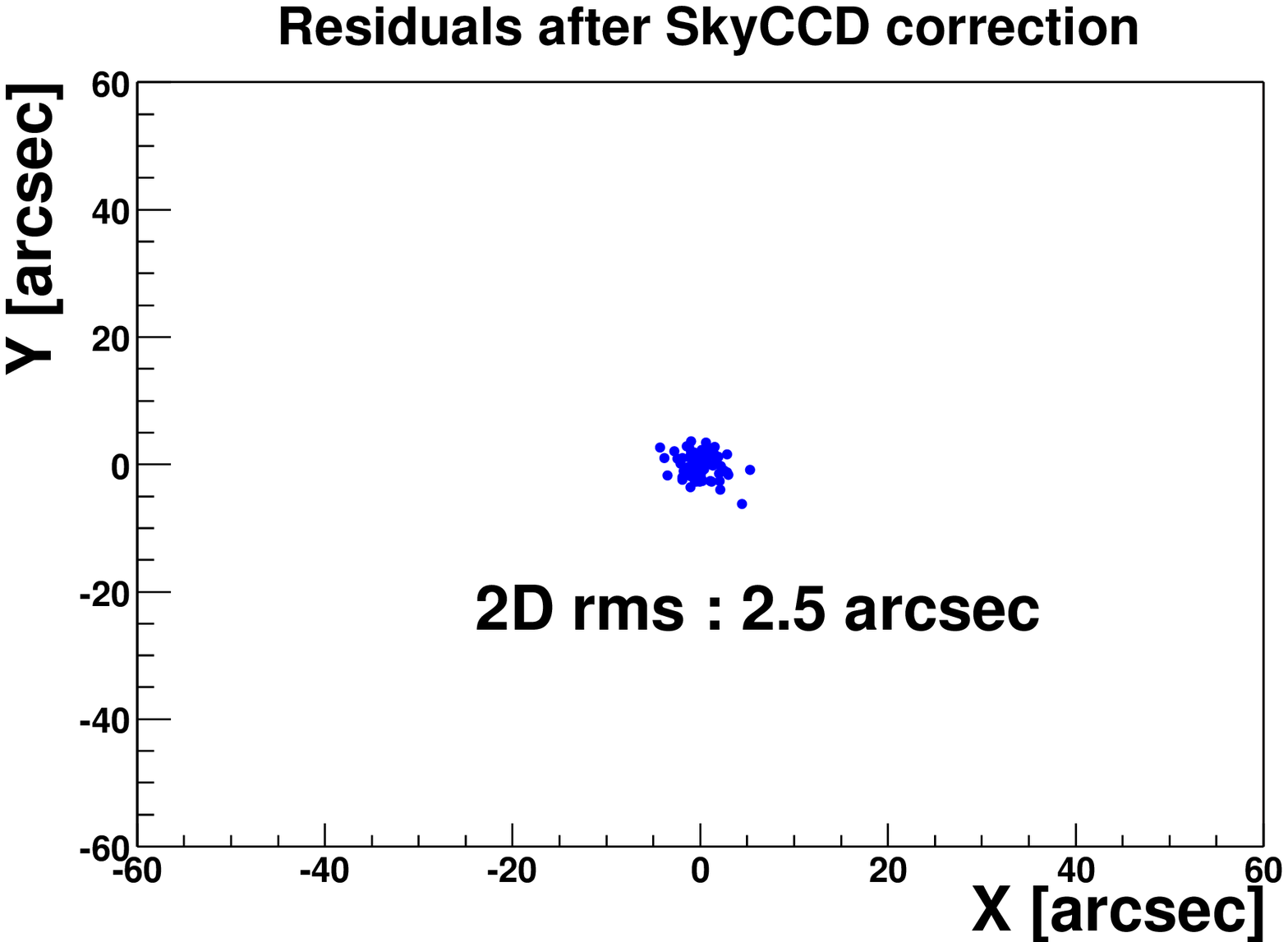}
  \end{center}
  \vspace{-0.5pc}
  \caption{Left: (Vertical) residual in the focal plane vs. 
  residual in SkyCCD\newline Right: Positions in the
  focal plane after the fine correction (same scale as Fig.~1)}
\end{figure}
\section{Conclusion}
The described two-step procedure allows the offline
pointing control of the H.E.S.S. telescopes to attain the 
desired accuracy of 2.5 arcseconds. It will
allow H.E.S.S. to address many morphological questions in TeV astronomy.

\vspace{\baselineskip}

\re
1.\ F.A. Aharonian et al. (HEGRA collaboration) \ 2003, A\&A 403, L1-L5
\re
2.\ R. Cornils et al. \  2003, The optical system of the H.E.S.S. 
imaging atmospheric Cherenkov telescopes, 
Part II, submitted to APP
\endofpaper

%% file: 008564-1.tex
%
%
%
%



%


\chapter{A Novel Alternative to UV-Lasers Used in Flat-fielding VHE {\bf $\gamma$}-ray Telescopes}

\author{K.-M. Aye,$^1$ P.M. Chadwick,$^1$ C. Hadjichristidis,$^1$ I.J. Latham,$^1$ R. Le Gallou,$^1$ \\
 A. Noutsos,$^1$ T.J.L. McComb,$^1$ J. McKenny,$^1$ J.L. Osborne,$^1$ S.M. Rayner$^1$ \\
 for the H.E.S.S. collaboration \\ and J.E. McMillan.$^2$ \\
{\it (1) Department of Physics,  Rochester Building,  Science Laboratories,  University of Durham,
 Durham,  DH1 3LE,  UK \\
(2) Department of Physics and Astronomy,  University of Sheffield,  Sheffield,  S3 7RH,  UK} \\
}

\section*{Abstract}
Preliminary tests of an alternative calibration system for the H.E.S.S. telescope array show that it
is possible to replace the currently operating UV-Laser device with an optical LED apparatus. Together with complementary optics, it is able to simulate the Cherenkov flashes, while at the same time illuminating the whole of the telescope's camera uniformly. The device in question is capable of driving a fixed number of specifically chosen LEDs to produce frequent flashes of very short duration similar to the Cherenkov emission generated by electromagnetic cascades. The design of the system continues to be refined. We describe the components and the operation of the device as developed so far.

\section{Introduction}

Calibration and monitoring of H.E.S.S. cameras is crucial for the post-processing of the recorded events. Exact knowledge of the individual gain for each photomultiplier tube (PMT) is required to translate the number of photoelectrons (p.e.) produced by the PMTs into the air shower's energy. This information can be acquired by using a well defined light source situated on the telescope, simulating Cherenkov flashes, and by recording the output from the camera's PMTs. One can summarise what specifications this light source should have to optimally match Cherenkov flashes. Chiefly, the light source's flashes should be as short as a few ns and have a spectrum that, ideally, matches that of the Cherenkov flash (see Fig.\,1). Furthermore, it will have to be wide enough so as to cover the camera completely. Last but not least, its light beam should be uniform so that each PMT receives the same amount of reference light during a flash.

The flat-fielding device used with the first H.E.S.S. telescope was based on a UV-LASER which would, in conjunction with a scintillator, produce short ($<$7~ns), frequent pulses [1]. Whilst this system performs well, there are several disadvantages, notably cost, a relatively low repetition rate, the problems inherent in the use of optical fibres and poor long-term stability. We therefore wanted to improve the device for later H.E.S.S. telescopes. 

While keeping our initial requirements for such a device within acceptable standards, we have constructed a much cheaper (6 times less), much higher repetition rate (40 times more) and much easier to maintain flat-fielding system.

\begin{figure}[t]
  \begin{center}
    \includegraphics[width=0.6\textwidth, angle=-90]{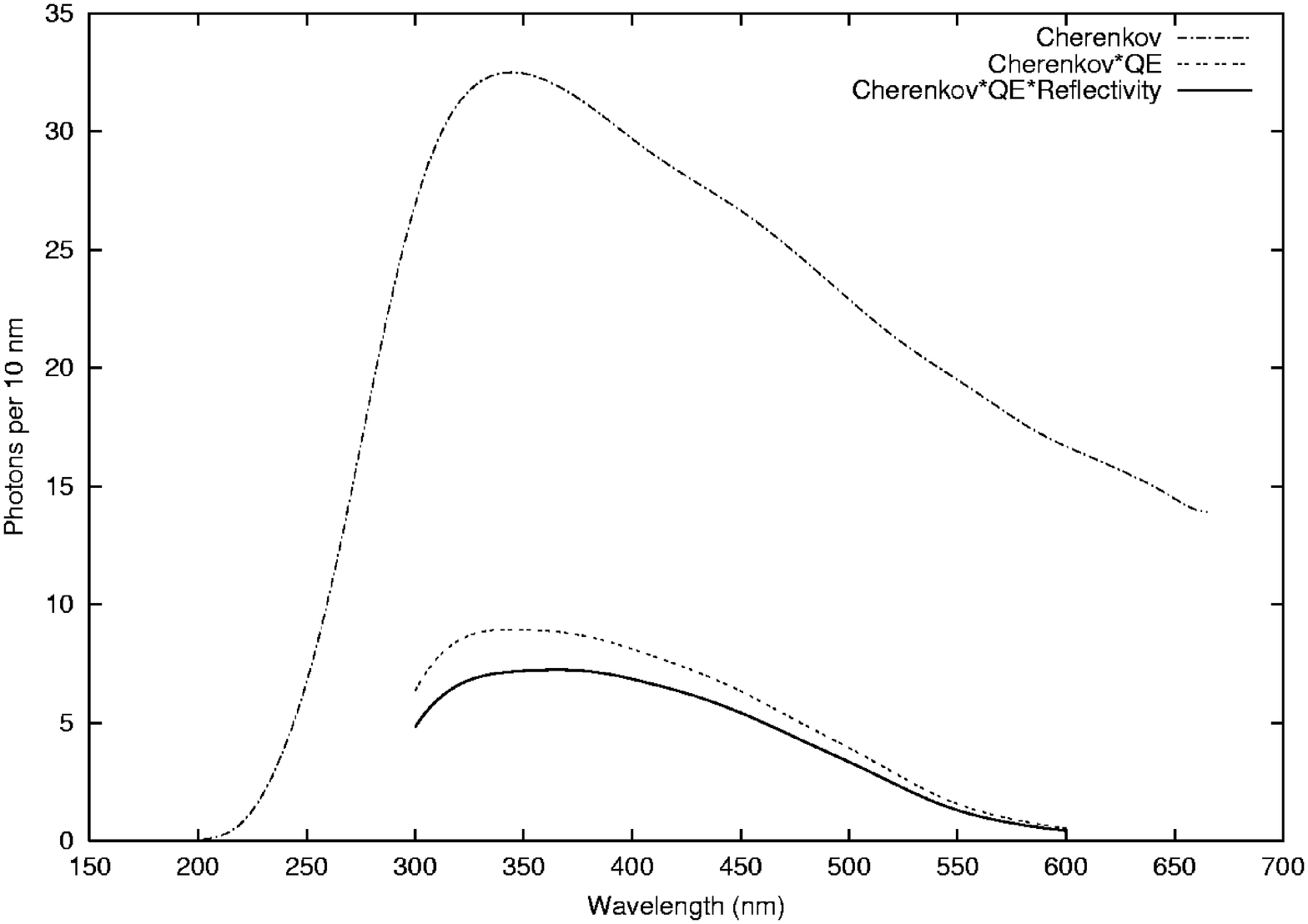}
  \end{center}
  \vspace{-0.5pc}
  \caption{The Cherenkov Spectrum for 1 TeV $\gamma$-rays at the Gamsberg plateau in Namibia (H.E.S.S. site). The dash-dotted line is derived from MOCCA simulations. The dotted line represents the same spectrum with the PMT's quantum efficiency (QE) included. The solid line is the simulated spectrum with both the PMT's QE and the mirror's reflectivity taken into account.} \end{figure}

\section{The System}

\subsection{General Description - Circuitry}
The flat-fielding device is composed of two independent circuits; the main circuit, which produces the actual flashes, and a monitoring sub-circuit, which monitors the light output coming from the LEDs. A complete schematic of the device is shown in Fig.\,2.

The main circuit is triggered via an RS232 interface controlled by the H.E.S.S. central data acquisition system via a W\&T Com-Server. The RS232 pulses are fed at the speed of 19600 bauds into an RS232-to-TTL converter that converts them to TTL gates. The UV-LASER system could only be triggered as fast as 25~Hz due to limitations of the laser itself. Our new system uses a trigger rate of 1~kHz, making it a useful tool for testing the response of the DAQ system under high incident event rates. 

Apart from the remote operation of the device, a built-in switch allows us to feed in TTL pulses from a local pulse generator (TTL-in mode). When set to the opposite position, the switch sends the TTL signature to an external display, thus allowing the monitoring of the remote operation (TTL-out mode).

The TTL signal, either coming from a remote or a local source, is directed into a driver, which is responsible for transforming the former into a signal compatible with the pulser circuits. These pulser circuits have been built by Sheffield University for the Antares collaboration, for timing calibration purposes [3]. There are 3 of these pulsers connected in parallel in the current configuration, all conveniently embedded on one board and each carrying a single LED. In front of the LED/pulser array there are a filterwheel, a monitoring photodiode, a diffuser and a UV-transparent window (Fig.\,2).

\begin{figure}[t]
  \begin{center}
    \includegraphics[width=1\textwidth]{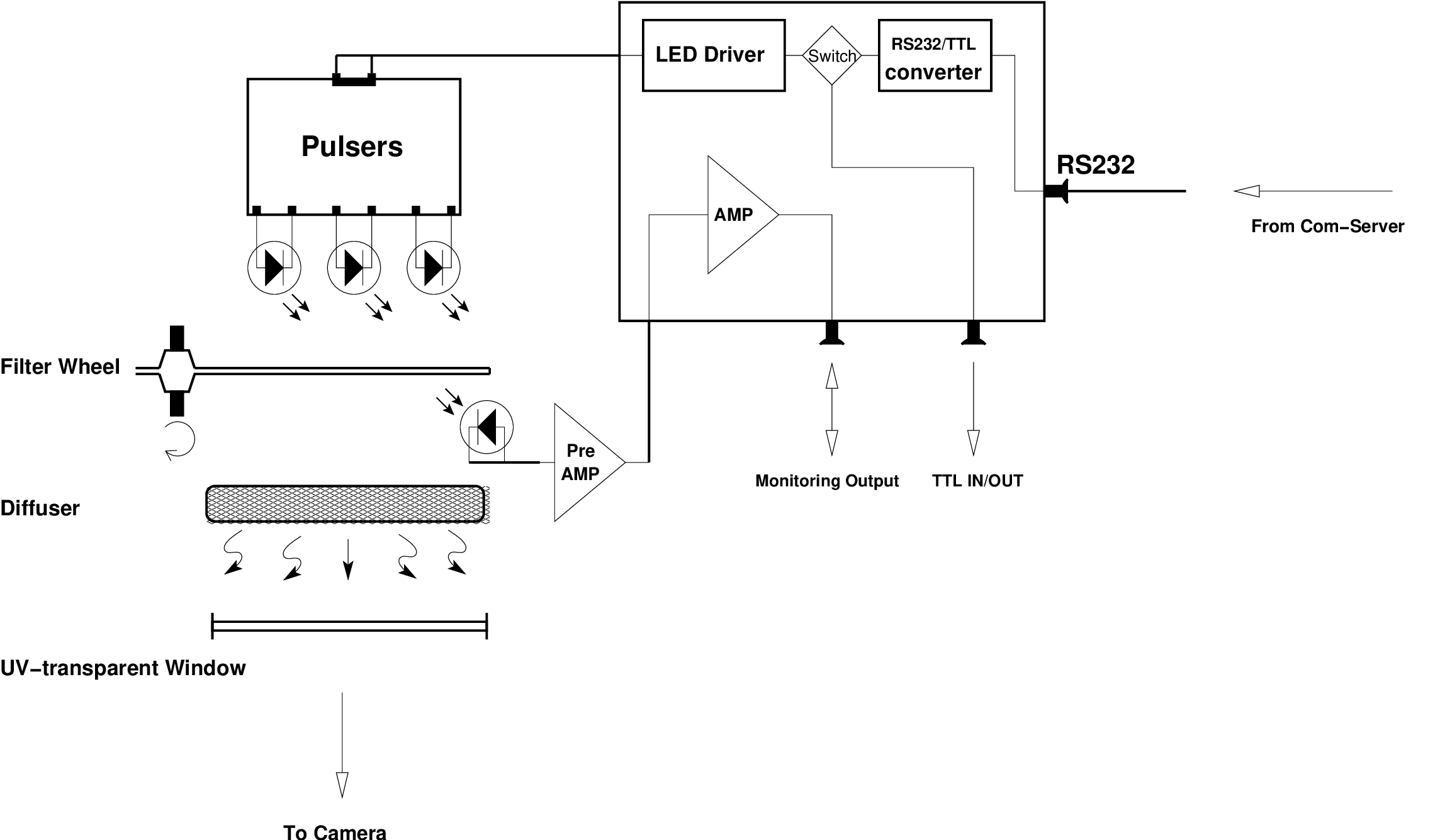}
  \end{center}
  \vspace{-0.5pc}
  \caption{The layout of the flat-fielding device.}
\end{figure}

\subsection{System components}
The `heart' of the flat-fielding system is the LED pulser developed by Sheffield University. A complete description of the device can be found in [2]. It exhibits low jitter ($<$0.5~ns) and a pulse drift rate $<$0.25~ns/y [3]. Our tests confirmed that this pulser produces pulses with a FWHM of $\simeq$5~ns and a rise time of about 2.5~ns, a significant improvement from the UV-LASER system, whose pulse rise time is 3.3~ns [1]. We also found that the pulse intensities deviate by less than 5\% RMS.
 
Most of the LEDs we tested were found to be compatible with the Sheffield pulser. The current configuration at the H.E.S.S. site uses the HUVL400-520 LEDs by HERO Electronics. Their spectrum ranges from 390 to 410~nm and their half-intensity angle is 20$^\circ$.

The response of a PMT varies with intensity and the calibration process should take into account these gain variations to result in more accurate flat-fielding. Hence, an automated filter wheel is installed in front of the LEDs to control the intensity of the light output. It is supplied by Elliot Scientific and is capable of holding 6 filters. Currently, it holds 5 neutral density filters ---supplied by Coherent-Ealing---, graded as follows: 0.5, 1.0, 1.3, 1.5 and 2.0; while one position is left empty to transmit the full intensity. By analysing the data recorded during the calibration runs, we can estimate the number of p.e. that the H.E.S.S. PMTs detected from the flat-fielding apparatus. Using the filter wheel settings we can get a photon flux per pulse ranging from 2.5 to 250~ph~cm$^{-2}$ at a distance of 15~m.

A complementary accessory that allows the monitoring of the LED intensity without having to access the dish is also part of the flat-fielding system. It is based on a photodiode and allows us to monitor the light pulses with, e.g. an oscilloscope. 

Since it is crucial to have the same photon intensity on all of the PMTs when calibrating them, it was decided to place a diffuser before the UV-window in the flat-fielding apparatus. We used a 25~mm diameter holographic diffuser by Coherent-Ealing that produces a circular diffusing output $20^\circ$ wide, for collimated incident light beams.

In order to protect the flat-fielding apparatus from the weather, we shielded it with a UV-transparent window made of borosilicate glass supplied by Edmund Optics. The window's transmissivity, in the frequency range between 350 and 800~nm, is $\approx$90\%.

\section{References}

\re 1.\ Chadwick, P.M.\ et al.\ 2001, Flat-fielding of H.E.S.S. phase I, Proc. 27th Int. Cosmic Ray  Conf., 2919

\re 2.\ McMillan, J.E.\ 2001, Using the Sheffield Pulser, private communication

\re 3.\ McMillan, J.E. on behalf of the ANTARES Collaboration \ 2001, Calibration Systems for the ANTARES Neutrino Telescope, Proc. 27th Int. Cosmic Ray  Conf., 2919

\endofpaper

%% file: 008546-2.tex
%
%
%
%






\chapter{
Atmospheric Monitoring For The H.E.S.S. Project}

\author{
%
%
K.-M. Aye, P.M. Chadwick, C. Hadjichristidis, M.K. Daniel$^1$, I.J. Latham, 
R. Le Gallou, J.C. Mansfield, T.J.L. McComb, J.M. McKenny, A. Noutsos, 
K.J. Orford, J.L. Osborne, and S.M. Rayner for the H.E.S.S. Collaboration\\
{\it Dept. of Physics, University of Durham, DH1 3LE Durham, United Kingdom}\\
\it (1) now at Department of Physics and Astronomy, Iowa State University, Ames, IA 50011-3160, USA.\\
}

\section*{Abstract}
Several atmospheric monitoring devices have been installed at the H.E.S.S. site in
Namibia. Firstly, Heitronics KT19 infrared radiometers, aligned paraxially with the  
H.E.S.S. telescopes, measure the infrared radiation of the water in clouds
crossing the telescope field of view. Correlations between the trigger rate of the 
telescope and these IR measurements are shown in this paper.
For a general judgment of the atmosphere's transmittance, i.e. the detection of any
light-attenuating aerosols, a ceilometer -- a LIDAR with built-in atmospheric data
reduction code -- is being used.
The overall status of the weather is monitored by a fully automated weatherstation.

\section{Overview}
The main causes of extinction of Cherenkov light are absorption and Rayleigh scattering by molecules, and Mie scattering by aerosols. The H.E.S.S. photomultipliers and mirrors are sensitive to light between 250 and 700 nm. In this range the only light-\emph{absorbing} molecule is ozone, but the most significant loss of Cherenkov light in the case of a `clear' sky is caused by Rayleigh \emph{scattering} off all atmospheric molecules dominant at lower wavelengths due to its $\lambda^{-4}$ dependance, and Mie scattering on aerosols which becomes dominant above approximately 400 nm [2].

\section{Weatherstation}
A UK Meteorogical Office approved weatherstation from Campbell Scientific has been installed at the H.E.S.S. site. It records air temperature, relative humidity, atmospheric pressure, wind speed, wind direction and rainfall 24 hours a day. The data acquisition (DAQ) is integrated in the standard DAQ scheme for the camera data and therefore allows efficient cross-checking of atmospheric conditions and camera data. The weather data are especially important for providing input values for atmospheric models constructed with the commercially available MODTRAN package in connection with radiometer and LIDAR (``\emph{LI}ght \emph{D}etection \emph{A}nd \emph{R}anging'') data.

\begin{figure}[t]
  \begin{center}
    \includegraphics[width=25pc]{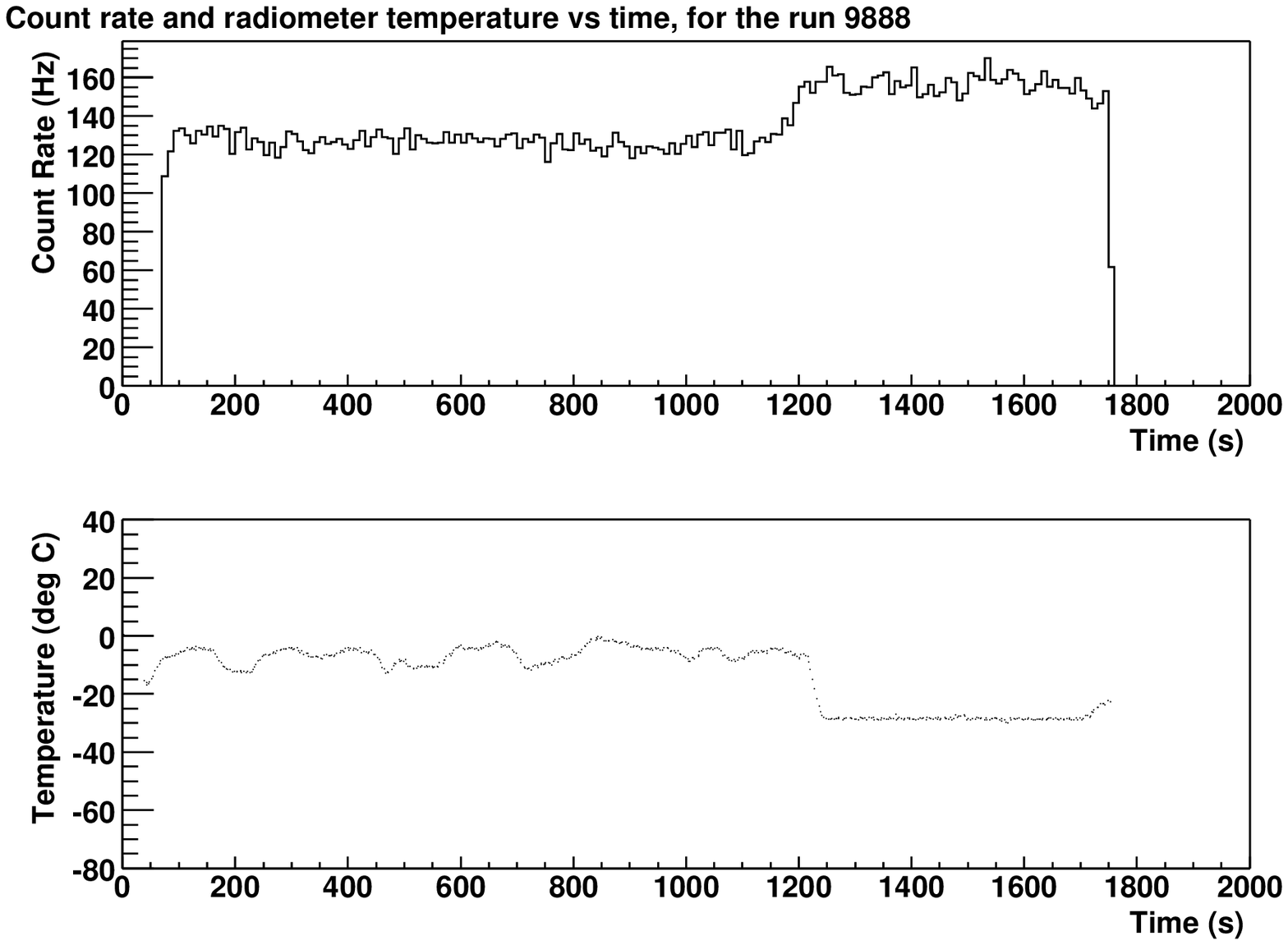}
  \end{center}
  \vspace{-0.5pc}
  \caption{Correlation between telescope count rate and radiometer temperature due to Cherenkov light absorption by cloud.}
\end{figure}

\section{The Ceilometer}
A Vaisala CT25K Ceilometer has been installed at the site. A Ceilometer is a LIDAR with cloud detection and ranging facility and built in algorithms to invert the received light power to backscatter values in units of $\mathrm{(km\cdot sr)^{-1}}$. Using an InGaAs diode laser working at $(905\pm5)$nm, the Ceilometer detects backscatter mainly due to aerosol scattering out to 7.5 km. The backscatter profile can be inverted to recreate the optical density profile for the atmosphere. This profile can be compared to profiles from model atmospheres which then can be used to calculate the extinction in the wavelength range of interest, i.e. 250 to 700 nm. Results of preliminary studies can be found in a parallel paper [1].  
  
\begin{figure}[t]
  \begin{center}
    \includegraphics[height=14.5pc]{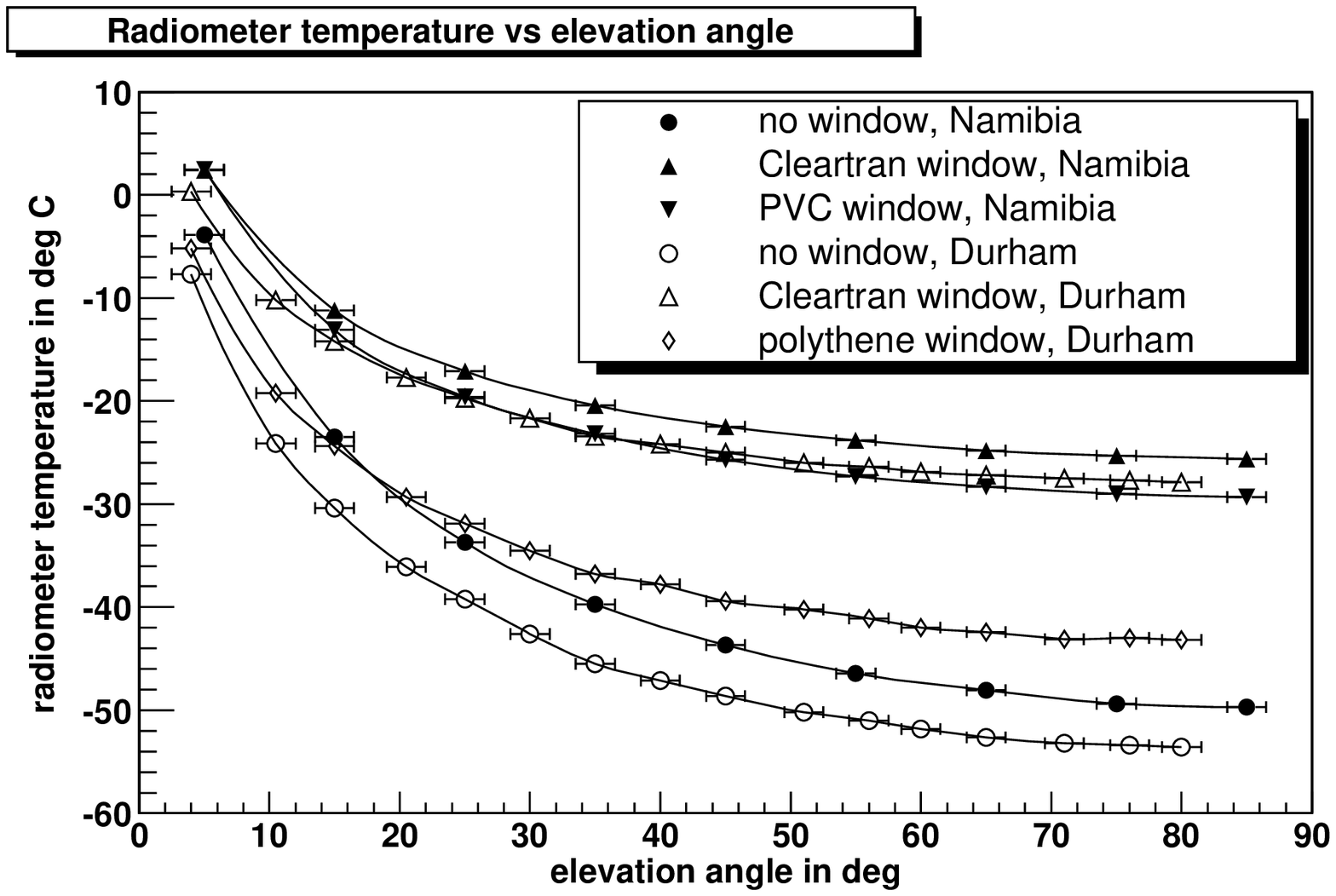}
  \end{center}
  \vspace{-0.5pc}
  \caption{The relation between sky temperature and elevation angle of the radiometer with different winow materials. Ambient conditions in Namibia: nighttime, $T=16^\circ C$, rel.Humidity 41\%; Durham: afternoon, $T=5^\circ C$, rel.~Humidity $(70\pm10)\%$  }
\end{figure}
\label{fig:zenithAngleDependence}

\section{Infrared radiometer}
The Heitronics KT19.82A Mark II is a radiometer designed for measuring the infrared radiation in  the transmission window between 8 and 14$\mathrm\mu$m [4]. We use it to measure the infrared radiation from the sky in its field of view of 2.9$^\circ$. By comparing the observed quantity to a blackbody spectrum, the radiometer then calculates the `radiative' temperature of the sky. It has been shown [3] that the measured sky temperature is very sensitive to the presence of clouds and water vapour which is crucial for determining the cause of a variation in the count rate of an IACT. Although clouds are not significantly warmer than the surrounding atmosphere, they are more effective emitters of blackbody radiation than the atmosphere in this wavelength range. If there are no clouds, the temperature still can vary from night to night due to relative humidity and temperature changes which may induce ice crystallisation on aerosols and therefore change the scattering phase function of Mie scattering.

Two of the planned four telescopes of H.E.S.S Phase 1 are presently operational and on each of them a radiometer is installed paraxially to provide an immediate means of cloud detection in the field of view of the camera. Figure 1 shows the detection of the clearance of the sky after a period of high cloud. Furthermore a scanning radiometer is installed to give the shift crew an immediate overview of the sky for any presence of clouds or approaching weather fronts.  

In addition to detecting clouds, the radiometer data can
be used to determine the amount of water vapour contained in the atmosphere,
a quantity on which the transmissivity of the latter for Cherenkov light depends.
Such a measurement is not trivial. 
The temperature measured by the radiometer depends on several parameters:
the temperature and water vapour profile of the atmosphere, the observing zenith angle, 
and the material of the window used to protect the instrument from the weather.
Semi-empirical models like the one from Idso [5] try to relate the infrared flux detected by the radiometer to the temperature and 
the water vapour pressure measured at ground level in a quantitative way. 
Indeed, we have measured such a correlation in our data, but nevertheless this model is not
satisfactory and a suitable one has yet to be found.

\subsection{Zenith angle and window material dependencies}
The temperature measured with the radiometer for a clear sky increases with the zenith angle due to a thicker section of the warm atmosphere being sampled [3]. 
In figure 2 one can see the zenith angle dependence for different window materials in front of the radiometer lens. This window protects the lens of the radiometer from weather, but as it emits in the infrared, its influence on the measured value of the radiometer is quite significant, depending on the chosen material. As one can see it not only increases the measured temperatures, but also alters the sensitivity ($T(\theta_{max})-T(\theta_{min})$). For this reason, the thin polyethylene film, whilst less robust than Cleartran\texttrademark, is the chosen material for the protective window. 

Moreover, the parametrization of the zenith angle dependence of the radiometer measurement can provide a differential estimate of the water vapour content of the atmosphere, which can in turn be related to its profile.

\section{Conclusion}
To conclude, we can say that the atmospheric monitoring instruments for the HESS experiment 
are now installed and running. Their observations have yet to be understood in some details and exploited to a fuller degree in order to let them allow us to estimate the actual transmissivity of the atmosphere to Cherenkov light to a better extent than so far. 

\section{References}
\vspace{\baselineskip}
\re
1.\ Aye et.al.,\ \emph{Implications of LIDAR observations at the H.E.S.S. site in Namibia
for energy calibration of the atmospheric Cherenkov telescopes}, this conference
\re
2.\ Bernlohr K.\ 2000, Astroparticle Physics 12, 255-268
\re
3.\ Buckley et.al.\ 1999, Experimental Astronomy 9, p237-249
\re
4.\ http:\slash\slash www.heitronics.de, \ Accessed 2003-05-13
\re
5.\ Idso S.B.\ 1981, Water Resources Research 17, p295-304
\re

\endofpaper

%% file: 008949-1.tex





\chapter{
Implications of LIDAR Observations at the H.E.S.S. Site in Namibia for
Energy Calibration of the Atmospheric Cherenkov Telescopes}

\author{
%
%
K.-M. Aye, P.M. Chadwick, C. Hadjichristidis, M.K. Daniel$^1$, I.J. Latham, 
R. Le Gallou, T.J.L. McComb, J.M. McKenny, S.J. Nolan$^2$, A. Noutsos, 
K.J. Orford, J.L. Osborne, and S.M. Rayner for the H.E.S.S. Collaboration\\
{\it Dept. of Physics, University of Durham, DH1 3LE Durham, United Kingdom}\\
{\small $^1$ now at Iowa State University USA.  $^2$ now at Purdue University USA}
}

\section*{Abstract}
Scattering values returned by a LIDAR installed at the 
H.E.S.S. site may be used for calculating optical depth profiles. One 
may then construct model atmospheres which fit these  using
the commercially available MODTRAN atmospheric radiation transfer
code. Examples of the results of Monte Carlo simulations of the
telescope response for two characteristic atmospheres are shown.

\section{Introduction}
As an instrument for ground-based astronomy the atmospheric Cherenkov 
telescope (ACT) is unusual in that the atmosphere is an integral part
of the detection system.  Its stucture, essentially the
pressure-altitude profile, controls the particle shower
development and the Cherenkov emission.  The atmospheric
attenuation, as a function of altitude and wavelength,
determines the probability of the Cherenkov photons reaching the
ACT. Monitoring the properties of the atmosphere is therefore
essential for the interpretation of the Cherenkov signal in
terms of the energy spectrum of the gamma-rays and source flux variation.
 The atmospheric attenuation in the 250 to 700 nm range depends on 
molecular absorption, Rayleigh scattering and Mie scattering by aerosols [1].
Mie scattering is expected to
be the most time-variable component.  We describe a LIDAR instrument
which gives information on this for 7.5 km above the ACT.  
The generation of model
atmospheres to fit the LIDAR measurements is still in
development but an example of the differences in ACT response
for two characteristic atmospheres is given in the final section.

\begin{figure}[t]
  \begin{center}
       \includegraphics[height=14.5pc]{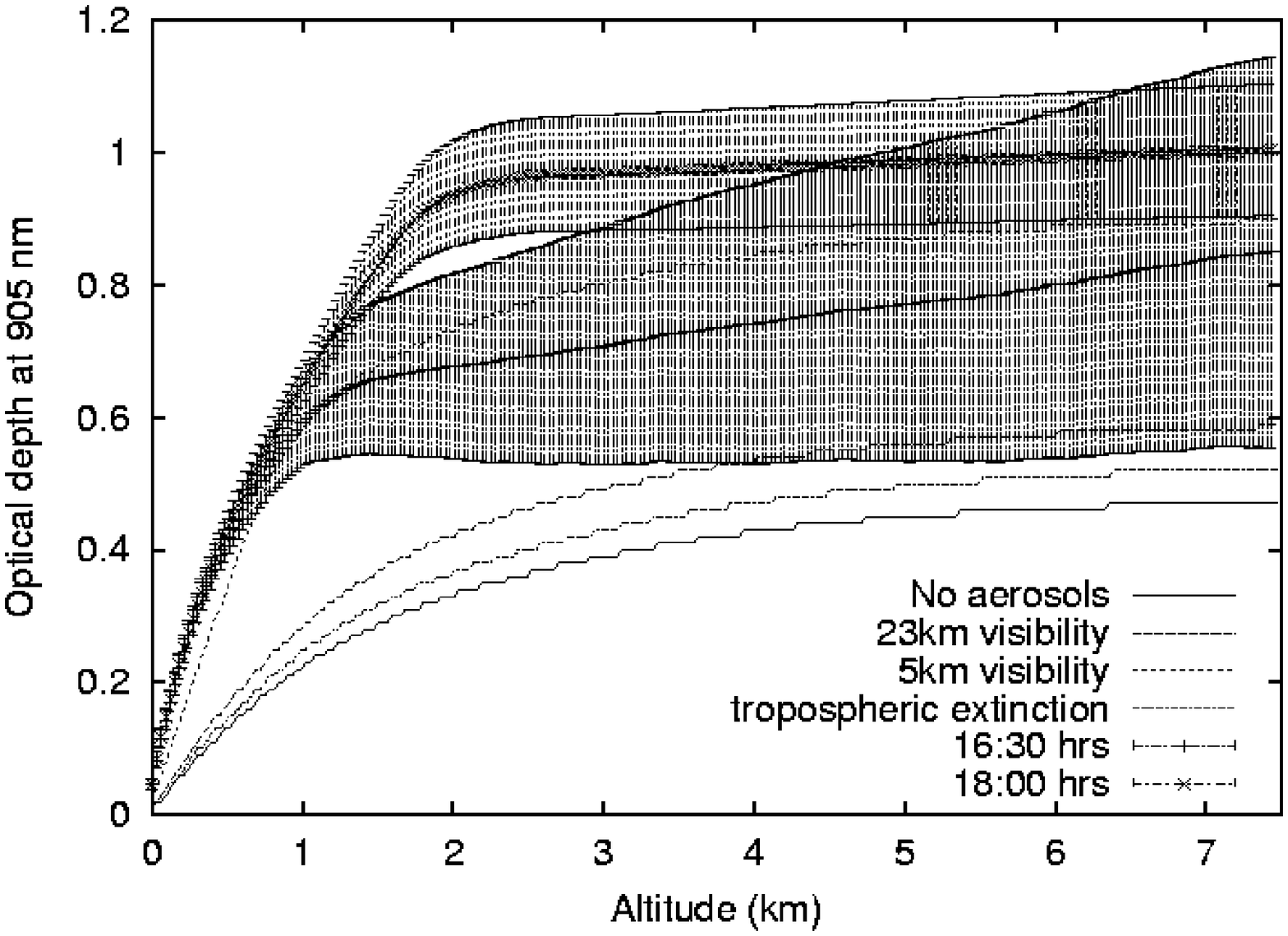}
 \end{center}
  \vspace{-0.5pc}
  \caption{Optical depth generated by the ceilometer backscatter data 
(the 2 graphs with large error bars) and calculated by MODTRAN for 
different visibilities}
\end{figure}

\section{The Ceilometer}
The Vaisala CT25K Ceilometer is a commercial LIDAR (LIght Detection 
And Ranging)
which measures cloud heights and vertical visibilities by sending out
light pulses at $(905\pm10)$nm with a pulsed InGaAs diode laser. As
the laser pulse traverses the sky, the resulting backscatter profile
signal strength over height is stored and processed with a sampling 
interval of 100 ns, resulting in a spatial resolution of 30m.

The general description of the instantaneous return signal strength is
given as the LIDAR equation
\begin{equation}
P_r(z)=E_0\frac{c}{2}\frac{A}{z^2}\beta(z)e^{-2\tau(z)}
\end{equation}
where $P_r(z)$ is the instantaneous power received from distance $z$,
 $E_0$ the effective pulse energy in Joules (taking
all optical attenuation into account, measured by internal 
monitoring),
$c$ the speed of light, $A$ the receiver aperture (in m$^2$sr), 
$\beta(z)$ the volume backscatter coefficient at distance $z$ 
(in m$^{-1}$sr$^{-1}$) and
$\tau(z)=\int_0^z\alpha(z')dz'$ the optical depth (OD) at
 $z$ produced by an attenuation coefficient $\alpha (z)$, the 
factor 2 accounting for the travel out and back.

\subsection{The Backscatter Coefficient}
The volume backscatter coefficient $\beta(z)$ is a result of combined
molecular (Rayleigh) and aerosol (Mie) scattering. Although the 
attenuation
$\alpha(z)$ is unknown, it can be assumed to
correlate the backscatter with the attenuation via the Lidar Ratio $k$: 
$\beta(z)=k\alpha (z)$ [3].
The Lidar Ratio can take values between 0.02 in high humidities to
0.05 in low humidities, but is mostly assumed to be 0.03 [2].
Figure 1 shows OD profiles which have
been generated from the backscatter data by summing the backscatter
values over the sampled time-bins like
$\tau(z)=\sum_z\frac{\beta(z)}{k}$.
Also plotted are model OD profiles generated by MODTRAN 4 [4] for a
default tropical atmosphere with differing aerosol profiles ranging from a
minimal OD for no aerosols included (solid line) to a rural profile
with 5 km visibility (dashed line inside the errors of the lower
ceilometer profile).

Fig.\,2 Shows preliminary OD profiles generated from backscatter data 
taken under clear sky conditions for 2 different values of the lidar ratio 
$k$ (0.05
and 0.02 respectively). These curves lie beneath the minimum
curve for the OD resulting only from Rayleigh scattering and molecular
absorption of Fig.\,1 (no aerosols). This is because the ceilometer's
internal algorithm's primary purpose is to identify significant 
backscatter from
cloud layers. Distances between 100 and 900 m from the receiver have 
the
optimum signal to noise for determining this and when no significant 
scatter is
detected the machine automatically sets scatter to zero in this 
region. This
drawback can be compensated for with the modelling done with the 
MODTRAN package, this is ongoing work and results are expected to be 
shown on the conference poster accompanying this paper.
Once out of the optimum signal to noise range the algorithm once again 
starts to give non-zero, albeit noisy, values.
Starlight extinction measurements and the observation of a calibrated 
light source on a distant hill will help resolve uncertainties in the 
backscatter coefficient.

\begin{figure}[t]
  \begin{center}
    \includegraphics[height=13.5pc]{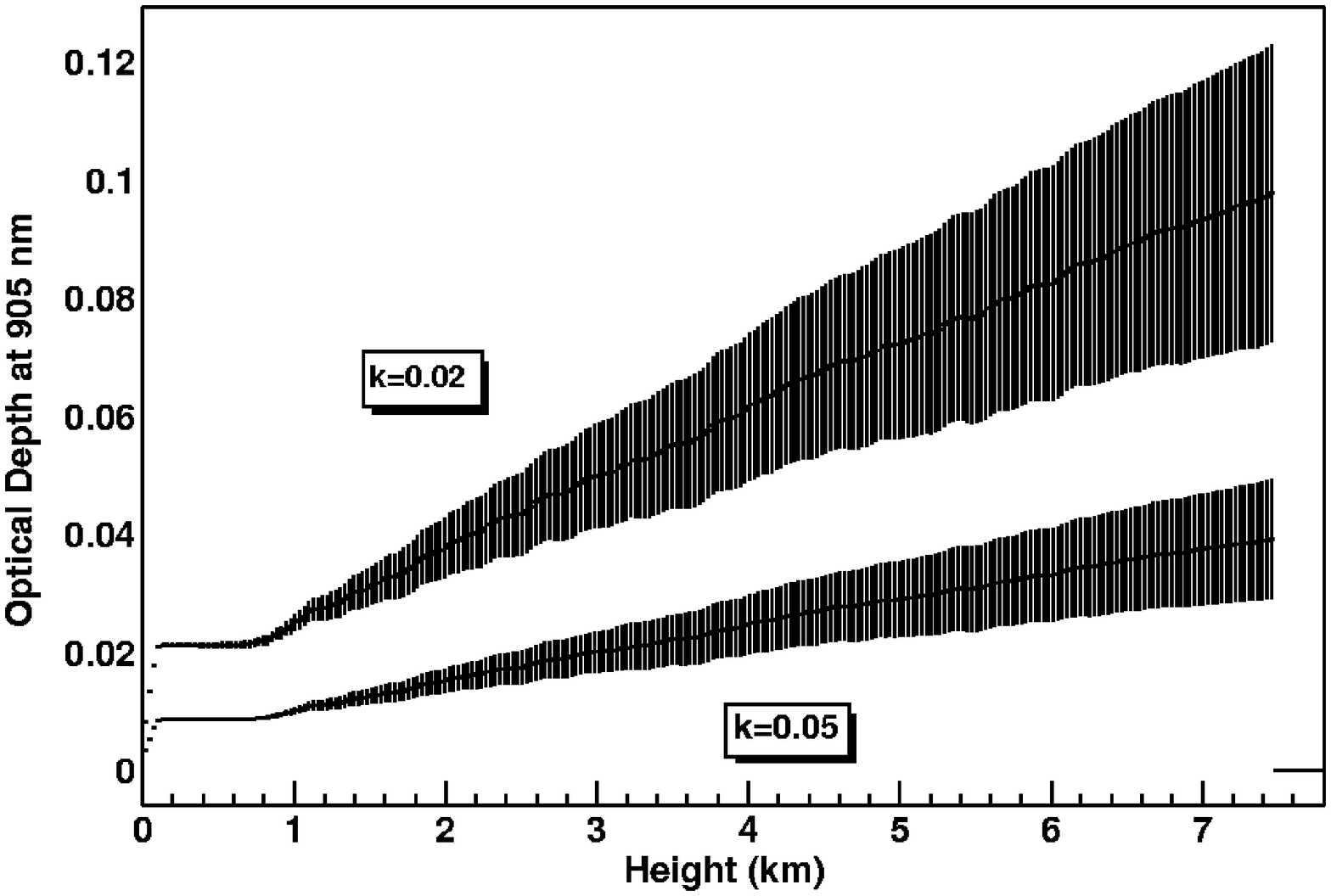}
  \end{center}
  \vspace{-0.5pc}
  \caption{Optical densities created from ceilometer backscatter data
on a very clear night for 2 different values of the lidar ratio}
\end{figure}

\section{Monte Carlo Simulations}
Radio sonde measurements at Windhoek show that the annual mean atmospheric
pressure profile is close to the MODTRAN `tropical' model.  Monte Carlo 
simulations using the MOCCA shower code and an adaptation of the CAMERAHESS
ACT simulation code  show that the seasonal variation 
from this model have a significantly smaller effect on the ACT response than
likely variations in the atmospheric attenuation.  
The current standard atmospheric attenuation model for Monte Carlo simulations of 
H.E.S.S. ACT response is a maritime haze aerosol model from 0 to 2 
km above sea level, a tropospheric spring-summer model for 2 to
10 km and a layer of default stratospheric dust.  This is
referred to as Atm.8 in figure 3. The zenith angle is typical of that for
observations of the Crab nebula. The site, being at an altitude
of 1.8 km, is above most of the model's maritime haze as was the
the HEGRA site on the island of La Palma, for which the model was
chosen.  As H.E.S.S. is on a plateau more than 100 km from the sea
this model could be regarded as being among the lower of the
levels of attenuation likely to be encountered.
A straightforward variant of this model 
is Atm.11 which puts the base of the 2 km of maritime haze at 1.8 km.
Both have the `tropical' structure. 
The reduction in effective area for triggering is shown in figure 3.

\begin{figure}[t]
  \begin{center}
    \includegraphics[height=20.0pc]{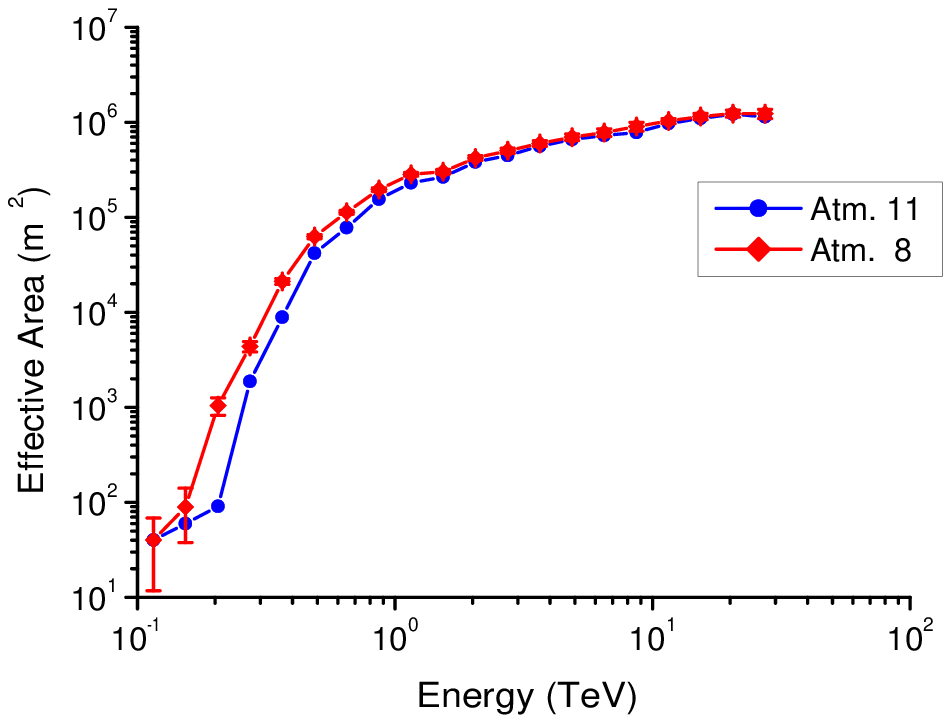}
  \end{center}
  \vspace{-0.5pc}
  \caption{The effective area of a single H.E.S.S. ACT for
triggering by gamma-rays incident at $50^\circ$ zenith angle for two
characteristic aerosol profiles.}
\end{figure}

\section{References}
\re
1.\ Bernl\"{o}hr K.\ 2000, Astropart. Phys. 12, 255
\re
2.\ Filipcic A., Horvat M.,  Veberic D., Zavrtanik D., Zavrtanik M.\ 
2003, Astropart.Phys. 18, 501-512
\re
3.\ Klett J.D.\ 1981, Appl. Optics 20, 211; 1985, ibid 24, 1638 
\re
4.\ MODTRAN\ patented US software, US Air Force Phillips Laboratories
\re

\endofpaper

%% file: 008971-2.tex
%
%
%
%



%


\chapter{
Optical Observations of the Crab Pulsar using the first H.E.S.S. 
Cherenkov Telescope
}

\author{
A. Franzen,$^1$ S.Gillessen,$^1$ G.Hermann,$^1$ and J. Hinton$^1$ \\
for the H.E.S.S. Collaboration\\
{\it (1) Max-Planck-Institut f\"ur Kernphysik, PO Box 103980, D-69029 Heidelberg, Germany}\\
}

\section*{Abstract}

For the understanding of the mechanisms of particle acceleration in 
pulsars, it is necessary to determine the high energy cutoff of the 
pulsed emission. In order to derive upper limits (or detections) on 
the pulsed emission in the TeV energy range using Cherenkov 
telescopes, typically data taken over periods of months or even years 
are superimposed. It is therefore necessary to use a time capture 
system and analysis tools which provide a base for pulsar phase analysis
which is stable over a time-scale of years. We have built a device consisting 
of a photomultiplier tube with a fast current digitization system and 
components of the timing system of the H.E.S.S. experiment, 
which allows to measure pulsed optical emission making use of 
the large mirror area of Cherenkov telescopes.
The system was installed into the first H.E.S.S. Cherenkov 
telescope in January 2003, where data was taken over 8 nights on
the Crab and Vela pulsars. Optical pulsation from the Crab pulsar
could be measured with observation times as short as a second. This system 
can therefore be used to determine the pulsar phase and to monitor
the short term and long term stability of the timing system of the
H.E.S.S. experiment.

\section{Introduction}

The new generation of ground based gamma-ray experiments currently 
under construction will provide improved sensitivity and thresholds
in the sub-100 GeV region. One of the science goals of these experiments
(H.E.S.S., MAGIC, CANGAROO, VERITAS) is to investigate the upper energy 
range for pulsed emission from EGRET pulsars.

Detections or upper limits on pulsed emission will likely involve the
combination of data taken over months and even years.  It is therefore
necessary to prove the long term stability of the timing systems used
by these experiments and also the proper performance of the analysis
software. It should also be noted that contemporary ephemerides are
not always available and that many of the pulsars of interest suffer 
frequent glitches. There is, therefore, a clear motivation for 
regular optical monitoring of pulsars by gamma-ray observatories.

The goal of the experiment described here was to develop and 
demonstrate an apparatus for fast timing measurement of optical signals
to be used for monitoring the timing system of the H.E.S.S. experiment.
Previous optical fast timing measurements are described in [1,2]. 

\section{Experimental Method}

The H.E.S.S. Experiment is an array of four Cherenkov telescopes
with 15 m focal length and 107~m$^{2}$ mirror area, 
located in Namibia ($23^{\circ}16'$ S, $16^{\circ}30'$ E, at 1800~m).
The optical properties of the H.E.S.S. telescopes are described in [3].

The experiment described here utilized the first 
complete H.E.S.S. telescope and a custom-built detector 
installed on the closed lid of the Cherenkov camera.
A silver coated plane mirror mounted at $45^{\circ}$ to the telescope axis
was used to place a lid-mounted photomultiplier tube (PMT) 
in the centre of the focal plane of the primary mirror. An aperture stop
limited the field of view of the PMT to 23 mm (equivalent to $5'$),
matched conservatively to the point-spread-function (PSF) of the
telescope. The average spectral sensitivity (quantum efficiency 
$\times$ reflectivity of primary and secondary mirrors) in the 
300-600 nm range was 0.11.

The PMT (Photonis XP2960 with passive base)
signal was read out via a shaping amplifier (FWHM of response
100$\,{\mu}$s) and digitized using a 16-bit ADC, sampling at 20~kHz.
Each sample is accompanied by a timestamp derived from a GPS
clock (Meinberg 167, precision $<1\,\mu$s) via a custom-built VME counter module.
The anode signal was DC coupled in order to easily measure background
light levels as well as pulsation. See Figure~1 for details.

\begin{figure}[t]
\parbox{0.72\linewidth} {\centering     \includegraphics[height=12.5pc]{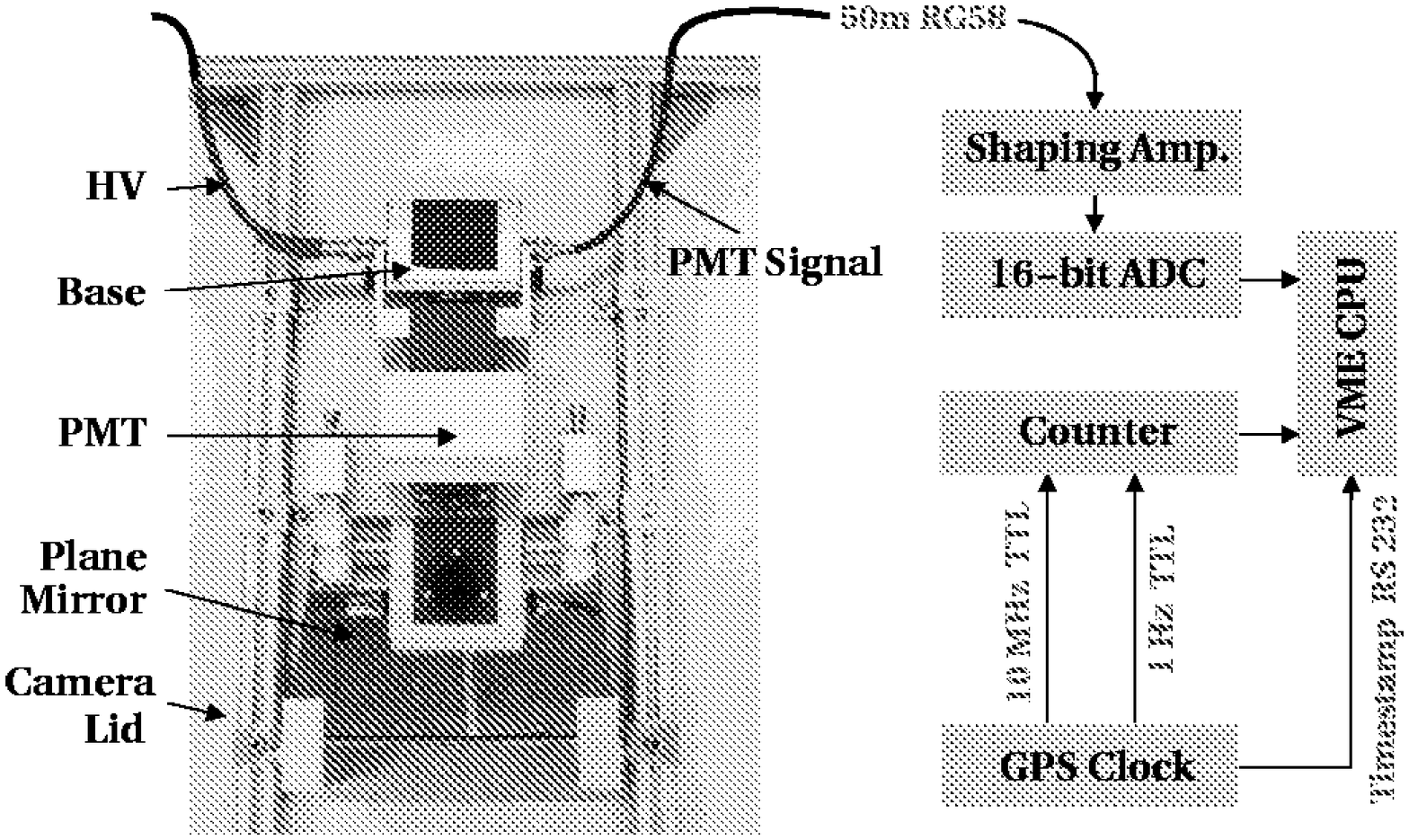}}
\parbox{0.27\linewidth} {  \caption{\re Mechanical and electronic setup of the experiment.}  }
\vspace{-0.5pc}
\end{figure}

\section{Measurements}

To achieve the required tracking precision of $\sim 1'$ it
was necessary to make online corrections for atmospheric 
refraction and bending in the arms of the telescope.
To verify the absolute pointing of the instrument, several
offset pointing runs and drift scans were made of bright stars
(with an 0.01 neutral density filter in place). 
These runs verified that $>80$\% of the light in the PSF
was collected onto the PMT.
Observation runs were made in a cycle of HV off (to monitor 
the ADC pedestal position) and on-source runs. Several empty
fields were also observed as background references.
The typical PMT anode current was 2-6 ${\mu}$A depending on the
region observed.

After removing data compromised by unstable weather conditions,
6~hours of data on the Crab (spread over 8 nights from the
21- 29 January) remain.
10~hours of data were obtained on the Vela pulsar.

\section{Results and Discussion}

The first step of the analysis is to barycentre the 
time associated with each sample using the standard H.E.S.S. software scheme.
The barycentred time is then folded using radio ephemerides from the Jodrell Bank Observatory~[4] 
(with frequency and first derivative from 15.1.03 and second derivative from the period 12.02-2.03).
 After a phase position
is established, 4 consecutive samples are summed to provide independent
current measurements for the light-curve. The data are then split into 
10-second slices, each of which is tested for stability by examining the RMS
of the DC signal. 
Figure~2 shows the phasogram extracted from a single 10 second slice. 
The expected double-peaked structure is clearly visible on top of a large DC background.

\begin{figure}[t]
\parbox{0.72\linewidth} {\centering     \includegraphics[height=13pc]{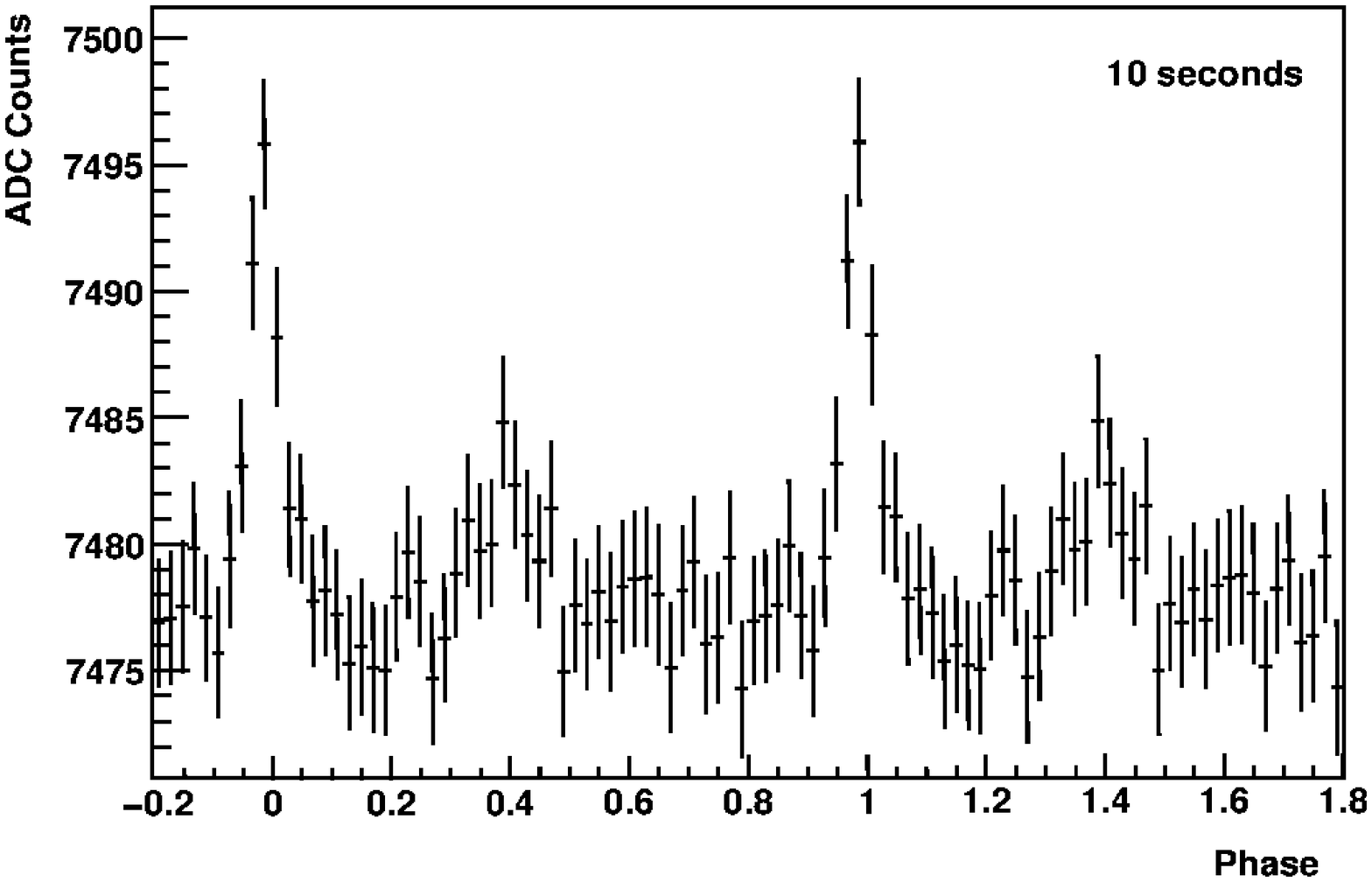}}
\parbox{0.27\linewidth} {  \caption{Optical$\,$ signal from the Crab pulsar 
from a 10 second observation} }
\vspace{-0.5pc}
\end{figure}

The sum of slices passing the RMS cut (with the DC component subtracted) is shown 
in Figure~3. The position, width, and relative heights of the two peaks
are consistent with the measurement made by the Hubble Space Telescope [1].
The significance of the pulsation is $\approx 4\sigma / \sqrt{t/\mathrm{seconds}}$.
Each complete phase contains $\approx 800$ photoelectrons, seen against a background 
of $\approx 2\times 10^{6}$ pe. 
The phase position of the mean peak can be determined with a precision of $30\,\mu$s  
in $\sim$30 minutes and is stable over the period of observations 
(see Figure~4 right).

\begin{figure}[t]
\parbox{0.6\linewidth} {\centering     \includegraphics[height=13pc]{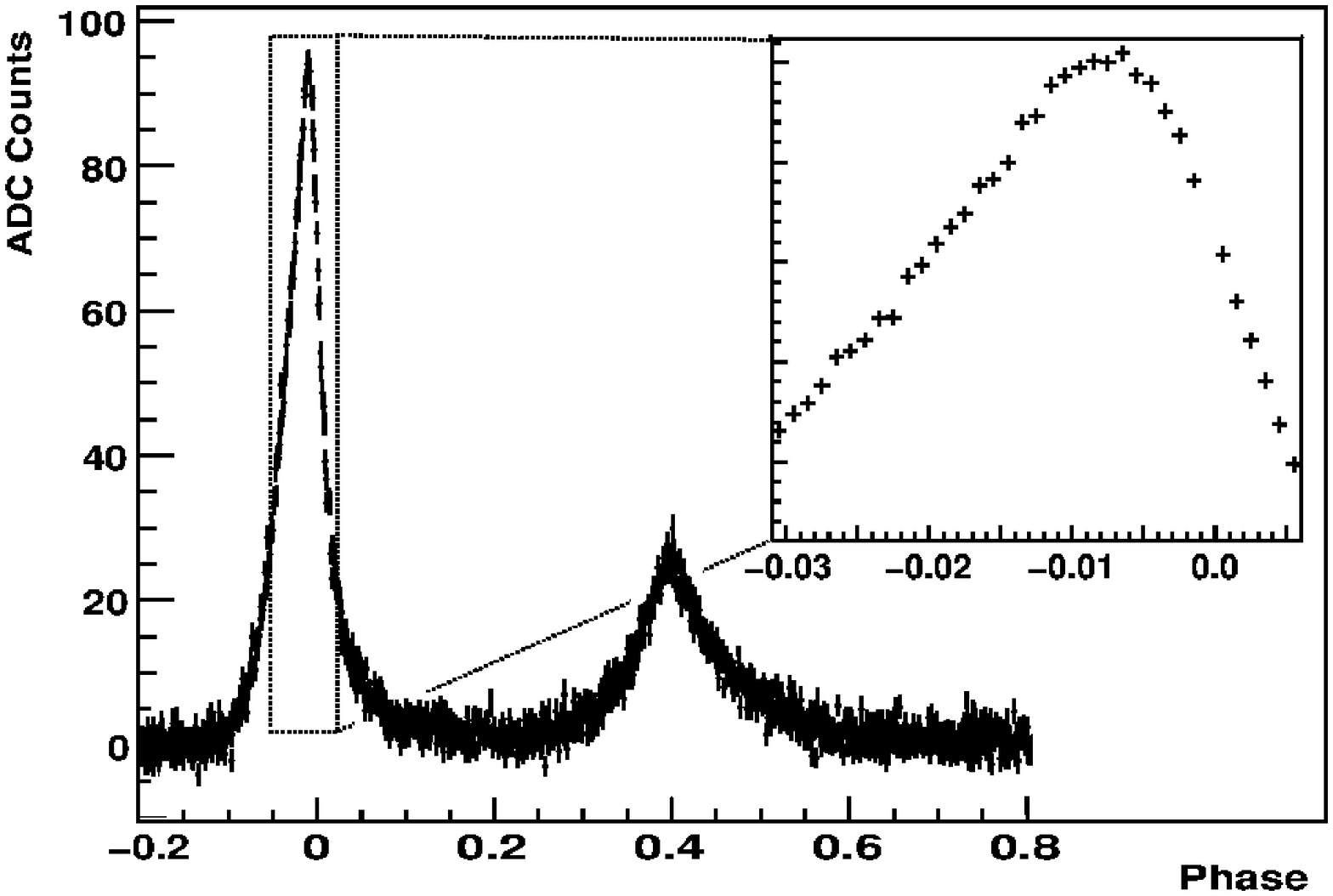}}
\parbox{0.4\linewidth} {\centering     \includegraphics[height=12pc]{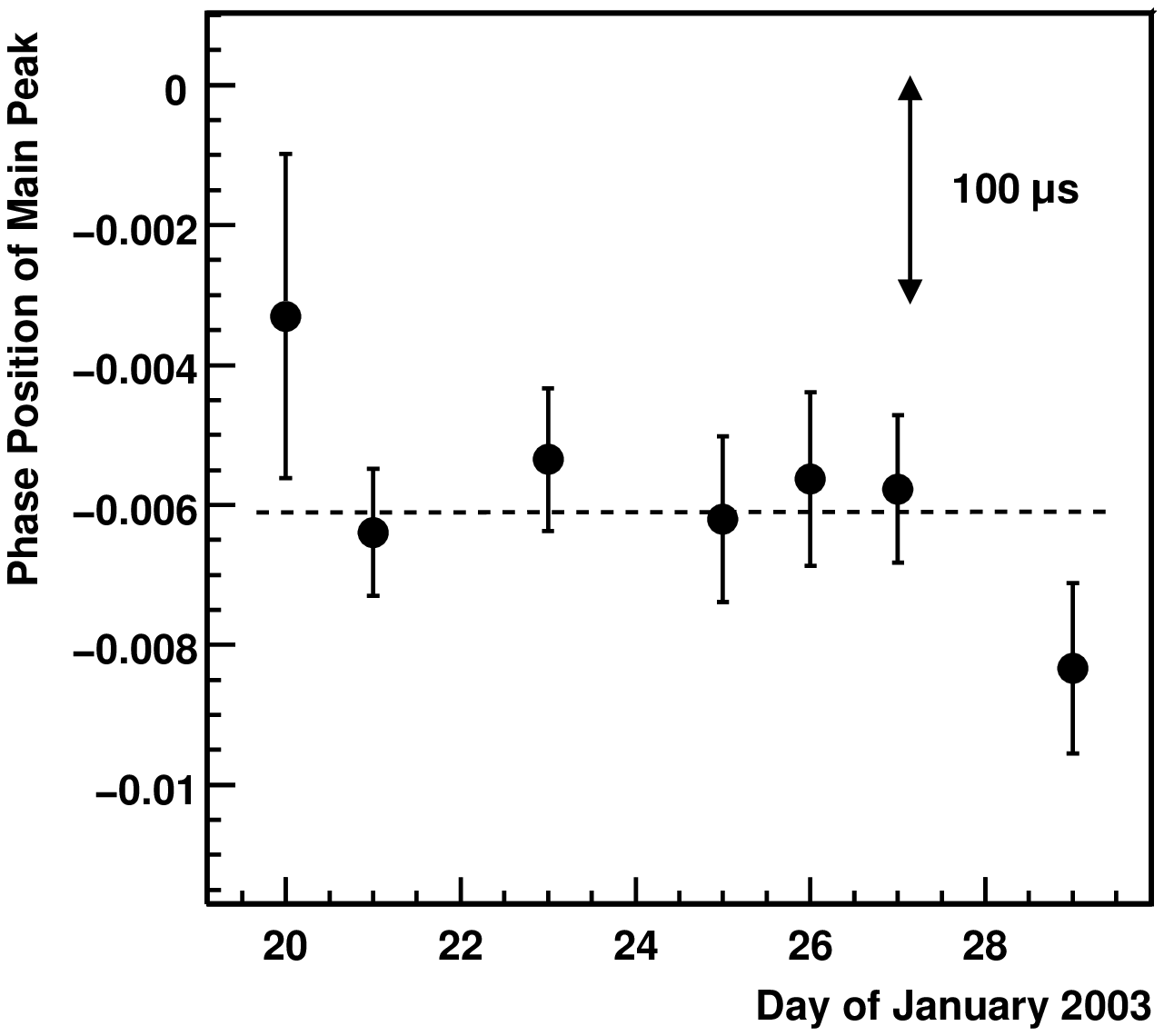}}  
\vspace{-0.5pc}
\caption{Left: the Crab pulsar phasogram extracted from the full data-set of 3~hours
of best quality observations. Right: The position of the main peak as a function of 
time during the measurement. The mean position of $-0.0061\pm 0.0004$ is consistent with
the result of $-0.0058$ given by [2].} 
\vspace{-0.5pc}
\end{figure}

Observations of the Vela pulsar were also made during this period, however,
with the 10~hours obtained, no significant pulsation was observed (using 
ephemerides from [5]). We estimate $\sim 30$ hours would be required for a
$5\sigma$ detection with this instrument.

\section{Outlook}

This prototype device has demonstrated the utility of an optical 
monitoring device for high energy gamma-ray experiments. We are 
currently developing an improved device with multiple light sensors
(for the exclusion of optical transients, for example meteorites) and 
possibly higher quantum efficiency light sensors.
A search for optical Giant Pulses in our data-set is under way.
 
\vspace{\baselineskip}
We would like to thank our technical support staff in Namibia, 
T. Hanke, E. Tjingaete and M. Kandjii, for their excellent support
and D. Lewis for providing ephemerides for the Vela pulsar.

\section{References}

\re 1.\ Percival J. W.\ et al.\ 1993, ApJ 407, 276
\re 2.\ Straubmeier C.\ et al.\ 2001, Exp. Astron. 11, 157
\re 3.\ Cornils R.\  et al.\ 2003, Submitted to Astropart. Phys.
\re 4.\ Jodrell Bank pulsar group page, {\it http://www.jb.man.ac.uk/$\sim$pulsar/}
\re 5.\ Lewis, D.\ (University of Tasmania), private communication.

\endofpaper